\documentclass[12pt]{article}
\usepackage{amsmath}
\usepackage{mathtools}
\usepackage{amssymb}
\usepackage{mathrsfs}
\usepackage{graphicx}
\usepackage{appendix}
\usepackage{hyperref}
\numberwithin{equation}{section}
\allowdisplaybreaks

\def\be{\begin{equation}}
\def\ee{\end{equation}}
\def\ba{\begin{eqnarray}}
\def\ea{\end{eqnarray}}

\begin{document}
\begin{flushright}
\mbox{\texttt{UTTG-16-18}}
\end{flushright}
\author{Jacques Distler\thanks{distler@golem.ph.utexas.edu}, Raphael Flauger\thanks{flauger@physics.ucsd.edu} and Bart Horn\thanks{bhorn01@manhattan.edu}}
\title{\bf\Large Double-soft graviton amplitudes\\ and the extended BMS charge algebra}
\date{$^{*}$ \normalsize\textit{Weinberg Theory Group, Department of Physics,\\ University of Texas at Austin, Austin, TX 78712, USA}\\
$^{\dagger}$ \normalsize\textit{High Energy Theory Group, Department of Physics,\\ University of California at San Diego, La Jolla, CA 92093, USA}\\
$^{\ddagger}$ \normalsize\textit{Department of Physics, School of Science, \\ Manhattan College, New York, NY 10471, USA}}

{\let\newpage\relax\maketitle}
\thispagestyle{empty}
\vskip 1.5cm
\begin{abstract}
We discuss how scattering amplitudes in 4d Minkowski spacetime which involve multiple soft gravitons realize the algebra of BMS charges on the null boundary.  In particular, we show how the commutator of two such charges is realized by the antisymmetrized consecutive soft limit of the double soft amplitude.  The commutator is found to be robust even in the presence of quantum corrections, and the associated Lie algebra has an extension, which breaks the BMS symmetry if the BMS algebra is taken to include the Virasoro algebra of local superrotations.  We discuss the implications of this structure for the existence of a 2d CFT dual description for 4d scattering amplitudes. 
\end{abstract}
\newpage
\tableofcontents
\thispagestyle{empty}
\newpage
\setcounter{page}{1}
\section{Introduction}

Asymptotic symmetries in gauge and gravitational theories have seen a resurgence of interest in recent years, both for studying the structure of cosmological observables and for investigating the formal structure of scattering amplitudes in field theory and gravity. Asymptotic symmetries are residual gauge or diffeomorphism symmetries of the gauge-fixed action that do not fall off at infinity, and since asymptotic symmetries do not leave the wavefunction invariant (i.e., they are spontaneously broken) they can lead to physical Ward identities involving the associated Goldstone bosons.  These identities constitute a generalization of soft-pion theorems for internal symmetries in field theory to the case of spontaneously broken spacetime symmetries, and the associated Goldstones are gauge bosons or gravitons.  For a general discussion of asymptotic symmetries and the construction of the associated Noether charges, see for instance \cite{Barnich:2001jy, Avery:2015rga, Banados:2016zim}.  

A specific application of this formalism is to the derivation of consistency relations for in-in correlation functions for cosmological perturbations, performed in unitary gauge in \cite{Hinterbichler:2013dpa} and in conformal Newtonian gauge in \cite{Horn:2014rta}.  Here the associated Ward identities of the residual symmetries are phrased in terms of relations between the soft limit of an $(N+1)$-point function on the one hand and a symmetry transformation acting on an $N$-point function on the other.  In the soft momentum limit a Goldstone boson will become locally indistinguishable from an asymptotic symmetry transformation, and can therefore be transformed away.  Schematically the Ward identities take on the form
\begin{equation}
\langle \left[Q, \mathcal{O}\right]\rangle = -i \langle \delta \mathcal{O} \rangle\,,
\end{equation}
where the charge $Q$ creates the soft Goldstone boson that realizes the nonlinear symmetry transformation, and $\delta$ denotes the part of the symmetry that acts linearly on observables.  Another choice of notation (see for instance \cite{Strominger:2013jfa, He:2014laa}) which we will follow is
\begin{equation}
\langle \left[Q_{S}, \mathcal{O}\right]\rangle = -\langle \left[Q_{H}, \mathcal{O}\right]\rangle\,,
\end{equation}
where $Q_{S}$ creates the soft mode realizing the nonlinear part of the symmetry, and $Q_{H}$ is the linear transformation acting on the hard modes.  Strictly speaking, $Q_S$ is not well-defined for spontaneously broken charges since it is not normalizable, but its commutator with local operators is.  The full charge $Q = Q_{S} + Q_{H}$ then commutes with the operator $\mathcal{O}$.

It was recently shown in \cite{Strominger:2013jfa, He:2014laa} that Weinberg's soft graviton theorem for scattering amplitudes \cite{Weinberg:1965nx} arises as the Ward identities of the BMS symmetries \cite{Bondi:1962px, Sachs:1962wk, Sachs:1962zza} of asymptotically Minkowski spacetimes, with the soft graviton playing the role of the Goldstone boson.  This was shown to hold at subleading order in the soft momenta as well \cite{Kapec:2014opa}, and has been further generalized to include asymptotic gauge and fermionic symmetries \cite{Strominger:2013lka, He:2014cra, Dumitrescu:2015fej} and to the scattering of massive particles \cite{Campiglia:2015kxa}, which travel out to timelike infinity.  A more comprehensive and pedagogical review of these ideas can be found in \cite{Strominger:2017zoo}.  The BMS symmetries enlarge the Poincar\'e algebra to an infinite-dimensional algebra consisting of supertranslations and superrotations, and it remains to be fully understood whether they contain novel information about the structure of the gravitational S-matrix in flat space, or whether they repackage the known symmetry content of the theory in an illuminating way.  In \cite{Hawking:2016msc} it was proposed that the soft charges mediate transitions between degenerate vacua in quantum gravity and may help resolve the problem of information loss in black hole evaporation.\footnote{See however \cite{Mirbabayi:2016axw} for a discussion of why the soft modes may be insufficient to encode the information loss at leading order.} BMS symmetry has also been studied as the starting point for defining a holographic dual to Minkowski space which would live on the null boundary (see for instance \cite{Arcioni:2003xx, Dappiaggi:2005ci, Bagchi:2010eg, Bagchi:2012cy} for early works on this subject).  Further evidence for a 2d CFT structure dual to the 4d scattering amplitudes was found e.g. in  \cite{Lipstein:2015rxa, Kapec:2016jld, Pasterski:2016qvg, Cheung:2016iub}.  It is fair to say, however, that whether it is possible to have a well-defined holographic theory living on the null boundary, and how such a theory dual to Minkowski space should behave, is still not well understood.

In the current work our goal is to understand how the asymptotic BMS symmetry algebra is realized in terms of the scattering amplitudes.  This generalizes the work of \cite{Strominger:2013jfa,He:2014laa, Kapec:2014opa} to amplitudes where more than one graviton is taken to be soft, and a particular combination of soft limits corresponds to the commutator of the BMS charge algebra.  The general structure of the BMS algebra and the corresponding Dirac bracket in three and four dimensions was analyzed by studying the form of the classical symmetry transformations and charges in \cite{Barnich:2011mi} (see also \cite{Barnich:2017ubf}), and it was found that while the global subalgebra in 4d has no central charges, there is a nontrivial extension of the classical algebra by a generalized 2-cocycle when the BMS algebra is promoted to include local (singular) superrotations.  The extension term breaks the symmetry, similar to the breaking of conformal invariance by a nonzero central charge.  In the current work we will show how the symmetry algebra at null infinity is realized in the language of scattering amplitudes as a particular limit of the double soft amplitude.  (See also \cite{Anupam:2018} for previous work relating the double consecutive soft amplitude to nested Ward identities, in which many similar issues are discussed.)   What makes this more subtle than the single-soft case is that the Goldstones themselves are charged under the symmetry; therefore, transforming away one soft mode will shift the second as $i\delta_1 Q_{2S} = \left[Q_{2S},Q_{1H}\right] \neq 0$, and this shift needs to be accounted for when transforming away the second soft mode\footnote{This is already familiar from the case of two soft pions -- see Appendix \ref{softpions} for a review.}.  We will see that this shifting of the single-soft amplitudes is crucial for realizing the asymptotic BMS algebra, and is related to the structure of contact terms between single-soft factors that arise in the antisymmetrized consecutive double-soft limit.

Our main results can be written schematically in terms of the S-matrix elements as 
\begin{equation}
\begin{split}
\lim_{[\omega_2 \to 0}\lim_{\omega_1 \to 0]}\sum\limits_{\lambda_1,\lambda_2}\int d^2 z_1d^2z_2\Psi_1(q_1)\Psi_2(q_2)\langle out \, q_1,q_2 | \mathcal{S} | in \rangle 
&=\langle out | [ Q_{\left[1,2\right]} + K_{(1,2)}, \mathcal{S} ] | in \rangle\,,
\end{split}
\end{equation}
where the $q_{1,2}$ collectively denote the energies $\omega_{1,2}$, the directions $z_{1,2}$ and the helicities $\lambda_{1,2}$ of the soft gravitons, and the weights $\Psi_{1,2}$ are appropriately chosen for the BMS transformations corresponding to $Q_1$ and $Q_2$. $Q_{\left[1,2\right]}$ refers to the charge associated with the commutator in the unextended BMS algebra, and $K_{(1,2)}$ contains the extension, which agrees to leading order with the expression found in \cite{Barnich:2011mi}. In general, $K_{(1,2)}$ is non-zero and does not commute with $\mathcal{S}$ if one of the transformations is a supertranslation and the other is a singular superrotation, but vanishes otherwise. We will perform the calculation at the level of scattering amplitudes, and also at the level of the commutators of the charge operators.  The term $K$ transforms under the BMS algebra and satisfies the generalized cocycle condition 
\begin{equation}
\begin{split}
i[Q_{3}, K_{(1,2)}] + K_{(\left[1,2\right],3)} + (\mbox{cyclic permutations}) = 0\,,
\end{split}
\end{equation}
so the Jacobi identity is satisfied and the BMS algebra has a nontrivial extension.



This paper is organized as follows: in the next section we will review the form of the BMS transformations and the structure of the algebra.  
In Section \ref{singleSoft} we review and rederive the definition of the integrated charges and the connection between the BMS Ward identities and the single-soft graviton theorems, and in Section \ref{chargeAlgebra} we demonstrate step by step how the BMS algebra is realized in the double-soft graviton amplitudes.  The results of this section, which comprise the main results of the paper, are summarized in \ref{summary}.  We discuss the possibility of relating these asymptotic charges to local currents and operators in a dual picture in Section \ref{Jacobi}, although we stress that we still do not know whether we have the necessary ingredients for an understanding of flat space holography.  We conclude and indicate further directions in Section \ref{discussion}, and compare the soft pion and asymptotic Yang-Mills calculations in the Appendices.


\section{BMS transformations and algebra}\label{BMS}

The BMS transformations (named for Bondi, van der Burg, Metzner and Sachs \cite{Bondi:1962px,Sachs:1962wk,Sachs:1962zza}) arise as residual diffeomorphism symmetries of asymptotically flat spacetime in Bondi gauge which do not fall off at infinity.  While the metric may be quite complicated in a localized spatial region, we will assume that it looks like Minkowski at large $r$, and the Penrose diagram is therefore the same as for Minkowski space.  The ${\rm BMS}^{+}$ symmetries apply near the future null boundary $\mathscr{I}^{+}$, and there is a corresponding set of ${\rm BMS}^{-}$ symmetries associated with the past null boundary $\mathscr{I}^{-}$.  We will focus on ${\rm BMS}^{+}$ in what follows; although the actual symmetry operating on gravitational scattering amplitudes is the diagonal subgroup of ${\rm BMS}^{+} \times {\rm BMS}^{-}$ \cite{Strominger:2013jfa}, the generalization to the appropriate linear combination of symmetries involving the full null boundary is straightforward. Near $\mathscr{I}^{+}$, we can write the Minkowski metric in the advanced coordinates $\left\{u, r, z , \bar{z}\right\}$ as
\begin{equation}
ds^2 = -du^2 -2du dr + 2 r^2 \gamma_{z\bar{z}}dz d\bar{z}\,,
\end{equation}
where $\gamma_{z\bar{z}} = \frac{2}{(1+z\bar{z})^2}$ is the round metric on the sphere.  Allowing fluctuations around this metric, Bondi gauge is defined by 
\begin{equation}
g_{rr}=0\,, \qquad g_{rz}=0\, , \qquad g_{r\bar{z}}=0\,, \qquad \det g_{AB} = 4r^4 \gamma^2_{z\bar{z}}\,.
\end{equation}
The first three conditions ensure that outgoing radial trajectories are geodesics for massless particles, and the final condition links the radial coordinate to the volume of the 2-sphere.  Bondi gauge is well adapted to studying the interaction of gravitational radiation with an isolated system in an otherwise flat space, for which it was originally developed. 

The metric is also required to satisfy certain asymptotic flatness conditions, which keep the metric close to Minkowski up to corrections at higher order in a $1/r$ expansion\footnote{A more coordinate invariant definition of asymptotic falloff conditions exchanges the coordinate expansion in $1/r$ to powers of the scalar function $\Omega$ which appears in the formal definition of the conformal compactification.  A little work shows that the specific coordinate choice above, which is much more convenient for calculations, is in fact equivalent -- see for instance Chapter 11 of \cite{Wald}.}.  The exact definition of asymptotic flatness under consideration is not a gauge condition, but is an additional choice depending on the level and type of structure one wishes to consider.  In Bondi gauge, near the future null boundary we take the metric of an asymptotically Minkowski spacetime to leading order in metric perturbations to have the form (in the notation of \cite{Bondi:1962px,Sachs:1962wk,Sachs:1962zza, Barnich:2011mi})
\begin{equation}
ds^2 = -e^{2\beta}\left(\left(1-\frac{2m}{r}\right)du^2 + 2dudr\right) -2U_{A}dx^{A}du + g_{AB}dx^{A}dx^{B}
\end{equation}
where to $\mathcal{O}(1/r^2)$, the corrections have the form
\begin{equation}
\begin{split}
e^{2\beta} &= 1 - \frac{1}{16r^2} C_{AB}C^{AB} + \cdots\\
U_{A} &= -\frac{1}{2}D^{B}C_{AB} - \frac{2}{3r}\left(\frac{1}{4}C_{AB}D_{C}C^{BC} + N_{A}\right) + \cdots\\
g_{AB} &= rC_{AB} +r^2 \gamma_{AB} + \frac{1}{4}C_{CD}C^{CD}\gamma_{AB} + \cdots
\end{split}
\end{equation}
The form of the metric is fixed by the gauge and flatness conditions, and we have also applied the constraint equations to derive the form of $U_{A}$.  The quantities $m, N_{A}$ are respectively the Bondi mass and the Bondi angular momentum, and $N_{AB} = \partial_{u}C_{AB}$ is the Bondi news, which is related to the energy carried out to null infinity by gravitational radiation.  Here and afterwards, raised indices $A,B$ mean raised only with the round metric $\gamma_{z\bar{z}} = \frac{2}{(1+z\bar{z})^2}$ on the two-sphere, and $D_{A}$ refers to the covariant derivative with respect to $\gamma_{z\bar{z}}$.  A similar parametrization holds for the metric perturbations around $\mathcal{I}^{-}$, and appropriate matching conditions between $\mathcal{I}^{+}$ and $\mathcal{I}^{-}$ can be defined (see for instance \cite{Strominger:2013jfa, Mirbabayi:2016xvc}).


Although the gauge condition does not allow transformations $x^{\mu} \to x^{\mu} + \xi^{\mu}$ which fall off at infinity, there are residual symmetries which consist of diffeomorphisms $x^{\mu} \to x^{\mu}+\xi^{\mu}$ that do not fall off at infinity.  The gauge conditions restrict them to have the form
\begin{equation}
\begin{split}
\xi^{u} &= f\,, \qquad \xi^{A} = Y^{A} - \frac{1}{2r}D^{A}f + \frac{1}{2r^2}C^{AB}\partial_{B}f\,, \\
\qquad \xi^{r} &= -\frac{1}{2}r D_{A}\xi^{A} + \frac{1}{2r}U^{A}\partial_{A}f \\ &\approx -
 \frac{1}{2}rD_{A}Y^{A} + \gamma^{z\bar{z}}\partial_{z}\partial_{\bar{z}}f - \frac{1}{4r}C^{AB}D_{A}D_{B}f + \frac{1}{r}\gamma^{AB}U_{A}\partial_{B}f\,,
\end{split}
\end{equation}
where $A$ runs over the complex spherical coordinates $\left\{z, \bar{z}\right\}$, $D$ is the covariant derivative with respect to the spherical metric $\gamma_{z \bar{z}} = \frac{2}{(1+z \bar{z})^2}$, and $f$, $Y^{A}$ depend on the coordinates $\left\{u, z, \bar{z}\right\}$.


The BMS symmetries are further required to obey the asymptotic falloff conditions at large $r$; that is, they must preserve the form of the asymptotic Minkowski metric above.  Equivalently, the BMS symmetries are asymptotic solutions to Killing's equation, meaning that they satisfy $\mathcal{L}_{\xi}g_{\mu \nu}= 0$ up to a certain order in a $1/r$ expansion around Minkowski space.  Requiring that the BMS transformations preserve this form of the metric further restricts $Y^{A}$ to be a conformal Killing vector on the sphere, and $f, Y^{A}$ to have the form  
\be
f = T(z, \bar{z}) + \frac{1}{2}u D_{A}Y^{A}\,, \qquad Y^{z} = Y^{z}(z)\,, Y^{\bar{z}} = Y^{\bar{z}}(\bar{z}).
\ee
The $T(z,\bar{z})$ piece is called a supertranslation, and the $Y^{A}$ piece is a superrotation.  Only the modes $Y^{z} \supset \left\{1, z, z^2\right\}$, $Y^{\bar{z}} \supset \left\{1, \bar{z}, \bar{z}^2\right\}$ are nonsingular on the sphere, and these define the global subalgebra of BMS.  Including the singular configurations, where the symmetry breaks down at a set of isolated poles on the sphere, the superrotations are enlarged to an infinite-dimensional Virasoro symmetry.  The physical significance of the local Virasoro symmetries is more subtle, but it was proposed in \cite{Strominger:2016wns} that they are related to topological transitions between asymptotically locally flat spacetimes with stringlike defects.


In order to derive the BMS algebra, we must remember that performing a transformation will alter the metric, which will backreact on any other asymptotic Killing vectors present.  The Lie bracket will therefore pick up improvement terms and is generalized to the Dirac bracket
\begin{equation}
\left\{ \xi^{\mu}, \xi^{\nu}\right\} = \left[\xi_1, \xi_2\right] - \delta^{g}_{\xi_{1}}\xi_{2} + \delta^{g}_{\xi_{2}}\xi_1\,,
\end{equation}
where $\xi$ is considered to be an implicit function of the metric, and $\delta^{g}_{\xi} \xi'$ is given by applying the chain rule and using $\delta^{g}_{\xi}g_{\mu \nu} = \mathcal{L}_{\xi}g_{\mu \nu}$.  The result is found in \cite{Barnich:2011mi} and is itself a BMS transformation with\footnote{We will refer to this as the ``commutator'' $\left[s_1, s_2\right]$ of the BMS transformations $s_1$ and $s_2$, since this is a well-defined Lie bracket structure for the algebra $BMS_{4}$ , but the bracket on the corresponding vector fields $\xi^{\mu}$ is the Dirac bracket.  Hopefully this will not lead to too much confusion.}
\be
T_{[1,2]} = Y_1^{A}\partial_{A}T_2 - \frac{1}{2}D_{A}Y_{1}^{A} T_2 - (1 \leftrightarrow 2)\,, \qquad Y^{A}_{[1,2]} = Y_1^{B}\partial_{B}Y_{2}^{A} - Y_2^{B}\partial_{B}Y_{1}^{A}\,.
\ee

Another prescription for extending the global BMS algebra is to consider the set of all smooth functions $Y^{A} \in C^{\infty}(S^2)$ \cite{Campiglia:2014yka}; this, however, does not preserve the same asymptotic falloff conditions and may therefore not be as well suited to the same physical situations, such as to symmetries of the S-matrix.  Another prescription is to apply the BMS formalism to the asymptotic symmetries of the near-horizon limit of a Schwarzschild black hole \cite{Donnay:2015abr, Donnay:2016ejv}.  Here the superrotations have the same Virasoro structure, but the form of the commutator between a  supertranslation and a Virasoro transformation is instead
\begin{equation}
T_{\left[1, 2\right]} = Y_{1}^{A}\partial_{A}T_2 - Y_{2}^{A}\partial_{A}T_1 \,.
\end{equation}
The associated group manifold in this case is parameterized by ${\rm SDiff}(S^2) \ltimes C^{\infty}(S^2)$, which is the semidirect product of volume-preserving diffeormorphisms of the two-sphere with the set of smooth functions living there.  This group also arises as the set of symmetries of a compressible fluid on the two-sphere and may be relevant for a deeper understanding of the membrane paradigm for black holes horizons~\cite{Penna:2017bdn}\footnote{We thank Robert Penna for discussions on this point.}.  The algebra for near-horizon BMS can also be extended to include a second set of supertranslation generators; see \cite{Donnay:2016ejv} for details.

Our goal is to explore how the algebra at null infinity, which arises from the specific asymptotic flatness prescriptions appropriate to this case, is realized in the language of soft graviton amplitudes.  First, however, we will review how the BMS transformations are related to soft graviton theorems by considering the single-soft theorem(s) as a warmup.


\section{Review of the single-soft limits}\label{singleSoft}

It was shown in \cite{Strominger:2013jfa, He:2014laa, Kapec:2014opa} that the Ward identities of the BMS symmetries are equivalent to the soft-graviton identities with one soft graviton.  We will review these calculations here, and for the most part we follow the same notation.  The difference between our discussion and \cite{Strominger:2013jfa, He:2014laa, Kapec:2014opa} is that we also explicitly expand the terms quadratic in the boundary data in terms of creation and annihilation operators, which as we will see generates the linear transformation of the hard modes.  

To $O(q)$ (which is NNLO or sub-subleading order) the amplitude for the emission of a soft graviton with momentum $q$ in an underlying hard process involving quanta with momenta $p_1\,\dots, p_n$ is given by
\begin{equation}
\begin{split}
\lim_{q \to 0}\bar{\epsilon}_{\mu \nu}\mathcal{M}^{\mu \nu}(q;p_1, \cdots, p_n) &= \sum_{k}\frac{\kappa}{2}\Bigg[\frac{(\bar{\epsilon} \cdot p_k)^2}{(p_k \cdot q)} + \frac{(p_k \cdot \bar{\epsilon})(\bar{\epsilon}_{\mu}q_{\nu}J^{\mu \nu}_{k})}{(p_k \cdot q)}\\ &+ \frac{1}{2}\frac{(\bar{\epsilon}_{\mu}q_{\nu}J^{\mu \nu}_{k})(\bar{\epsilon}_{\rho} q_{\sigma} J^{\rho \sigma}_k)}{(p_k \cdot q)} + \cdots\Bigg]\mathcal{M}(p_1, \cdots, p_n)\\
&= \left( S^{(0)}(q) + S^{(1)}(q) + S^{(2)}(q) + \cdots \right)\mathcal{M}(p_1, \cdots , p_n)\,.
\end{split}
\end{equation} 
Here $\bar{\epsilon}_{\mu\nu}\mathcal{M}^{\mu \nu}$, $\mathcal{M}$ refer to the amplitudes with and without the soft graviton, respectively, and $\kappa^2 = 32\pi G$.  We have written the graviton polarization tensor as $\bar{\epsilon}_{\mu \nu} = \bar{\epsilon}_{\mu}\bar{\epsilon}_{\nu}$ for a graviton of definite helicity, and $J_{k}^{\mu \nu} = \left(p_k^{\mu} \partial_{p_k}^{\nu}-p_k^{\nu} \partial_{p_k}^{\mu} + \Sigma_{k}^{\mu \nu}\right)$ is the angular momentum\footnote{Note that this is $-i$ times the usual angular momentum operator, which is Hermitian.  We will adopt this convention instead for the sake of convenience.}, which further decomposes into an orbital piece and a spin piece.  The derivatives and spin matrices act on the hard amplitude $\mathcal{M}(p_1, \cdots, p_n)$. We take all momenta to be outgoing, and the generalization to an arbitrary S-matrix amplitude follows simply by applying the LSZ formula and crossing symmetry.  In the last line we have written $S^{(0)}(q), S^{(1)}(q), S^{(2)}(q)$ to refer to the leading, subleading, and subsubleading parts of the soft factor respectively.  The leading term in the soft factor is gauge invariant by conservation of global energy-momentum, the subleading term is gauge invariant by the conservation of global angular momentum, and the subsubleading piece is automatically gauge invariant because $J_{k}^{\mu \nu}$ is anti-symmetric.  The above expression can be derived diagrammatically at tree level using gauge invariance and the graviton coupling to external lines \cite{Bern:2014vva}; loop corrections begin for generic momenta at $\mathcal{O}(q)$ and at $\mathcal{O}(1)$ in the collinear limit~\cite{Bern:2014vva, Larkoski:2014bxa}.

We will show that the leading and subleading parts of the soft graviton theorem imply the Ward identity
\begin{equation}
\langle out |\left[Q, \mathcal{S}\right] | in \rangle = 0\,,
\end{equation}
where as usual $\mathcal{S}$ is the operator whose matrix elements encode the S-matrix and $Q = Q_{S} + Q_{H}$ is the Noether charge associated with the asymptotic BMS symmetry.  The soft part of the charge operator creates a soft Goldstone boson associated with the symmetry (in this case, a graviton) and the hard part acts on the hard modes in the $| in \rangle$ and $\langle out |$ states.  In other words, the soft charge is the nonlinearly realized part of the spontaneously broken symmetry, and the hard charge is the linearly realized part\footnote{In the language of Noether currents and soft pions, as in Appendix \ref{softpions}, the soft charge contains the LSZ pole, and the hard charge contains the regular terms.}.  The general procedure for defining and computing the asymptotic charges is discussed in~\cite{Barnich:2001jy}, \cite{Avery:2015rga}, and the Noether current is integrated over initial and final Cauchy surfaces that are to be sent to $\mathscr{I}^{\pm}$.  Comparing the notation of \cite{He:2014laa} and \cite{Avery:2015rga} with the notation we are using here, the above expression becomes
\begin{equation}
\begin{split}
\langle out | \left[Q_{S}, \mathcal{S}\right] | in \rangle &= -\langle out | \left[Q_{H}, \mathcal{S}\right] | in \rangle\, = -i \langle out | \delta \mathcal{S} | in \rangle\, , \\
\langle out | (Q_{S}^{+}\mathcal{S} - \mathcal{S}Q_{S}^{-}) | in \rangle &= - \langle out | (Q_{H}^{+}\mathcal{S} - \mathcal{S}Q_{H}^{-}) | in \rangle \qquad \left[7\right]\\
\Bigg\langle \left(\int_{\Sigma^{\pm}} *j\right) \Phi_1 \cdots \Phi_n \Bigg\rangle &= \delta \langle \Phi_1 \cdots \Phi_n \rangle \qquad \left[2\right]
\end{split}
\end{equation}
where on the left hand side of each equation we have the nonlinear part of the transformation, and on the right hand side we have the linear part.  The correlators are always taken to have the usual time ordering.  The second line, which uses the notation of \cite{He:2014laa}, makes explicit the difference between the BMS symmetries at future and past null infinity. We will not make this distinction in what follows but implicitly assume that the full symmetry is indeed the diagonal combination ${\rm BMS}^{+} \times {\rm BMS}^{-}$.  The final line is in the notation of \cite{Avery:2015rga}, where $\Sigma^{\pm}$ are initial and final Cauchy surfaces (which for the S-matrix elements will be taken to $\pm \infty$), and $j^{\mu}$ is the Noether current associated with the symmetry\footnote{We should note that the total charge is gauge invariant, but the soft and hard charges separately are not.  Under a gauge transformation $\bar{\epsilon}^{\mu} \to \bar{\epsilon}^{\mu} + \lambda q^{\mu}$, the polarization of the soft graviton created by $Q_{S}$ will be shifted longitudinally, and the hard charge is shifted by a global transformation $Q_{H} \to Q_{H} + Q_{0}$ which will commute with the S-matrix.  Since we have already restricted our attention to Bondi gauge we will not need to worry about gauge transformations, but we mention this issue anyway for the sake of completeness.}.

A general discussion of how to derive the Noether charges for an asymptotic symmetry is given e.g.\ in~\cite{Barnich:2001jy, Avery:2015rga, Banados:2016zim}, building on the work of \cite{Katz:1985, Iyer:1994ys}.  The integrated charge can be expressed as an integral of the Noether current 3-form over a Cauchy surface $\Sigma$:
\begin{equation}
Q = \int d\Sigma_{\mu \nu \rho} J^{\mu \nu \rho}=\int d\Sigma_{\mu}\,J^{\mu}
\end{equation} 
where $J^{\mu} = S^{\mu} + \partial_{\nu}K^{\mu \nu}$ consists of a part $S^{\mu}$ which vanishes on-shell plus improvement terms which have the form of a divergence of an antisymmetric two-form field.  We have absorbed a factor of $\sqrt{-g}$ into the definition of the one-form, so it is the flat-space divergence of the two-form field, not the covariant one, that appears.  We can derive the Noether current including the improvement terms $K^{\mu \nu}$ from applying the Noether procedure to the following action \cite{Katz:1985},
\begin{equation}\label{Katzaction}
S = \frac{1}{16\pi G}\int d^{4}x\, (\sqrt{-g} R - \sqrt{-\bar{g}}\bar{R} + \partial_{\mu}k^{\mu})
\end{equation}
where the unbarred and barred quantities refer to the quantities associated with the metric $g_{\mu \nu}$ and the background metric $\bar{g}_{\mu \nu}$ respectively, which is this context is taken to be the Minkowski metric $\eta_{\mu \nu}$.  Both metrics are evaluated with respect to the same coordinates.  The vector $k^{\mu}$ is given by
\begin{equation}
k^{\mu} = \frac{1}{\sqrt{-g}}\partial_{\nu}(\sqrt{-g}g^{\nu \mu}) = \sqrt{-g}(g^{\mu \nu}\delta \Gamma^{\rho}_{\nu \rho} - g^{\nu \rho}\delta \Gamma^{\mu}_{\nu \rho})\,,
\end{equation}
where $\delta \Gamma = \Gamma - \bar{\Gamma}$ is a tensor even though individual Christoffels are not.  It can be shown that the term $\partial_{\mu}k^{\mu}$, which is a boundary term (and not the usual Gibbons-Hawking-York term) effectively removes all terms from the Ricci scalar which involve second derivatives of the metric tensor.  The action \eqref{Katzaction} may appear problematic from the perspective of quantization due to the wrong-sign kinetic term for $\bar{g}_{\mu \nu}$, but this term (which vanishes anyway for $\bar{g}_{\mu \nu} = \eta_{\mu \nu}$) should be considered merely as a formal trick for covariantizing the boundary term.
For a derivation using a more covariant formalism, see for instance \cite{Iyer:1994ys}, or \cite{Barnich:2011mi}.  Note that this formalism is not background independent, but this is understandable given that the BMS transformations are defined with respect to the Minkowski metric.  

Performing the Noether procedure on this Lagrangian, we find the Noether current $J^{\mu} = S^{\mu} + \partial_{\nu}K^{\nu \mu}$, where $K^{\nu \mu}$ is given by
\begin{equation}\label{Kexpression}
K^{\mu \nu} = \frac{1}{16\pi G}(\sqrt{-g}\nabla^{\left[\mu\right.}\xi^{\left.\nu \right]} - \sqrt{-\bar{g}}\bar{\nabla}^{\left[\mu\right.}\xi^{\left.\nu \right]} + \sqrt{-g}\xi^{\left[\mu\right.}k^{\left.\nu \right]})
\end{equation}
The first term is the Komar formula, \cite{Komar:1958wp} the second is the Komar formula associated with the Minkowski metric, and the third is the boundary term.  In the derivation of \eqref{Kexpression} we have used the property that $\xi^{\mu}$ is a Killing vector, but not the equations of motion.  The bulk contribution $S^{\mu}$ to the current vanishes identically, consistent with the fact that there are no local observables in a gravitational theory.

The BMS charge is then given by
\begin{equation}
\int_{\mathcal{I}^{+}} \ast J = \int_{\mathcal{I}^{+}_{\pm}} \ast K = \int_{\mathcal{I}^{+}_{\pm}} K^{ru}\,,
\end{equation}
which can be expanded perturbatively around flat space in terms of the boundary data in Bondi gauge.  The Cauchy surface $\Sigma$ near the future null boundary consists of the null rays fibered over a sphere at large $r$, and its boundary consists of a pair of spheres at large $|u|$, which we then send to $u = \pm \infty$.  The radius $r$ is also taken to infinity, so $\Sigma \to \mathscr{I}^{+}$ and $\partial \Sigma \to \mathscr{I}^{+}_{\pm}$.  It is straightforward to calculate the form of the charges in terms of the Bondi boundary data $m, N_{A}, N_{AB} = \partial_{u}C_{AB}$ and the BMS transformations $T, Y^{A}$, and the result is
\footnote{This is closely related to the charges defined~\cite{Barnich:2011mi}, but differs from it. We work with the difference between the negative of the charges defined there as $u\to\infty$ and $u\to -\infty$. This implies that our hard charges compute a weighted sum of the energy/angular momentum carried out to $\mathscr{I}^+$ by massless states that, for example, reduces to the total energy for $T=1, Y^A=0$. The negative sign was introduced to achieve the usual convention for the Poisson bracket that an infinitesimal transformation of $Q_1$ generated by $Q_2$ is given by $\delta_2Q_1=\{Q_1,Q_2\}$. }
\begin{equation}
\begin{split}
Q &=- \frac{1}{16\pi G}\int_{\mathcal{I}^{+}_{\pm}}d^2 z \, \gamma_{z\bar{z}}\Big[4mf + 2N_{A}Y^{A} + \frac{1}{16}Y^{A}D_{A}(C_{CD}C^{CD})\Big]\\
&= -\frac{1}{16\pi G}\int_{\mathcal{I}^{+}_{\pm}}d^2 z \, \gamma_{z\bar{z}}\Big[4mT  + 2m uD_{A}Y^{A}+ 2N_{A}Y^{A} + \frac{1}{16}Y^{A}D_{A}(C_{CD}C^{CD})\Big]\,.
\end{split}
\end{equation}

\subsection{Leading symmetry}

First we show that the leading order soft graviton theorem is equivalent to the BMS Ward identity for supertranslations.  For the supertranslations, the charge at positive null infinity is given up to terms that vanish as $r \to \infty$ by
\begin{equation}
Q^{(0)} =- \frac{1}{4\pi G} \int_{\mathscr{I}^{+}_{\pm}} d^2 z \,\gamma_{z\bar{z}}T m\,,
\end{equation}
where $m$ is the Bondi mass aspect. Using the constraint equation
\begin{equation}
\partial_u m=\frac14D_AD_BN^{BA}-\frac18N^A_BN^B_A-4\pi G\lim_{r\to\infty}r^2 T_{uu}\,,
\end{equation}
where $N_{AC}$ is the Bondi news, we can equivalently write this as
\begin{equation}
Q^{(0)} = -\frac{1}{16\pi G} \int_{\mathscr{I}^{+}} d^{2}z du \, \gamma_{z\bar{z}}\, T \left[D_A D_C N^{AC} - \frac{1}{2}N_{AC}N^{AC} - 16 \pi G\lim_{r \to \infty} r^2 T_{uu}^{M}\right]\,.
\end{equation}
The first term contains the LSZ pole for the soft graviton insertion,
\begin{equation}\label{eq:Q0}
Q^{(0)}_{S} =-\frac{1}{16\pi G} \int du d^2z \,\gamma^{z \bar{z}}\left[D_{\bar{z}}^2 T\, N_{zz} + D_{z}^2 T\, N_{\bar{z}\bar{z}}\right]\,,
\end{equation}
where we have integrated by parts, assuming appropriate fall-off conditions on $T(z,\bar{z})$ as $z \to \infty$\footnote{More generally, we can split a general function $T(z, \bar{z})$ on the sphere into functions with compact support using a partition of unity, and then add the results at the end.}. There is a corresponding integral over the boundary at negative null infinity, but we will not worry about this here, since the soft gravitons can be moved from the $in$- to the $out$-state using crossing symmetry\footnote{This can be expressed in terms of antipodal boundary conditions~\cite{He:2014laa}, or in terms of what are the appropriate adiabatic modes in the asymptotic limit~\cite{Mirbabayi:2016xvc}.}. Expanding the Bondi news in terms of creation and annihilation operators for the graviton and evaluating the integrals using the method of steepest descent gives
\begin{equation}\label{Nzz}
\begin{split}
N_{zz} &= -\frac{\kappa}{8\pi^2}\gamma_{z\bar{z}}\int_{0}^{\infty}d\omega\, \omega \left[a^{out}_{+}(\omega \hat{x})e^{-i\omega u}+a^{out}_{-}(\omega \hat{x})^{\dagger}e^{i\omega u}\right]\,,\\
\end{split}
\end{equation}
as well as\footnote{A factor 1/2 arises because the integral over $\omega$ only extends over the positive half-line.}
\begin{equation}\label{Nzz2}
\begin{split}
\int du \,N_{zz} &=-\frac{\kappa}{8\pi}\gamma_{z\bar{z}} \lim_{\omega \to 0} \omega \left[a^{out}_{+}(\omega \hat{x}) + a^{out}_{-}(\omega \hat{x})^{\dagger}\right]\,.
\end{split}
\end{equation}
and a corresponding contribution from $N_{\bar{z}\bar{z}}$.  Although we will not need to change gauges in what follows, it is straightforward to check using the stationary phase approximation that this expression is invariant under a general gauge transformation.  

Considering the $N_{zz}$ contribution first, the left hand side of the Ward identity becomes 
\begin{equation}
\begin{split}
\langle out | [Q^{(0)}_{S}, \mathcal{S}] | in \rangle \supset& \hskip 0.5cm\frac{1}{4\pi\kappa}\int d^2 z D_{\bar{z}}^2 T \lim_{\omega \to 0} \omega  \langle out |a^{out}_{+}(\omega \hat{x}) \mathcal{S}| in \rangle \\
&-\frac{1}{4\pi\kappa}\int d^2 z D_{\bar{z}}^2 T \lim_{\omega \to 0} \omega \langle out | \mathcal{S}a^{in}_{-}(\omega \hat{x})^\dagger| in \rangle\,,
\end{split}
\end{equation}
where we have implicitly used that the charge $Q^{(0)}_{S}$ is an element of the diagonal subalgebra of \mbox{${\rm BMS}^+\times{\rm BMS}^-$} in writing $a^{out}_{+}$ on the first and ${a^{in}_{-}}^\dagger$ on the second line. Using crossing symmetry to relate the amplitude with an incoming negative helicity graviton to the corresponding amplitude with an outgoing positive helicity graviton, and including the contribution from $N_{\bar{z}\bar{z}}$, we find 
\begin{equation}\label{singlesoftleft}
\begin{split}
\langle out | [Q^{(0)}_{S}, \mathcal{S}] | in \rangle = & \hskip 0.5cm \frac{1}{2\pi\kappa}\int d^2 z\,D_{\bar{z}}^2 T \lim_{\omega \to 0}  \omega \langle out |a^{out}_{+}(\omega \hat{x}) \mathcal{S}| in \rangle \\
&+\frac{1}{2\pi\kappa}\int d^2 z\,D_{z}^2 T \lim_{\omega \to 0} \omega \langle out |a^{out}_{-}(\omega \hat{x}) \mathcal{S}| in \rangle\,.
\end{split}
\end{equation}
With the understanding that any creation and annihilation operators that act on $out$-states are $a^{out}_{\pm}$, $a{^{out}_{\pm}}^\dagger$ and those acting on $in$-states are $a^{in}_{\pm}$, $a{^{in}_{\pm}}^\dagger$, we will from now on drop the $in$ and $out$ labels.

We can now apply Weinberg's soft-graviton theorem. It will be convenient to express the momenta and polarization vectors in holomorphic coordinates 
\begin{equation}
p_k^\mu=\frac{E_k}{1+z_k\bar{z}_k}\big(1+z_k\bar{z}_k,z_k+\bar{z}_k,-i(z_k-\bar{z}_k),1-z_k\bar{z}_k\big)\,,
\end{equation}
and using the choice of gauge in \cite{He:2014laa}
\begin{equation}\label{Stromingergauge}
\bar{\epsilon}^{\mu}_{+} = \frac{1}{\sqrt{2}}(\bar{z}, 1, -i, -\bar{z})\, , \qquad \bar{\epsilon}^{\mu}_{-} = \frac{1}{\sqrt{2}} (z, 1, i, -z)\, ,
\end{equation}
so that the relevant expressions are\footnote{For simplicity we will assume the hard quanta are massless.}
\begin{equation}
\begin{split}\label{basicExpressions}
(p_k \cdot \bar{\epsilon}^{+}) = -\frac{\sqrt{2}E_{k}(\bar{z} - \bar{z}_k)}{(1+z_k \bar{z}_k)}\, , &\qquad (p_k \cdot \bar{\epsilon}^{-}) =- \frac{\sqrt{2}E_{k}(z - z_k)}{(1+z_k \bar{z}_k)}\\
(p_k \cdot q) &= -\frac{2E_{k}\omega|z - z_k|^2}{(1+z \bar{z})(1+z_k \bar{z}_k)}\,,
\end{split}
\end{equation}
where we are working in the ``mostly plus'' convention for the metric.  Inserting the resulting expression for the soft factor
\begin{equation}
\frac{\kappa}{2}\frac{(p_k \cdot \bar{\epsilon}^{+})^2}{(p_k \cdot q)} =- \frac{\kappa}{2}\frac{E_{k}}{\omega}\frac{(\bar{z}-\bar{z}_k)}{(z-z_k)}\frac{(1+z \bar{z})}{(1+z_k \bar{z}_k)}
\end{equation}
into \eqref{singlesoftleft}, we have
\begin{eqnarray}
\langle out|[Q_S^{(0)},\mathcal{S}]|in\rangle&=&- \frac{1}{4\pi} \int d^2 z \,D_{\bar{z}}^2 T \sum_{k} E_k \frac{(\bar{z}-\bar{z}_k)(1+z\bar{z})}{(z-z_k)(1 + z_k\bar{z}_k)}\langle out | \mathcal{S}| in \rangle\nonumber\\
&&- \frac{1}{4\pi} \int d^2 z \,D_{z}^2 T \sum_{k} E_k \frac{(z-z_k)(1+z\bar{z})}{(\bar{z}-\bar{z}_k)(1 + z_k\bar{z}_k)}\langle out | \mathcal{S}| in \rangle\,.
\end{eqnarray}
Undoing the integration by parts in the definition of the charge and using the Cauchy-Pompeiu formula
\be
\partial_{\bar{z}}\left(\frac{1}{z-z_k}\right) = (2\pi)\delta^{(2)}(z-z_k)\,,
\ee
leads us to
\begin{equation}
\begin{split}
\langle out|[Q_S^{(0)},\mathcal{S}]|in\rangle=
&\hphantom{-}\frac{1}{4\pi} \int d^2 z \, \left[\partial_{\bar{z}} T \sum_{k} \frac{E_k(1 + z \bar{z}_k)}{(z-z_k)(1+z_k \bar{z}_k)}+c.c.\right]\langle out | \mathcal{S}| in \rangle \\
=&-\frac{1}{4\pi} \int d^2 z \, T \sum_k \left[ E_k  (2\pi)\delta^{(2)}(z-z_k)\frac{(1+z \bar{z}_k)}{(1+z_k \bar{z}_k)}+c.c\right]\langle out | \mathcal{S} | in \rangle\\
=& - \sum_k E_k T(z_k)\langle out | \mathcal{S} | in \rangle\,.
\end{split}
\end{equation}
so that
\begin{equation}
\langle out | [Q^{(0)}_{S}, \mathcal{S}] | in \rangle=-\sum_k E_k T(z_k)\langle out | \mathcal{S} | in \rangle\,.
\end{equation}
 Note that keeping both helicities is important here.  In \cite{He:2014laa}, the calculation focused on a single helicity, but the factor of two was preserved by taking the boundary conditions $N_{zz} = D_{z}^{2}N, N_{\bar{z}\bar{z}} = D_{\bar{z}}^{2}N$ at future null infinity.  It is also worth noting, as emphasized in \cite{He:2014laa}, that one linear combination of the helicities decouples in the leading soft limit -- this can be thought of as the statement that there are two graviton polarizations but only one Goldstone boson. 


What remains is to show that the remaining terms in the charge
\begin{equation}
Q^{(0)}_H=\frac{1}{16\pi G} \int_{\mathscr{I}^+} du d^{2}z\gamma_{z\bar{z}}\, T \left(\frac{1}{2}N_{AC}N^{AC} + 16 \pi G\lim_{r \to 0} r^2 T_{uu}^{M}\right)
\end{equation}
generate the same contribution with opposite sign so that $\langle out | \left[Q, \mathcal{S} \right] | in \rangle=0$.  Focusing on the $N_{AC}N^{AC}$ terms, this part of the charge is given in terms of graviton creation and annihilation operators by
\begin{equation}
\begin{split}
Q^{(0)}_H&\supset\frac{1}{16 \pi G}\int_{\mathscr{I}^{+}} du d^{2}z\, \gamma^{z\bar{z}} T :\!N_{zz}N_{\overline{z}\overline{z}}: \\
&= \frac{1}{16\pi^3}\int d^{2}z\, \gamma_{z\bar{z}} T \int_{0}^{\infty} d\omega \, \omega^2 \left[ a_{+}(\omega \hat{x})^{\dagger}a_{+}(\omega \hat{x}) + a_{-}(\omega \hat{x})^{\dagger}a_{-}(\omega \hat{x}) \right] + \cdots\,.
\end{split}
\end{equation}
Here we have made use of the expansion \eqref{Nzz} for $N_{zz}$ in terms of creation and annihilation operators, and the ellipses indicate terms of higher order in terms of the number of creation and annihilation operators\footnote{These can in principle be found order by order in a more careful treatment of the stationary phase approximation for products of the boundary fields.}.  The commutator of this expression with a graviton operator is straightforward to calculate and is given by $[a_\pm(\mathbf{k}),Q^{(0)}_H]=E_k T(z_k) a_\pm(\mathbf{k})$ so that
\begin{equation}
\langle out | [Q^{(0)}_{H}, \mathcal{S}] | in \rangle=\sum_k E_k T(z_k)\langle out | \mathcal{S} | in \rangle\,,
\end{equation}
which is equal and opposite in sign to the result for the soft part of the charge as expected. Because we have only considered the contribution from gravitons, the sum so far only runs over all outgoing hard gravitons. However, the terms involving the stress-energy tensor of the matter field provide the same contribution for each of the matter lines so that $\langle out | \left[ Q, \mathcal{S} \right] | in \rangle = 0.$

\subsection{Subleading symmetries}\label{superrotation}

The subleading symmetry arises from the parts of the charge that are of higher order in $u$.  To keep track of these, we must consider the angular momentum contribution to the charge as well. We will work with the following subleading charge
\begin{equation}\label{subleadingSoftCharge} 
\begin{split}
Q^{(1)} = &-\frac{1}{16\pi G} \int du d^{2}z \,\gamma_{z\bar{z}} \, \Big[\frac{1}{2}u D_{B}Y^{B} D_A D_C N^{AC} \\
&\qquad \qquad \qquad \qquad \qquad - \frac{1}{2}u N^{C}_{A}\Big[D_C D^B D_B Y^A - D_{C}D_{B}D^{A}Y^{B}\Big]\Big]\\
&-\frac{1}{16\pi G} \int du d^{2}z\, \gamma_{z\bar{z}}\Big[-\frac{1}{4}u D_{A}Y^{A}N_{zz}N^{zz} + \frac{1}{4}Y^{z}\partial_{z}\partial_{u}(C_{zz}C^{zz})\\ &- \frac{1}{2}Y^{z}N^{zz}D_{z}C_{zz} -\frac{1}{2}Y^{z}N_{zz}D_{z}C^{zz} - \frac{1}{2}Y^{z}\partial_{z}(C^{zz}N_{zz}-C_{zz}N^{zz}) \Big] + h.c.\\
&+ \int du d^{2}z\, \gamma_{z\bar{z}}\left[\frac12 u D_{A}Y^{A}\lim_{r \to \infty}r^2 T_{uu} + 2 Y^{A}\lim_{r\to \infty}r^2 T_{uA}\right]\,.
\end{split}
\end{equation}
The first two lines comprise the soft graviton insertion, and the last three lines contain the terms which rotate the hard particles. This charge can be obtained from the subleading part of the charge introduced earlier 
\be
Q^{(1)} = -\frac{1}{16\pi G} \int_{\mathscr{I}^{+}_{\pm}} d^{2}z\, \gamma_{z\bar{z}}\, \Big[2N_A Y^A + 2 u D_{A}Y^{A} m + \frac{1}{16}Y^{A}D_{A}(C_{CD}C^{CD})\Big]\,,
\ee
by using the constraint equations
\begin{eqnarray}
\partial_u m&=&\frac14D_AD_BN^{BA}-\frac18N^A_BN^B_A-4\pi G\lim_{r\to\infty}r^2 T_{uu}\,,\\
\partial_uN_A&=&\partial_Am-\frac14D_B\left(D^BD_CC^C_A-D_AD_CC^{BC}\right)\nonumber\\
&&+\frac{1}{16}\partial_A\left(N^B_CC^C_B\right)-\frac14N^B_CD_AC^C_B-\frac14D_B\left(C^B_CN^C_A-N^B_CC^C_A\right)\\
&&-8\pi G\lim_{r\to\infty}r^2 T_{uA}\nonumber\,,
\end{eqnarray}
and dropping the total $u$-derivative
\begin{eqnarray}
\Delta Q_S^{(1)} &=& \frac{1}{16\pi G} \int_{\mathcal{I}^{+}} du d^{2}z\, \gamma_{z\bar{z}}\, \frac{\partial}{\partial u} \left[\frac12 u Y^A D_B\left(D^BD_C C^C_A-D_A D_C C^{BC}\right) \right]\label{eq:QS1}\,.
\end{eqnarray}
The last step implies that our charge differs from that in~\cite{Barnich:2011mi} by $\Delta Q_S^{(1)}$, but the definition~(\ref{subleadingSoftCharge}) is appropriate in the context of soft graviton theorems. 
To see this, notice that in terms of creation and annihilation operators $\Delta Q_S^{(1)}$ contains contributions of the form
\begin{equation}
\int_{-\infty}^\infty du\, \partial_u(u C_{zz})=\frac{i\kappa}{4\pi(1+z\bar{z})^2}\lim_{\omega\to 0}\left[\omega\partial_\omega a_+(\omega\hat{x})-\omega\partial_\omega a_-(\omega\hat{x})^\dagger\right]\,,
\end{equation}
which leads to $\langle out|[Q_S^{(1)},\mathcal{S}]|in\rangle$ that are singular in the soft limit. Such contributions cannot arise in $\langle out|[Q_H^{(1)},\mathcal{S}]|in\rangle$ so we must drop the the total $u$-derivative and work with~(\ref{subleadingSoftCharge}) to bring the soft graviton theorem into the form $\langle out|[Q^{(1)},\mathcal{S}]|in\rangle$. 

Just like for the leading soft theorem, our goal will now be to determine $\langle out|[Q_S^{(1)},\mathcal{S}]|in\rangle$ and $\langle out|[Q_H^{(1)},\mathcal{S}]|in\rangle$ to show that $\langle out|[Q^{(1)},\mathcal{S}]|in\rangle=0$. We first focus on the soft graviton insertion. Integrating by parts, making use of the fact that $Y^{z}, Y^{\bar{z}}$ are holomorphic and antiholomorphic, respectively, and using the identity $D_{\bar{z}}^3Y^{\bar{z}}=\partial_{\bar{z}}^3 Y^{\bar{z}}$, we can write it as
\begin{equation}
Q^{(1)}_{S} =- \frac{1}{16\pi G} \int du d^{2}z \gamma^{z\bar{z}} \, u \left[\partial_{\bar{z}}^3 Y^{\bar{z}} N_{zz}+\partial_{z}^3 Y^{z} N_{\bar{z}\bar{z}} \right]\,.
\end{equation}
We can express the integral of $N_{zz}$ over $u$ in terms of creation and annihilation operators
\begin{equation}\label{Nzz3}
\int du\, u\,N_{zz} = \frac{i\kappa}{8\pi}\gamma_{z\bar{z}} \lim_{\omega \to 0} \left[(1+\omega\partial_\omega)a_{+}(\omega \hat{x}) -(1+\omega\partial_\omega) a_{-}(\omega \hat{x})^{\dagger}\right]\,,
\end{equation}
and see that the charge is given by
\begin{equation}\label{eq:Q1}
Q^{(1)}_{S} =- \frac{i}{4\pi\kappa}\int d^{2}z \partial_{\bar{z}}^3 Y^{\bar{z}} \lim_{\omega \to 0} \left[(1+\omega\partial_\omega)a_{+}(\omega \hat{x}) -(1+\omega\partial_\omega) a_{-}(\omega \hat{x})^{\dagger}\right]+h.c. 
\end{equation}
The subleading contribution is then given by
\begin{eqnarray}\label{subleadingLHS}
&&\hskip -0.8cm\langle out | [ Q^{(1)}_{S}, \mathcal{S} ] | in \rangle \supset- \frac{i}{2\pi\kappa} \int d^{2}z\, \partial_{\bar{z}}^3 Y^{\bar{z}}\lim_{\omega \to 0}(1+\omega \partial_{\omega})\langle out|a_{+}(\omega \hat{x})\mathcal{S}|in \rangle \\
&&\hskip -0.5cm= -\frac{i}{4\pi}\int d^{2}z\,  \partial_{\bar{z}}^3 Y^{\bar{z}}\sum_{k}\Bigg[\frac{(\bar{z}-\bar{z}_k)(1+\bar{z}z_k)}{(z_k - z)(1+z_k \bar{z}_k)}E_k \partial_{E_k} + \frac{(\bar{z}-\bar{z}_k)^2}{(z_k - z)}\partial_{\bar{z}_k} +h_k \frac{(\bar{z}-\bar{z}_k)}{(z_k - z)}\Bigg]\langle out|\mathcal{S}|in\rangle\nonumber\,,
\end{eqnarray}
where $h_k$ is the helicity of the $k^{th}$ particle, and in the last line we have applied the subleading soft theorem with the subleading soft factor given in terms of the holomorphic coordinates by\footnote{This form of the soft factor disagrees with the forms given in holomorphic coordinates in \cite{Kapec:2014opa} and \cite{Kapec:2016jld}, but it agrees with the soft factor in spinor helicity variables given in \cite{Bern:2014vva}, and we have confirmed that it agrees with explicit perturbative calculations.}
\begin{equation}
\frac{(p_k \cdot \bar{\epsilon}^{+})(\bar{\epsilon}^{+}_{\mu}q_{\nu} J_{k}^{\mu \nu})}{(p_k \cdot q)} = \frac{(\bar{z}-\bar{z}_k)(1+\bar{z}z_k)}{(z_k - z)(1+z_k \bar{z}_k)}E_k \partial_{E_k} + \frac{(\bar{z}-\bar{z}_k)^2}{(z_k - z)}\partial_{\bar{z}_k} +h_k \frac{(\bar{z}-\bar{z}_k)}{(z_k - z)}\,.
\end{equation}
We have again included an overall factor of two from applying crossing symmetry to the corresponding expression at $\mathscr{I}^{-}$.  

We can evaluate the $\int d^{2}z$ integral in \eqref{subleadingLHS} by integrating by parts and applying the Cauchy-Pompeiu theorem. The final result is
\begin{eqnarray}
&&\hskip -0.8cm\langle out | [Q^{(1)}_{S}, \mathcal{S}]|in\rangle \supset i\sum_{k}\Big[\frac12(D_{\bar{z}_k}Y^{\bar{z}_k})E_k \partial_{E_k} - Y^{\bar{z}_k}\partial_{\bar{z}_k} + \frac{h_k}{2} \partial_{\bar{z}_k} Y^{\bar{z}_k}  \Big]\langle out|\mathcal{S}|in\rangle\\
&&\hskip 2.25cm=i\sum_{k}\Big[\frac12(D_{\bar{z}_k}Y^{\bar{z}_k})\left(E_k \partial_{E_k} + h_k\right)-Y^{\bar{z}_k}\left(\partial_{\bar{z}_k}+h_k\Omega_{\bar{z}_k}\right)\Big]\langle out|\mathcal{S}|in\rangle\,,\nonumber
\end{eqnarray}
where $\Omega_{\bar{z}}=\frac12\Gamma_{\bar{z}\bar{z}}^{\bar{z}}$ is the spin connection. Including the contribution from $N_{\bar{z}\bar{z}}$, which creates a negative helicity graviton, the result is therefore
\begin{equation}\label{eq:QsS}
\begin{split}
&\hskip -0.3cm\langle out | [ Q^{(1)}_{S}, \mathcal{S}] | in \rangle= 
i\sum_{k}\Big[\frac{1}{2}D_{A}Y^{A}(z_k)E_k \partial_{E_k}  + \frac{h_k}{2}(D_{\bar{z}_k}Y^{\bar{z}_k}-D_{z_k}Y^{z_k}) \\
&\hskip 4.65cm - Y^{z_k}(\partial_{z_k} - h_k\Omega_{z_k})- Y^{\bar{z}_k}(\partial_{\bar{z}_k}+h_k\Omega_{\bar{z}_k})\Big]\langle out|\mathcal{S}|in\rangle\,.
\end{split}
\end{equation}
As before it was crucial to include both helicities in the soft charge in order to derive this expression.

Let us now consider the part of the charge that is quadratic in the boundary data -- this will perform a rotation on any hard gravitons in the initial and final states.  The piece involving only gravitons becomes:
\begin{equation}
\begin{split}
Q^{(1)}_{H} =- \frac{1}{16\pi G} \int du d^{2}z\, \gamma_{z\bar{z}}\Big[&-\frac{1}{4}u D_{A}Y^{A}:N_{zz}N^{zz}: + \frac{1}{4}Y^{z}D_{z}\partial_{u}(:C_{zz}C^{zz}:) \\ &- \frac{1}{2}Y^{z}:N^{zz}D_{z}C_{zz}: -\frac{1}{2}Y^{z}:N_{zz}D_{z}C^{zz}: \\&- \frac{1}{2}Y^{z}D_{z}(:C^{zz}N_{zz}:-:C_{zz}N^{zz}:)\Big] + h.c.
\end{split}
\end{equation}
We can express the charge in terms of creation and annihilation operators, using the expressions
\begin{equation}
\begin{split}
N_{zz} &= -\frac{\kappa}{8\pi^2}\gamma_{z\bar{z}}\int_{0}^{\infty}d\omega\, \omega \left[a_{+}(\omega \hat{x})e^{-i\omega u}+a_{-}(\omega \hat{x})^{\dagger}e^{i\omega u}\right]\,,\\
C_{zz} &= -\frac{i\kappa}{8\pi^2}\gamma_{z\bar{z}}\int_{0}^{\infty}d\omega\, \left[a_{+}(\omega \hat{x})e^{-i\omega u}-a_{-}(\omega \hat{x})^{\dagger}e^{i\omega u}\right]\,,
\end{split}
\end{equation}
and similarly for the complex conjugates.  Keeping only the terms quadratic in creation and annihilation operators, this leads to
\begin{eqnarray}
Q_H^{(1)}&=&-\frac{i}{16\pi^3} \int_{\mathcal{I}^{+}} d^{2}z\, \gamma_{z\bar{z}}\int_0^\infty d\omega\,\omega\Big\{\Big[\frac14D_AY^A  a_+(\omega\hat{x})^\dagger \partial_\omega(\omega a_+(\omega\hat{x}))\nonumber\\
&&\hskip 2cm-\partial_z Y^z a_+(\omega\hat{x})^\dagger a_+(\omega\hat{x})- Y^z a_+(\omega\hat{x})^\dagger\partial_z a_+(\omega\hat{x})\nonumber\\ 
&&\hskip 2cm-\frac12 D_zY^z  \left(a_+(\omega\hat{x})^\dagger a_+(\omega\hat{x})\right)\Big]-(a\leftrightarrow a^\dagger,+\leftrightarrow -)\Big\}+h.c.
\end{eqnarray}
Including the contribution from the complex conjugate explicitly, this can be written as
\begin{eqnarray}
&&\hskip -2.2cm Q_H^{(1)}=-\frac{i}{16\pi^3} \int_{\mathcal{I}^{+}} d^{2}z\, \gamma_{z\bar{z}}\int_0^\infty d\omega\,\omega\Big\{\Big[\frac12D_AY^A  a_+(\omega\hat{x})^\dagger \omega\partial_\omega( a_+(\omega\hat{x}))\nonumber\\
&&\hskip 4.2cm-\partial_z Y^z a_+(\omega\hat{x})^\dagger a_+(\omega\hat{x})- Y^z a_+(\omega\hat{x})^\dagger\partial_z a_+(\omega\hat{x})\nonumber\\ 
&&\hskip 4.2cm +\partial_{\bar{z}} Y^{\bar{z}} a_+(\omega\hat{x})^\dagger a_+(\omega\hat{x})- Y^{\bar{z}} a_+(\omega\hat{x})^\dagger\partial_{\bar{z}} a_+(\omega\hat{x})\Big]\nonumber\\
&&\hskip 8.5cm-(a\leftrightarrow a^\dagger,+\leftrightarrow -)\Big\}\,,
\end{eqnarray}
and together with the contribution for negative helicity gravitons, we can bring the contribution of the hard charge that is of second order in creation and annihilation operators into the form
\begin{eqnarray}
Q_H^{(1)}&=&-\frac{i}{16\pi^3} \int_{\mathcal{I}^{+}} d^{2}z\, \gamma_{z\bar{z}}\int_0^\infty d\omega\,\omega\\
&&\hskip-1.0cm\times \Big\{a_+(\omega\hat{x})^\dagger\Big[\frac12D_AY^A\omega\partial_\omega+(D_{\bar{z}} Y^{\bar{z}}-D_z Y^z)-Y^z(\partial_z-2\Omega_z)-Y^{\bar{z}}(\partial_{\bar{z}}+2\Omega_{\bar{z}})\Big]a_+(\omega\hat{x})\nonumber\\
&&\hskip -0.7cm+a_-(\omega\hat{x})^\dagger\Big[\frac12D_AY^A\omega\partial_\omega-(D_{\bar{z}} Y^{\bar{z}}-D_z Y^z)-Y^z(\partial_z+2\Omega_z)-Y^{\bar{z}}(\partial_{\bar{z}}-2\Omega_{\bar{z}})\Big]a_-(\omega\hat{x})\Big\}\,.\nonumber
\end{eqnarray}
The commutator of this with a hard graviton operator is given by
\begin{equation}
\left[a_\pm(E_k \hat{x}_{k}),Q_H^{(1)} \right]=-i\Big[\frac12D_AY^AE_k\partial_{E_k}\pm(D_{\bar{z}_k} Y^{\bar{z}_k}-D_{z_k} Y^{z_k})-Y^{z_k}D_{z_k}-Y^{\bar{z}_k}D_{\bar{z}_k})\Big]a_\pm(E_k\hat{x}_k)\,,
\end{equation}
where for later convenience we have introduced the notation 
\begin{equation}
D_za_\pm(E\hat{x})=(\partial_z\mp2\Omega_{z})a_\pm(E\hat{x})\qquad\text{and}\qquad D_{\bar{z}}a_\pm(E\hat{x})=(\partial_{\bar{z}}\pm2\Omega_{\bar{z}})a_\pm(E\hat{x})\,.
\end{equation}
Comparing to equation~(\ref{eq:QsS}), we see that the contributions from hard graviton legs in $\langle out | [ Q_H^{(1)}, \mathcal{S} ] | in \rangle$ and $\langle out | [ Q_S^{(1)}, \mathcal{S} ] | in \rangle$ are equal and opposite.  
 The matter contribution to the hard charge quadratic in creation and annihilation operators similarly generates the appropriate rotation of the matter fields, concluding the proof that $\langle out | \left[ Q, \mathcal{S} \right] | in \rangle = 0.$  

We have integrated by parts on the sphere several times, and assumed that there are no boundary terms at $z = \infty$.  It is worth discussing this point in further detail.  Some (but not all) of the integrations by parts are purely a matter of convenience, since we integrated by parts several times in going from \eqref{subleadingSoftCharge} to \eqref{subleadingLHS}, and then undid many of these same integrations again when deriving the Ward identity.  Whether dropping the boundary terms is fully justified, however, is more of an issue here than it was for supertranslations: since we have chosen falloff conditions on the metric that restrict $Y^{z}$, $Y^{\bar{z}}$ to be holomorphic and antiholomorphic, if we do not restrict to the global subalgebra, we will introduce singular points on the sphere.  We can avoid this by choosing the extended $Y^{A}$ to be smooth, of course, as in \cite{Campiglia:2014yka}, but this will not preserve the same falloff conditions, so the application to S-matrix elements is less clear, and the definition of the soft charges will need to be modified.  

For the extended algebra involving Virasoro transformations, if we impose that the quantities $Y^{z}$, $Y^{\bar{z}}$ fall off sufficiently fast at infinity that there are no boundary terms in the integrals, (anti)holomorphy means that we necessarily introduce singularities at finite $z$.  For the derivation of \eqref{subleadingLHS} and the subleading soft theorem, these additional singularities will not contribute, since no holomorphic derivatives act on the antiholomorphic poles in $Y^{\bar{z}}$ (or vice versa for $Y^{z}$) but this is not always the case\footnote{In Section 4.5, we will find that in the calculation of the commutator of two BMS transformations, the only time when the derivatives may act on the poles in $Y^{A}$ to produce a spurious delta function contribution arises in the commutator of the two superrotation charges, for the part of the charges where the two soft gravitons are of opposite helicities.}, and it would be interesting to know whether these poles can have more subtle consequences.  A similar set of questions arises when deriving Ward identities in a 2d CFT: here, although the Virasoro generators give rise to an infinity of locally conserved currents, only specific choices of wavefunction and contour lead to meaningful global Ward identities for the correlation functions, and the rest generate spurious results involving the value of the correlator at the introduced poles\footnote{In particular, we may find interesting Ward identities when one of the hard operators has a null vector which vanishes at a given pole.}.  It might be interesting to pursue these issues further for the case at hand, and to understand whether these poles can have nontrivial physical consequences.  It may be the case, however, that they indicate that the interpretation of the charges is in fact more subtle, and that the integrals should be considered as formal objects in order to drop the boundary terms -- such subtleties can arise for instance in the case of a vertex operator algebra~\cite{Barnich:2017ubf}.

Another complication is that the subleading soft theorem may receive quantum corrections arising at one-loop level in the collinear limit~\cite{Bern:2014vva, Larkoski:2014bxa}.  It was found in \cite{He:2017fsb}, however, that there is nevertheless a Virasoro symmetry still at one-loop order, and that this symmetry can be generated by adding only local corrections to the subleading charge.  The corrections to the soft charge are given by the expression
\begin{equation}
\Delta Q_{S} = \frac{i}{16\pi^2 G \bar{\epsilon}}\int d^{2}z\,\gamma^{z\bar{z}}Y^{z}\Bigg[N^{(0)}_{zz} D_{z}N^{(0)}_{\bar{z}\bar{z}} + D_{z}\left(N^{(0)}_{zz} N^{(0)}_{\bar{z}\bar{z}}\right)\Bigg] + h.c.
\end{equation}
where $N^{(0)}_{zz} = \int du N_{zz}$ and $\bar{\epsilon} = 4-d$ comes from the UV divergence terms in dimensional regularization\footnote{This is presumably just the charge in $4-\bar{\epsilon}$ dimensions written in terms of four-dimensional expressions -- it would of course be preferable to have an expression for the charges that works in all dimensions, or to have a better understanding of the regulator in order to derive this directly from the expression for $Q$.}.  The Virasoro symmetry may therefore persist at one-loop in terms of these dressed charges\footnote{Although we cannot rule out the possibility of finite corrections at one loop, as these have not been calculated explicitly -- see discussion in \cite{He:2017fsb}.}.  In what follows we will continue to work with the tree-level expressions for $Q$ for ease of calculation; however, we expect that our arguments will generalize in a straightforward manner to the corrected version of the charge, and the commutator should therefore continue to be robust in the presence of these one-loop collinear quantum corrections.

\section{Charge algebra and double-soft limit}\label{chargeAlgebra}

We will now study the structure of the charge algebra, generalizing the analysis of the previous section to include multiple insertions of the charge operator. 
The expression we would like to check, and the relevant limit of soft graviton amplitudes, is schematically
\begin{equation}\label{eq:BMSalg}
\begin{split}
\lim_{[\omega_2 \to 0}\lim_{\omega_1 \to 0]}\sum\limits_{\lambda_1,\lambda_2}\int d^2 z_1d^2z_2\Psi_1(q_1)\Psi_2(q_2)\langle out \, q_1,q_2 | \mathcal{S} | in \rangle\\
=\langle out | \left[ \left[Q_{1}, Q_{2}\right] - \left[Q_{1H}, Q_{2H}\right], \mathcal{S}\right]| in \rangle &\overset{?}=i \langle out | \left[ Q_{\left[1,2\right]}, \mathcal{S} \right] | in \rangle\,,
\end{split}
\end{equation}
where the $q_{1,2}$ collectively denote the energies $\omega_{1,2}$ and directions $z_{1,2}$ defining the 4-momenta of the gravitons, as well as their helicities $\lambda_{1,2}$. The charges associated with the BMS transformations $\xi^{\mu}_{1,2}$ and $\xi^{\mu}_{\left\{1,2\right\}}$ are denoted $Q_{1,2}$ and $Q_{\left[1,2\right]}$, respectively, and and the soft graviton weights $\Psi_{1,2}$ are chosen appropriately for the BMS transformations of interest.  For a general derivation of this expression, and to understand why the charge algebra is realized this way and not by $\langle out | \left[ \left[Q_1, Q_2\right], \mathcal{S}\right] | in \rangle = i \langle out | \left[ Q_{[1,2]}, \mathcal{S}\right] | in \rangle$ when the soft part of the charge operator is restricted to the creation and annihilation of on-shell states, see Appendix \ref{softpions}.

To evaluate the commutator using scattering amplitudes, we will split the commutator into a piece that changes the number of gravitons as it acts on the state, and a piece that does not. In the language of soft-pion theorems, these pieces correspond to terms in the current that give rise to LSZ poles for pions and terms that do not. As we did for the individual charges, we will denote these as $\left[Q_{1}, Q_{2}\right]_S$ and $\left[Q_{1}, Q_{2}\right]_H$, respectively. For soft pion theorems, current conservation relates the pole and non-pole pieces. 
In terms of the soft and hard parts of the charges $Q_1$ and $Q_2$, these are simply given by
\begin{eqnarray}
\langle out |\left[\left[Q_{1}, Q_{2}\right]_H, \mathcal{S}\right] | in \rangle = \langle out | \left[(\left[Q_{1H},Q_{2S}\right]_{H}+\left[Q_{1S},Q_{2H}\right]_H), \mathcal{S}\right] | in \rangle\,,\label{eq:q1q2hard}
\end{eqnarray}
\begin{eqnarray}
\langle out |\left[\left[Q_{1}, Q_{2}\right]_{S}, \mathcal{S}\right] | in \rangle = \langle out | \left[(\left[Q_{1S},Q_{2S}\right] +\left[Q_{1H},Q_{2S}\right]_{S} + \left[Q_{1S},Q_{2H}\right]_{S}), \mathcal{S}\right] | in \rangle\,.\label{eq:q1q2soft}
\end{eqnarray}
The first term $\left[Q_{1S},Q_{2S}\right]$ on the right hand side of equation~(\ref{eq:q1q2soft}) is associated with the commutator of two soft graviton operators and will be shown to vanish. The second and third terms $\left[Q_{H}, Q_{S}\right]_{S}$ create a single soft graviton, and these terms are present because gravitons are themselves charged under the broken symmetry generators. Such terms appear in the context of soft pion theorems if the coset is not a symmetric space, and we show this in more detail in Appendix \ref{softpions}.  In the language of general relativity, the presence of a single soft graviton will alter the metric, which will affect the action of the other charge.  This is familiar from the study of consistency relations for cosmological correlators, where the presence of a transverse traceless metric perturbation $\gamma_{ij}$ alters the consistency relations order by order in $\gamma$ \cite{Hinterbichler:2013dpa}.  

We might expect the soft and hard parts of the charges to obey the commutator algebra~(\ref{eq:BMSalg}) independently. However, we will see that while the hard charges do indeed obey $\langle out |\left[(\left[Q_{1}, Q_{2}\right]_{H} - \left[Q_{1H}, Q_{2H}\right]), \mathcal{S}\right] | in \rangle = i \langle out | \left[Q_{\left[1,2\right]H}, \mathcal{S}\right] | in \rangle $, the soft parts of the charges instead encode the extended structure found in \cite{Barnich:2011mi}
\begin{equation}\label{masterCommutator}
\begin{split}
\langle out |&\left[\left[Q_{1}, Q_{2}\right]_{S}, \mathcal{S}\right] | in \rangle \\
&= \langle out | \left[(\left[Q_{1S},Q_{2S}\right]+\left[Q_{1H},Q_{2S}\right]_{S} - \left[Q_{2H},Q_{1S}\right]_{S}), \mathcal{S}\right] | in \rangle\\
&= i\langle out | \left[(Q_{\left[1,2\right]S}+K_{(1,2)S}), \mathcal{S}\right] | in \rangle 
\end{split}
\end{equation}
with the extension term given (to leading order in $u$) by \cite{Barnich:2011mi}\footnote{Since our charges differ from those \cite{Barnich:2011mi} (see footnote 9) the extension term also differs from that given in~\cite{Barnich:2011mi} but is consistent with it up to the change in conventions.} 
\begin{equation}
\begin{split}
K_{(1, 2)} &=- \frac{1}{32\pi G}\int d^2 z\, \gamma_{z\bar{z}}C^{BC}\left(T_1 D_B D_C (D_{A}Y_2^{A}) - T_2 D_B D_C (D_A Y_1^{A})\right)\\
&=- \frac{1}{32\pi G}\int du d^2 z\, \gamma^{z\bar{z}}N_{zz}\left(T_1 D_{\bar{z}}^2 (D_{\bar{z}}Y_2^{\bar{z}}) - T_2 D_{\bar{z}}^2 (D_{\bar{z}}Y_1^{\bar{z}})\right) + h.c.\\
\end{split}
\end{equation} 
where in the last step we have assumed that $Y^z$ and $Y^{\bar{z}}$ at most have poles at infinity (see the discussion of subtleties involving poles at finite $z$ in the previous section). The extension term vanishes when we restrict to the global BMS algebra, as was already noted in \cite{Barnich:2011mi}.  This type of extended structure can arise when asymptotic symmetries act on manifolds with boundary, in which case the associated Noether charge can have a bulk and a boundary contribution.  We refer the reader to \cite{Banados:2016zim} for an instructive example in the context of Chern-Simons theory: there the integrated charges consist of corresponding bulk and boundary pieces, and when taking the commutator of two such charges, the commutator of the two boundary terms gives an additional boundary term that has no corresponding bulk piece. As also discussed in \cite{Avery:2015rga}, this term comes from the failure of the commutator of two asymptotic transformations to satisfy the gauge fixing conditions on the boundary as well as in the bulk. 
The extra boundary term is sometimes referred to as a central charge, or more precisely as a field-dependent central extension \cite{Barnich:2001jy, Barnich:2011mi, Barnich:2017ubf} since the corresponding bulk part of the charge is trivial.  While in the Chern-Simons example in \cite{Banados:2016zim} the extended terms can be thought of as a purely boundary effect, in gravity the situation is a little different, since here the current is a total derivative and there is no unambiguous definition of bulk and boundary terms\footnote{To guide our intuition and to connect with the example in \cite{Banados:2016zim}, however, we might choose to think of $Q_{H}$ as a bulk charge when it acts on hard momenta, since these should correspond to wavepackets with a finite extent in $u$, and to all effects creating or transforming a soft charge as boundary terms.}.  Here the ``bulk'' at future infinity is $\mathscr{I}^{+}$, and the ``boundary'' is given by the limiting two-spheres $\mathscr{I}^{+}_{\pm}$ without integrating over the null coordinate.

Since the extension term $K_{(1,2)}$ does not have a corresponding hard operator, this contribution break the symmetry.
The extension term was interpreted in \cite{Barnich:2011mi, Barnich:2017ubf} as a field-dependent central extension of the algebra giving rise to a Lie algebroid structure; because of the presence of the Bondi news in $K_{(1,2)}$, however, this operator does not commute with rest of the algebra.  Expressing the Bondi news in terms of creation and annihilation operators as before, we can write $K$ to leading order as
\begin{eqnarray}
K_{(1,2)S}&=&\frac{1}{8\pi \kappa} \int d^2z\left\{\bar{W}_{[1,2]}\lim_{\omega\to 0}\omega\left[a_+(\omega\hat{x})+a_-(\omega\hat{x})^\dagger\right] +h.c.\right\}\,,
\end{eqnarray}
where
\begin{equation}
\bar{W}_{[1,2]}=-4D_{\bar{z}}^2\bar{V}_{[1,2]}\,,
\end{equation}
with
\begin{equation}
\bar{V}_{[1,2]}=\frac{1}{8\pi}\int d^2w\frac{(1+w\bar{w})(\bar{w}-\bar{z})}{(1+z\bar{z})(w-z)}(T_2\partial^3_{\bar{w}}Y_1^{\bar{w}}-T_1\partial^3_{\bar{w}}Y_2^{\bar{w}})\,.
\end{equation}
If $V$ were real, this could simply be a leading soft charge with $T=-4V$, but since it is complex we cannot write it in this way.  Applying the soft-graviton theorem at leading order, writing only the terms involving $N_{zz}$, we have
\begin{equation}
\begin{split}
&\langle out | \left[K_{\left(1,2\right)}, \mathcal{S}\right] | in \rangle\supset\\ &-\frac{1}{8\pi}\int d^2 z \sum_{k}E_k\frac{(\bar{z}-\bar{z}_k)}{(z-z_k)}\frac{(1+z\bar{z})}{(1+z_k \bar{z}_k)}\left(T_1 \partial_{\bar{z}}^3 Y_2^{\bar{z}} - T_2 \partial_{\bar{z}}^3Y_1^{\bar{z}}\right)\langle out|\mathcal{S}|in\rangle + h.c.
\end{split}
\end{equation}
While this operator does not simply commute with the BMS transformations, we will confirm in \S \ref{cocyle} (and as found in \cite{Barnich:2011mi}) that the Jacobi identity continues to hold with the $K_{(1,2)}$ terms included, so the algebra is indeed well defined.  We will refer to the term $K_{(1,2)}$ as the extension term, since it indicates the existence of a modified Lie bracket for the algebra.  

The extension term we find here agrees with that in~\cite{Barnich:2011mi} to leading order in $u$, whereas an additional part subleading in $u$ found in \cite{Barnich:2011mi} does not appear.  It can be confirmed by explicit calculations at the level of the operators that this occurs precisely because the definition of the subleading charge in~(\ref{subleadingSoftCharge}) differs from that in~\cite{Barnich:2011mi}.  

\subsection{Operator commutators}

Before studying the charge algebra at the level of the amplitudes, we can attempt to evaluate the charge algebra directly at the level of the operators.
We can use the expressions for $Q = Q_S + Q_H$ in terms of creation and annihilation operators from the previous subsection and take the commutator.  The parts of the hard and soft charges which are leading and subleading in powers of $u$ are given by
\begin{eqnarray}
Q_{S}^{(0)} &=& \frac{1}{4\pi \kappa}\int d^2{z}\Big[D_{\bar{z}}^{2}T \lim_{\omega \to 0}\omega(a_{+}(\omega \hat{x}) + a_{-}(\omega \hat{x})^{\dagger}) + h.c. \Big]\,,\nonumber \\
Q_{H}^{(0)} &=& \frac{1}{16\pi^3}\int d^{2}z \, \gamma_{z\bar{z}} T \int_{0}^{\infty} d\omega\, \omega^2 \Big[a_{+}^{\dagger}(\omega \hat{x})a_{+}(\omega \hat{x})+a_{-}^{\dagger}(\omega \hat{x})a_{-}(\omega \hat{x})\Big] + \cdots \,, \nonumber \\
Q_{S}^{(1)} &=& -\frac{i}{4\pi \kappa}\int d^{2}z \Big[\partial_{\bar{z}}^{3}Y^{\bar{z}}\lim_{\omega \to 0}(1+\omega \partial_{\omega})\left[a_{+}(\omega \hat{x})-a_{-}(\omega \hat{x})^{\dagger}\right] - h.c.\Big]\,,\\
Q_{H}^{(1)} &=& -\frac{i}{16\pi^{3}}\int d^{2}z\,\gamma_{z\bar{z}}\int_{0}^{\infty} d\omega \Big\{\omega a_{+}^{\dagger}\Big[\frac12 D_{A}Y^{A}\omega \partial_{\omega} + (D_{\bar{z}}Y^{\bar{z}}-D_{z}Y^{z})\Big]a_{+}\nonumber\\*
&&\hskip 4cm+ \omega a_{-}^{\dagger}\Big[\frac12 D_{A}Y^{A}\omega \partial_{\omega} -( D_{\bar{z}}Y^{\bar{z}}-D_{z}Y^{z})\Big]a_{-}\nonumber\\*
&&\hskip 4cm- \omega a_{+}^{\dagger}Y^{A}D_{A}(a_{+})-\omega a_{-}^{\dagger}Y^{A}D_{A}(a_{-})\Big\} + \cdots \,. \nonumber
\end{eqnarray}
where the dots represent the contributions to the hard charges from matter as well as contributions that contain three or more creation and annihilation operators.  The commutators are straightforward to calculate, and the operators $Q_{H}$ act as supertranslations and superrotations on local operators such as $Q_{S}$.  The commutators of two soft charges $\left[Q_{1S}, Q_{2S}\right]$ are schematically
\begin{equation}
\begin{split}
[Q_{1S}^{(0)}, Q_{2S}^{(0)}] &\propto \int d^{2}z\, \gamma^{z\bar{z}} (D_{z}^2 T_1 D_{\bar{z}}^{2}T_2 - D_{\bar{z}}^2 T_1 D_{z}^{2}T_2)\,,\\
[Q_{1S}^{(0)}, Q_{2S}^{(1)}] &\propto \int d^{2}z\, \gamma^{z\bar{z}} (D_{z}^{2}T_{1}\partial_{\bar{z}}^{3}Y_{2}^{\bar{z}}- D_{\bar{z}}^{2}T_{1}\partial_{z}^{3}Y_{2}^{z})\,,\\
[Q_{1S}^{(1)}, Q_{2S}^{(1)}] &\propto \int d^{2}z\, \gamma^{z\bar{z}}\left(D_{z}^{2}(D_{A}Y_{1}^{A}))D_{\bar{z}}^{2}(D_{B}Y_{2}^{B}) - D_{z}^{2}(D_{A}Y_{2}^{A})D_{\bar{z}}^{2}(D_{B}Y_{1}^{B})\right)\,.
\end{split}
\end{equation}
These all vanish upon integration by parts in the angular variables.  Among the factors we have not written are delta functions in the soft momenta $\omega_1, \omega_2$, both of which are to be taken to zero.  To fix the order of soft limits, we take the commutator first before integrating in $u$, and this picks out the simultaneous double soft limit $\omega_1 = \omega_2 \to 0$.  

The remaining commutators (at leading order in the creation and annihilation operators) are given by
\begin{equation}\label{commutators}
\begin{split}
[Q_{1H}^{(0)}, Q_{2S}^{(1)}]_{S} &=\frac{i}{4\pi \kappa}\lim_{\omega \to 0} (1+\omega \partial_{\omega}) \int d^{2}z \, \Big[T_{1} \partial^3_{\bar{z}}Y_{2}^{\bar{z}}\omega(a_{+}+a_{-}^{\dagger})+T_{1} \partial^3_{z}Y_{2}^{z}\omega(a_{-}+a_{+}^{\dagger}) \Big]\,,\\
[Q_{1H}^{(1)}, Q_{2S}^{(0)}]_{S} &= \\
\frac{i}{4\pi \kappa}\lim_{\omega \to 0}\omega & \int d^{2}z\, \Big[D_{\bar{z}}^2 T_2\Big(\frac12D_{A}Y_{1}^{A}\omega \partial_{\omega} + ( D_{\bar{z}}Y_{1}^{\bar{z}}-D_{z}Y^{z}_{1}) - Y_{1}^{A}D_{A}\Big)a_{+}\\ 
&\hskip 0.9cm + D_{\bar{z}}^2 T_2\Big(\frac12 D_{A}Y_{1}^{A}\omega \partial_{\omega} + (D_{\bar{z}}Y_{1}^{\bar{z}}-D_{z}Y^{z}_{1}) - Y_{1}^{A}D_{A}\Big)a_{-}^\dagger\\
&\hskip 0.9cm + D_{z}^2 T_2\Big(\frac12 D_{A}Y_{1}^{A}\omega \partial_{\omega} - (D_{\bar{z}}Y_{1}^{\bar{z}}-D_{z}Y^{z}_{1}) - Y_{1}^{A}D_{A}\Big)a_{-}\\
&\hskip 0.9cm + D_{z}^2 T_2\Big(\frac12 D_{A}Y_{1}^{A}\omega \partial_{\omega} - (D_{\bar{z}}Y_{1}^{\bar{z}}-D_{z}Y^{z}_{1}) - Y_{1}^{A}D_{A}\Big)a_{+}^{\dagger}\Big]\,,\\
\end{split}
\end{equation}
\begin{equation}\nonumber
\begin{split}
[Q_{1H}^{(1)}, Q_{2S}^{(1)}]_{S} &= \\
\frac{1}{4\pi \kappa}\lim_{\omega \to 0}(1&+\omega \partial_{\omega})\int d^{2}z\, \Big[\partial_{\bar{z}}^{3}Y_{2}^{\bar{z}}\Big(\frac12D_{A}Y_{1}^{A}\omega \partial_{\omega} + (D_{\bar{z}}Y_{1}^{\bar{z}}-D_{z}Y^{z}_{1}) - Y_{1}^{A}D_{A}\Big)a_{+}\\ 
&\hskip 2.3cm + \partial_{\bar{z}}^{3}Y_{2}^{\bar{z}}\Big(\frac12D_{A}Y_{1}^{A}\omega \partial_{\omega} +(D_{\bar{z}}Y_{1}^{\bar{z}}-D_{z}Y^{z}_{1})- Y_{1}^{A}D_{A}\Big)a_{-}^{\dagger}\\
&\hskip 2.3cm + \partial_{z}^{3}Y_{2}^{z}\Big(\frac12D_{A}Y_{1}^{A}\omega \partial_{\omega} -(D_{\bar{z}}Y_{1}^{\bar{z}}-D_{z}Y^{z}_{1}) - Y_{1}^{A}D_{A}\Big)a_{-}\\
&\hskip 2.3cm + \partial_{z}^{3}Y_{2}^{z}\Big(\frac12D_{A}Y_{1}^{A}\omega \partial_{\omega} -(D_{\bar{z}}Y_{1}^{\bar{z}}-D_{z}Y^{z}_{1}) - Y_{1}^{A}D_{A}\Big)a_{+}^{\dagger}\Big]\,,\\
[Q_{1H}, Q_{2H}] &=i Q_{[1,2]H}\,.
\end{split}
\end{equation}
Combining the terms from $[Q_{1H}^{(0)}, Q_{2S}^{(1)}]_{S}$ and $[Q_{1H}^{(1)}, Q_{2S}^{(0)}]_{S}$, and integrating the $Y^{A}D_{A}$ terms by parts, we have
\begin{equation}\label{algebraFromOperators}
\begin{split}
&[Q_{1H}^{(0)}, Q_{2S}^{(1)}]_{S} + [Q_{1H}^{(1)}, Q_{2S}^{(0)}]_{S} - [Q_{2H}^{(0)}, Q_{1S}^{(1)}]_{S}- [Q_{2H}^{(1)}, Q_{1S}^{(0)}]_{S} = \\
&\frac{i}{4\pi \kappa}\lim_{\omega \to 0}\omega \int d^{2}z\,\Big( D_{\bar{z}}^{2}\Big(Y_{1}^{A}\partial_{A}T_{2} - \frac{1}{2}D_{A}Y_{1}^{A}T_{2}\Big)(a_{+} + a_{-}^{\dagger}) + D_{z}^{2}\Big(Y_{1}^{A}\partial_{A}T_{2} - \frac{1}{2}D_{A}Y_{1}^{A}T_{2}\Big)(a_{-} + a_{+}^{\dagger})\Big)\\
&-\frac{i}{8\pi \kappa}\lim_{\omega \to 0}\omega \int d^{2}z\,\Big(D_{\bar{z}}^{3}Y_{1}^{\bar{z}}T_2 (a_{+} + a_{-}^{\dagger}) +D_{z}^{3}Y_{1}^{z}T_2 (a_{-} + a_{+}^{\dagger}) \Big)\\
&+\frac{i}{8\pi \kappa}\lim_{\omega \to 0}\omega (1+ \omega \partial_{\omega}) \int d^{2}z\,\Big((D_{A}Y_{1}^{A}D_{\bar{z}}^{2}T_{2} + D_{\bar{z}}^{2}(D_{A}Y_{1}^{A})T_{2})(a_{+} + a_{-}^{\dagger}) \\& \qquad \qquad \qquad - (D_{A}Y_{1}^{A}D_{z}^{2}T_{2} + D_{z}^{2}(D_{A}Y_{1}^{A})T_{2})(a_{-} + a_{+}^{\dagger})\Big) - (1 \leftrightarrow 2)
\end{split}
\end{equation}
The first set of terms can be recognized as the leading (supertranslation) part of the operator $iQ_{[1,2]S}$, where $T(z,\bar{z})$ associated with the soft charge on the right hand side is given by
\begin{equation}
T_{[1,2]}=Y^A_1\partial_{A} T_2-\frac12T_2D_AY^A_1-(1\leftrightarrow 2)\,.
\end{equation}
The second set of terms corresponds to the leading part of the operator $iK_{(1,2)S}$, and the third set of terms will vanish when evaluated at the level of the amplitudes, because of the soft limit $\lim_{\omega \to 0}\omega(1+\omega \partial_{\omega})$.
Therefore, at subleading order in the charges, at the level of the amplitudes we have found
\begin{equation}
\langle out|[[Q_{1H}, Q_{2S}]_{S}+[Q_{1S}, Q_{2H}]_{S}, \mathcal{S}]|in\rangle=\langle out|[iQ_{[1,2]S}^{(0)},\mathcal{S}]|in\rangle+\langle out|[iK_{(1,2)}^{(0)},\mathcal{S}]|in\rangle\,,
\end{equation}

In the following subsections we will extract this commutator from double-soft scattering amplitudes and we will find that the two methods agree. The calculations here make it manifest
that this commutator can be derived from contact terms that arise when sequentially applying the single-soft limits.  First one soft graviton treats the other as hard, and the second soft graviton is then applied to the hard modes.  Here one single-soft factor acts on the soft momentum in the other.  The commutators $\left[Q_{H}, Q_{S}\right]_{S}$ therefore depend only on the sequential application of single-soft factors, which picks out a specific part of the double soft graviton amplitude that is singular in the collinear limit. 

We can similarly calculate the subsubleading commutators, and find
\begin{equation}
\begin{split}
&\langle out |[[Q_{1H}^{(1)}, Q_{2S}^{(1)}]_{S} + [Q_{1S}^{(1)}, Q_{2H}^{(1)}]_{S}, \mathcal{S}] | in \rangle\\ &\hskip 1cm= -\sum_{k}\left[\frac12D_{A}(Y_{1}^{B}\partial_{B}Y_2^{A}-Y_{2}^{B}\partial_{B}Y_1^{A})E_{k}\partial_{E_k}\right.\\
&\hskip 1.5cm \qquad + \frac{h_k}{2}\left[D_{\bar{z}}(Y_{1}^{A}\partial_{A}Y_{2}^{\bar{z}}-Y_{2}^{A}\partial_{A}Y_{1}^{\bar{z}}) - D_{z}(Y_{1}^{A}\partial_{A}Y_{2}^{z}-Y_{2}^{A}\partial_{A}Y_{1}^{z})\right] \\[.2cm]
&\hskip 1.5cm\qquad - \left. (Y_{1}^{B}\partial_{B}Y_{2}^{A}-Y_{2}^{B}\partial_{B}Y_{1}^{A})D_{A}\vphantom{\frac12}\right]\langle out | \mathcal{S} | in \rangle\,,
\end{split}
\end{equation}
consistent with
\begin{equation}
\langle out|[[Q_{1H}^{(1)}, Q_{2S}^{(1)}]_{S},\mathcal{S}]|in\rangle+\langle out|[[Q_{1S}^{(1)}, Q_{2H}^{(1)}]_{S},\mathcal{S}]|in\rangle=\langle out|[iQ_{[1,2]S}^{(1)},\mathcal{S}]|in\rangle\,,
\end{equation}
where the vector field associated with the charge on the right is
\begin{equation}
Y_{[1,2]}^B=Y_{1}^{A}\partial_AY_{2}^{B}-Y_{2}^{A}\partial_{A}Y_{1}^{B}\,.
\end{equation}
The soft parts of the commutators of the charges therefore realize the algebra
\begin{equation}
\langle out|[[Q_{1H}, Q_{2S}]_{S}+[Q_{1S}, Q_{2H}]_{S}, \mathcal{S}]|in\rangle=\langle out|[iQ_{[1,2]S},\mathcal{S}]|in\rangle+\langle out|[iK_{(1,2)},\mathcal{S}]|in\rangle\,,
\end{equation}
where the charge $Q_{[1,2]}$ is associated with the BMS transformation parametrized by
\begin{eqnarray}
T_{[1,2]}&=&\Big(Y^A_1\partial_{A} T_2-\frac12T_2D_AY^A_1\Big)-(1\leftrightarrow 2)\,,\\*
Y_{[1,2]}^B&=&Y_{1}^{A}\partial_AY_{2}^{B}-Y_{2}^{A}\partial_{A}Y_{1}^{B}\,.\nonumber
\end{eqnarray}
To derive the commutator for the hard part of the charges, we can either expand them in terms of creation and annihilation operators, or as a shortcut we can consider their action on other operators.  For the action on graviton operators, 
\begin{equation}
\begin{split}
\left[ a_{+}(E_k \hat{x}_k),\left[Q_{1H}, Q_{2H}\right]\right]& = -\left[\left[a_{+}(E_k \hat{x}_k),Q_{2H}\right],Q_{1H} \right] + \left[\left[a_{+}(E_k \hat{x}_k),Q_{1H}\right],Q_{2H}\right]\\
&= iE_k \left[Y_{1}^{A}\partial_{A}T_2 - \frac{1}{2}D_{A}Y_1^{A}T_{2} - (1 \leftrightarrow 2)\right]a_{+}(E_k \hat{x}_k) \\
&\quad+ \left[\frac12 D_{A}(Y_{1}^{B}\partial_{B}Y_2^{A})(E_{k}\partial_{E_k}) + D_{\bar{z}}(Y_{1}^{A}\partial_{A}Y_{2}^{\bar{z}}) - D_{z}(Y_{1}^{A}\partial_{A}Y_{2}^{z})\right.\\ &\qquad \left.-Y_{1}^{B}\partial_{B}Y_{2}^{A}D_{A} - (1 \leftrightarrow 2) \vphantom{\frac12}\right]a_{+}(E_k \hat{x}_k)\,,
\end{split}
\end{equation}
so that to subsubleading order
\begin{equation}
\langle out|[[Q_{1H}, Q_{2H},\mathcal{S}]|in\rangle=\langle out|[iQ_{[1,2]H},\mathcal{S}]|in\rangle\,.
\end{equation}

Among the commutators we have not derived directly from the operators are the hard pieces of the commutator $\left[Q_{H}, Q_{S}\right]_{H}$ -- these would arise from terms of cubic or higher order in $a$ and $a^{\dagger}$.  In the subsections to come we will show how the commutator algebra can be derived from the double soft amplitude, 
\begin{equation}
\begin{split}
&\hskip -1cm\lim_{[\omega_2 \to 0}\lim_{\omega_1 \to 0]}\sum\limits_{\lambda_1,\lambda_2}\int d^2 z_1d^2z_2\Psi_1(q_1)\Psi_2(q_2)\langle out \, q_1,q_2 | \mathcal{S} | in \rangle\\&\hskip 5cm= \langle out |\left[ \left[ Q_{1H}, Q_{2S}\right] + \left[ Q_{1S}, Q_{2H} \right],\mathcal{S}\right]  |in\rangle
\end{split}
\end{equation}
where the charge $\left[ Q_{1H}, Q_{2S}\right] + \left[ Q_{1S}, Q_{2H} \right]$ has soft and hard parts coming separately from the collinear and non-collinear parts of the amplitude.  In this way we will confirm that the double-soft graviton amplitude knows about both the commutator and the extension terms.

\subsection{Double soft graviton amplitude}

We will explore how the BMS commutator is realized by double soft graviton amplitudes, using the explicit expressions for the amplitude at tree level.  The relevant limit of the amplitude is primarily the antisymmetrized consecutive soft limit.  We have already seen how to write the single soft amplitudes in terms of the amplitude of the underlying hard process and soft factors, and we can similarly define the antisymmetrized consecutive double soft factor $S(q_1, q_2)$ as
\begin{equation}\label{Sdef}
\lim_{\left[\omega_2 \to 0\right.}\lim_{\left.\omega_1 \to 0\right]} \bar{\epsilon}_1^{\mu}\bar{\epsilon}_1^{\nu} \bar{\epsilon}_2^{\rho}\bar{\epsilon}_2^{\sigma}\mathcal{M}_{\mu \nu \rho \sigma}(q_1; q_2; p_1, \cdots p_n) = S(q_1, q_2)\mathcal{M}(p_1, \cdots, p_n)
\end{equation}
where $\mathcal{M}$ with and without indices refers to the matrix element with and without soft gravitons, and we are taking all of the hard momenta to be outgoing. 

To leading order in the soft momenta, the antisymmetrized consecutive double soft factor is given by
\begin{equation}\label{doublesoft}\nonumber
\begin{split}
S(q_1, q_2) &= S^{(1)}(q_1)\left\lbrace S^{(0)}(q_2)\right\rbrace - S^{(1)}(q_2)\left\lbrace S^{(0)}(q_1)\right\rbrace \\
& \qquad +\frac{\kappa}{2}\frac{(q_2 \cdot \bar{\epsilon}_1)^2}{(q_1\cdot q_2)} S^{(0)}(q_2) - \frac{\kappa}{2}\frac{(q_1 \cdot \bar{\epsilon}_2)^2}{(q_1\cdot q_2)} S^{(0)}(q_1)\\
\end{split}
\end{equation}
\begin{equation}
\begin{split}
&=\frac{\kappa^2}{4}\sum_{k}\Bigg[\frac{(p_k \cdot \bar{\epsilon}_1)^2}{(p_k \cdot q_1)}\left(\frac{2(p_k \cdot \bar{\epsilon}_2)(q_1 \cdot \bar{\epsilon}_2)}{(p_k \cdot q_2)} - \frac{(p_k \cdot \bar{\epsilon}_2)^2}{(p_k \cdot q_2)^2}(q_1 \cdot q_2)\right) \\&- (p_k \cdot \bar{\epsilon}_1)\left(\frac{2(p_k \cdot \bar{\epsilon}_2)(\bar{\epsilon}_1 \cdot \bar{\epsilon}_2)}{(p_k \cdot q_2)} - \frac{(p_k \cdot \bar{\epsilon}_2)^2}{(p_k \cdot q_2)^{2}}(\bar{\epsilon}_1 \cdot q_2)\right)\\
&+ \frac{(q_2 \cdot \bar{\epsilon}_1)^2(p_k \cdot \bar{\epsilon}_2)^2}{(q_1 \cdot q_2)(p_k \cdot q_2)}\left(1- \frac{(p_k \cdot q_1)}{(p_k \cdot q_2)}\right)-(q_2 \cdot \bar{\epsilon}_1)\left(-\frac{(p_k \cdot \bar{\epsilon}_2)^2(p_k \cdot \bar{\epsilon}_1)}{(p_k \cdot q_2)^2}\right)\\
&+\frac{(q_2 \cdot \bar{\epsilon}_1)(\bar{\epsilon}_1 \cdot \bar{\epsilon}_2)}{(q_1 \cdot q_2)}\left(\frac{2(p_k \cdot \bar{\epsilon}_2)(p_k \cdot q_1)}{(p_k \cdot q_2)}\right)\\
&-\frac{(q_2 \cdot \bar{\epsilon}_1)(q_1 \cdot \bar{\epsilon}_2)}{(q_1 \cdot q_2)}\left(\frac{2(p_k \cdot \bar{\epsilon}_2)(p_k \cdot \bar{\epsilon}_1)}{(p_k \cdot q_2)}\right)\Bigg] - (1\leftrightarrow 2)\,.
\end{split}
\end{equation}
This expression can be derived by taking the contact terms between single-soft factors. The last two lines make use of the fact that inside the soft factor and when acting on a gauge invariant amplitude we can take 
\begin{equation}
J^{\mu\nu}=p^\mu\frac{\partial}{\partial {p_\nu}}-p^\nu\frac{\partial}{\partial{p_{\mu}}} + \bar{\epsilon}^{\mu}\frac{\partial}{\partial{\bar{\epsilon}_\nu}} - \bar{\epsilon}^{\nu}\frac{\partial}{\partial{\bar{\epsilon}_\mu}}\,,
\end{equation}
where the derivatives with respect to the momenta only act on the explicit momentum dependence of the amplitude but not the momentum dependence of the polarization vectors~\cite{Bern:2014vva}. (See appendix~\ref{app:soft} for details.) This also shows that $S(q_1, q_2)$ is universal at this order, including quantum corrections, since the single-soft factors are\footnote{Remember the only possible loop corrections at this order come from the one-loop anomalous corrections to $S^{(1)}(q)$ in the collinear limit $q_1 \parallel q_2$.  The divergent piece can, however, be redefined away in the definition of the charges, as explained earlier in the single-soft limit case, and the dressed charge returns the tree-level contact term.}.  Note that this expression contains no terms of order $1/q^2$, which will be seen to be consistent with the fact that two supertranslations commute.  


For a given matter content, the expression for the antisymmetrized double-soft amplitude can, of course, also be derived by starting with the full tree-level amplitude to next to leading order (NLO) in the soft momenta, calculated using Feynman diagrams, and taking the appropriate consecutive soft limits. Here we explicitly provide a check for the scattering of $n$ scalars and two soft gravitons.  The full amplitude to NLO in the soft momenta is\footnote{See for instance \cite{BjerrumBohr:2004mz, Holstein:2006bh} for the explicit expressions for the graviton propagators and couplings.}:
\begin{align}\label{fullAmplitude}
&\bar{\epsilon}_1^{\mu}\bar{\epsilon}_1^{\nu} \bar{\epsilon}_2^{\rho}\bar{\epsilon}_2^{\sigma}\mathcal{M}_{\mu \nu \rho \sigma}(q_1; q_2; p_1, \cdots p_n) \nonumber \\
&= \Bigg[\sum_{j,k}\frac{\kappa^2}{4} \Bigg[\frac{(\bar{\epsilon}_{1} \cdot p_j)^2}{(p_j \cdot q_1)}\frac{(\bar{\epsilon}_2 \cdot p_k)^2}{(p_k \cdot q_2)} + \frac{(\bar{\epsilon}_{1} \cdot p_j)^2}{(p_j \cdot q_1)}\frac{(\bar{\epsilon}_2 \cdot p_k)(\bar{\epsilon}_{2\mu}q_{2\rho}J^{\mu \rho}_k)}{(p_k \cdot q_2)}  \nonumber \\
&+ \frac{(\bar{\epsilon}_{2} \cdot p_k)^2}{(p_k \cdot q_2)}\frac{(\bar{\epsilon}_1 \cdot p_j)(\bar{\epsilon}_{1\mu}q_{1\rho}J^{\mu \rho}_j)}{(p_j \cdot q_1)}\Bigg] \nonumber \\
&+\sum_{k} \Bigg[\frac{\kappa^2}{4}\frac{(\bar{\epsilon}_1 \cdot p_k)^2(\bar{\epsilon}_2 \cdot p_k)^2}{(p_k \cdot q_1)(p_k \cdot q_2)}\left(\frac{-q_1 \cdot q_2}{p_k \cdot (q_1 + q_2)}\right) \nonumber \\
&+ \frac{\kappa^2}{2}\left(\frac{(\bar{\epsilon}_1 \cdot p_k)^2 (\bar{\epsilon}_2 \cdot p_k)(\bar{\epsilon}_2 \cdot q_1)}{(p_k \cdot q_1)p_k \cdot(q_1 + q_2)} + \frac{(\bar{\epsilon}_2 \cdot p_k)^2 (\bar{\epsilon}_1 \cdot p_k)(\bar{\epsilon}_1 \cdot q_2)}{(p_k \cdot q_2)p_k \cdot(q_1 + q_2)} \right) \\
&-\frac{\kappa^2}{p_k \cdot (q_1 + q_2)}(\bar{\epsilon}_1 \cdot \bar{\epsilon}_2)(\bar{\epsilon}_1 \cdot p_k)(\bar{\epsilon}_2 \cdot p_k) \nonumber \\
&+\frac{\kappa^2}{4(q_1 \cdot q_2)p_k \cdot (q_1 + q_2)}\Bigg\{(\bar{\epsilon}_1 \cdot \bar{\epsilon}_2)^2\Big[(p_k \cdot q_1)^2 + (p_k \cdot q_2)^2 + (p_k \cdot q_1)(p_k \cdot q_2)\Big] \nonumber \\
&\qquad + (\bar{\epsilon}_1 \cdot p_k)^2 (q_1 \cdot \bar{\epsilon}_2)^2 + (\bar{\epsilon}_2 \cdot p_k)^2 (q_2 \cdot \bar{\epsilon}_1)^2 - 2(\bar{\epsilon}_1 \cdot q_2)(\bar{\epsilon}_2 \cdot q_1)(\bar{\epsilon}_1 \cdot p_k)(\bar{\epsilon}_2 \cdot p_k) \nonumber \\
&\qquad + (\bar{\epsilon}_1 \cdot \bar{\epsilon}_2)\Big[2(q_1 \cdot q_2)(\bar{\epsilon}_1 \cdot p_k)(\bar{\epsilon}_2 \cdot p_k) - 2(q_2 \cdot \bar{\epsilon}_1)(\bar{\epsilon}_2 \cdot p_k)(p_k \cdot q_2)\nonumber \\
&\qquad - 2(q_1 \cdot \bar{\epsilon}_2)(\bar{\epsilon}_1 \cdot p_k)(p_k \cdot q_1)\Big]\Bigg\}\Bigg]\Bigg]\mathcal{M}(p_1, \cdots, p_n)\, \nonumber.
\end{align}
Here the first two lines on the right hand side come from the insertions of external lines on separate external legs, and also from insertions on internal legs which are necessary to preserve gauge invariance.  The third and fourth lines come from the subleading contributions when two gravitons insert into separate points in the same external leg (``Born'' terms).  The fifth line comes from graviton seagull terms on the external legs, and the last four lines come from the graviton pole diagram, where a three-way graviton vertex inserts a single graviton into an external leg. As before, $\kappa^2 = 32\pi G$; and we have written the graviton polarization tensors as $\bar{\epsilon}^{\mu \nu} = \bar{\epsilon}^{\mu}\bar{\epsilon}^{\nu}$, which is always possible for gravitons of definite helicity; and the angular momentum operator for scalars is 
\begin{equation}
J_{k}^{\mu \nu} = p_k^{\mu}\frac{\partial}{\partial {p_{k\,\nu}}} - p_k^{\nu}\frac{\partial}{\partial {p_{k\mu}}}\,.
\end{equation}  

It is straightforward to check that the full amplitude reproduces the expression in equation \eqref{doublesoft} in the appropriate limits, and that the full amplitude is gauge invariant under the separate gauge symmetries $\bar{\epsilon}_{1,2}^{\mu} \to \bar{\epsilon}_{1,2}^{\mu} + \lambda_{1,2} q^{\mu}$.  Checking the gauge invariance $\bar{\epsilon}_{1}^{\mu} \to \bar{\epsilon}_{1}^{\mu} + \lambda q^{\mu}_{1}$  explicitly, we find that to linear order in $\lambda$,
\begin{eqnarray}
&&\hskip -1.cm\Delta\left( \bar{\epsilon}_1^{\mu}\bar{\epsilon}_1^{\nu} \bar{\epsilon}_2^{\rho}\bar{\epsilon}_2^{\sigma}\mathcal{M}_{\mu \nu \rho \sigma}(q_1; q_2; p_1, \cdots p_n)\right)\nonumber\\
&&=\sum_{j,k}\frac{\kappa^2}{2} (\bar{\epsilon}_1 \cdot p_j)\Bigg[\frac{(\bar{\epsilon}_2 \cdot p_k)^2}{(p_k \cdot q_2)} + \frac{(\bar{\epsilon}_2 \cdot p_k)(\bar{\epsilon}_{2\mu}q_{2\rho}J^{\mu \rho}_k)}{(p_k \cdot q_2)}\Bigg] + \sum_{j,k}\frac{\kappa^2}{4}(\bar{\epsilon}_{1\mu}q_{1\rho}J_{j}^{\mu \rho})\frac{(\bar{\epsilon}_2 \cdot p_k)^2}{(p_k \cdot q_2)}\nonumber\\
&&+\sum_{k}\frac{\kappa^2}{2}\Bigg[\frac{(\bar{\epsilon}_1 \cdot q_2)(\bar{\epsilon}_2 \cdot p_k)^2}{(p_k \cdot q_2)} - (\bar{\epsilon}_1 \cdot \bar{\epsilon}_2)(\bar{\epsilon}_2 \cdot p_k) + \frac{(p_k \cdot q_2)(q_1 \cdot \bar{\epsilon}_2)(\bar{\epsilon}_1 \cdot \bar{\epsilon}_2)}{(q_1 \cdot q_2)}\nonumber \\
&&- \frac{(q_1 \cdot \bar{\epsilon}_2)(q_2 \cdot \bar{\epsilon}_1)(\bar{\epsilon}_2 \cdot p_k)}{(q_1 \cdot q_2)}\Bigg]+\left(1 \leftrightarrow 2\right)\,.
\end{eqnarray}
These terms vanish by conservation of total momentum and angular momentum.  Note that the first of the subleading terms combines with the leading term to ensure total momentum conservation.  Equivalently, one can check that the expression for $S(q_1, q_2)$ in equation \eqref{doublesoft} is gauge invariant after antisymmetrization, although a single consecutive double soft limit need not be because the process of taking the soft limit does not necessarily commute with a general gauge transformation.

While the full amplitude for two soft graviton insertions is symmetric under exchange of the two soft graviton indices 1 and 2, as it must be for two identical bosons, the antisymmetrized consecutive double-soft limit in \eqref{doublesoft}, which involves the subtraction of different kinematic limits, retains the information about the commutator.  From the general form of \eqref{doublesoft}, the first two lines come from contact terms between the single-soft factors acting on the hard modes, and the last four lines come from the contact terms where a hard mode acts on the other soft graviton, treating it as a (relatively) hard mode.  The first set of terms therefore correspond to the terms $\left[Q_{1H}, Q_{2S}\right]_{H} + \left[Q_{1S},Q_{2H}\right]_{H}$, and the second set corresponds to $\left[Q_{1H}, Q_{2S}\right]_{S} + \left[Q_{1S},Q_{2H}\right]_{S}$.  Comparing to the expressions in~\eqref{eq:BMSalg} and reading off the weights for the leading~(\ref{eq:Q0}), (\ref{Nzz2}) and subleading part of the charge~(\ref{eq:Q1}), we therefore have
\begin{equation}\label{softantisymm}
\begin{split}
&\langle out | \left[(\left[Q_{1H}, Q_{2S}\right] + \left[Q_{1S}, Q_{2H}\right]), \mathcal{S}\right] | in \rangle\\
&= \frac{1}{4\pi^2\kappa^2}\lim_{\left[\omega_2 \to 0\right.}\lim_{\left.\omega_1 \to 0\right]}\int d^2 z_1 d^2 z_2  \\
&\hskip 1.2cm\Big\{ D_{\bar{z}_1}^2 T_1 D_{z_2}^2 T_2\omega_1 \omega_2 \langle out | a_{+}(\omega_1 \hat{x}_1)a_{-}(\omega_2 \hat{x}_2)\mathcal{S} | in \rangle \\
&\hskip 1cm - i D_{\bar{z}_1}^2 T_1 \partial_{z_2}^3Y_2^{z_2} \omega_1 (1 + \omega_2 \partial_{\omega_2})\langle out | a_{+}(\omega_1 \hat{x}_1)a_{-}(\omega_2 \hat{x}_2)\mathcal{S} | in \rangle \\
&\hskip 1cm -  i \partial_{\bar{z}_1}^3Y_1^{\bar{z}_1} D_{z_2}^2 T_2\omega_2 (1 + \omega_1 \partial_{\omega_1})\langle out | a_{+}(\omega_1 \hat{x}_1)a_{-}(\omega_2 \hat{x}_2)\mathcal{S} | in \rangle \\
&\hskip 1cm- \partial_{\bar{z}_1}^3Y_1^{\bar{z}_1}\partial_{z_2}^3 Y_2^{z_2}(1+\omega_1 \partial_{\omega_1})(1+\omega_2 \partial_{\omega_2})\langle out | a_{+}(\omega_1 \hat{x}_1)a_{-}(\omega_2 \hat{x}_2)\mathcal{S} | in \rangle \\
&\hskip 1cm+ \cdots \Big\}
\end{split}
\end{equation}
where we have expanded the charges order by order in $u$, and written only the $(1_{+}2_{-})$ helicity terms for illustrative purposes.  To sum over helicities, the other terms can be generated by switching between holomorphic and antiholomorphic expressions for the test functions\footnote{The factors of $i$ out front do not get conjugated, however, since they come from the Fourier transform in $u$.}.  We will show that \eqref{softantisymm} becomes
\begin{equation}\label{softantisymm2}
\begin{split}
\langle out | &\left[(\left[Q_{1H}, Q_{2S}\right] + \left[Q_{1S}, Q_{2H}\right]), \mathcal{S}\right] | in \rangle \\
&= \langle out | \left[(iQ_{[1,2]H} + iQ_{[1,2]S} + iK_{(1,2)S}), \mathcal{S}\right] | in \rangle
\end{split}
\end{equation}
with the separate hard and soft pieces corresponding corresponding to different pole structures in the double-soft amplitude.  We will therefore confirm the identities in \eqref{eq:BMSalg}, and also confirm the identity \eqref{masterCommutator} realizing the extended BMS algebra.  

The entire combination of terms in \eqref{softantisymm2} is gauge invariant, as it depends on a gauge-invariant amplitude.  The individual terms on the right hand side are not, but this is not a problem, and is even to be expected, since we are computing residual gauge symmetries for Bondi gauge \textit{after} having fixed the gauge.  Another technical point we should emphasize is that in order to derive an equivalence between non-gauge invariant quantities such as $[Q_{H}, Q_{S}]_{S}$ and $iQ_{[1,2]S} +iK_{(1,2)S}$ we must pick the same choice of gauge for all soft gravitons in the problem.  In particular the gauge choice in \cite{He:2014laa}, which makes the same choice of reference vector for all soft gravitons, is a good choice, but the choice $\bar{\epsilon}_1 \cdot q_2 = \bar{\epsilon}_2 \cdot q_1 = 0$, where we use gauge invariance separately for the first and second soft gravitons, is not.

The reader may want to consult the appendices first as a warm up: in Appendix \ref{softpions} we review the case of how soft pion amplitudes realize the corresponding algebra, and in Appendix \ref{YangMills} we study the case of asymptotic gauge Yang-Mills theory, which is conceptually similar to the gravitational case and technically much simpler.  
The leading term on the left hand side of \eqref{softantisymm} is then
\begin{equation}
\begin{split}
&\hskip -1.5cm\langle out | \left[(\left[Q_{1H}, Q_{2S}\right] + \left[Q_{1S}, Q_{2H}\right]), \mathcal{S}\right] | in \rangle \supset \\
&\frac{1}{4\pi^2 \kappa^2}\lim_{\left[\omega_2 \to 0\right.}\lim_{\left.\omega_1 \to 0\right]}\int d^2 z_1 d^2 z_2 \,  \Big\{ D_{\bar{z}_1}^2 T_1 D_{z_2}^2 T_2\omega_1 \omega_2 \langle out | a_{+}a_{-}\mathcal{S} | in \rangle \Big\}
\end{split}
\end{equation}
and will vanish after antisymmetrization, consistent with the fact that two supertranslations commute.  
The subleading terms can be written as
\begin{equation}\label{subleading}
\begin{split}
&\hskip -2cm\langle out | \left[(\left[Q_{1H}, Q_{2S}\right] + \left[Q_{1S}, Q_{2H}\right]), \mathcal{S}\right] | in \rangle \supset \\
-\frac{i}{4\pi^2 \kappa^2}&\lim_{\left[\omega_2 \to 0\right.}\lim_{\left.\omega_1 \to 0\right]}\int d^2 z_1 d^2 z_2 \, \\
&\Big\{ D_{\bar{z}_1}^2 T_1 \partial_{z_2}^3Y_2^{z_2} \omega_1 (1 + \omega_2 \partial_{\omega_2})\langle out | a_{+}a_{-}\mathcal{S} | in \rangle \\
&\hskip -0.2cm+   \partial_{\bar{z}_1}^3 Y_1^{\bar{z}_1} D_{z_2}^2 T_2\omega_2 (1 + \omega_1 \partial_{\omega_1})\langle out | a_{+}a_{-}\mathcal{S} | in \rangle\Big\} \\
\end{split}
\end{equation}
and will be related to the commutator of supertranslations and superrotations. The subsubleading terms can similarly be written as
\begin{equation}\label{subsubleading}
\begin{split}
&\hskip -0.5cm\langle out | \left[(\left[Q_{1H}, Q_{2S}\right] + \left[Q_{1S}, Q_{2H}\right]), \mathcal{S}\right] | in \rangle \supset
 \\
&-\frac{1}{4\pi^2 \kappa^2}\lim_{\left[\omega_2 \to 0\right.}\lim_{\left.\omega_1 \to 0\right]}\int d^2 z_1 d^2 z_2 \, \\
&\hskip 1.5cm\Big\{\partial_{\bar{z}_1}^3Y_1^{\bar{z}_1}\partial_{z_2}^3 Y_2^{z_2} (1+\omega_1 \partial_{\omega_1})(1+\omega_2 \partial_{\omega_2})\langle out | a_{+}a_{-}\mathcal{S} | in \rangle \Big\}
\end{split}
\end{equation}
and will be related to the commutator of two superrotations.

We should note that the antisymmetrized consecutive double-soft limit is also the relevant one for cosmological soft-pion theorems. In the case of double-soft limits for the adiabatic modes for the curvature $\zeta$ in unitary gauge in-in cosmological correlators~\cite{Joyce:2014aqa, Mirbabayi:2014zpa} only this limit satisfies all the necessary constraints to correspond to an adiabatic mode at second order\footnote{These are known as adiabatic mode conditions, and they ensure that the transformation satisfies the same constraint equations as a physical mode at small but nonzero momentum; a more complete discussion can be found e.g. in~\cite{Hinterbichler:2013dpa}.}.  More specifically, for cosmology in unitary gauge, performing a dilatation and then a special conformal transformation gives a configuration which is indistinguishable from a second order adiabatic mode and can be transformed away.  Performing the SCT and then the dilatation, however, we get the sum of this adiabatic mode and another SCT, which is a sum of adiabatic modes rather than a single mode.  The additional piece is consistent, however, with the expected commutator $\left[\mbox{D}, \mbox{SCT}\right] \propto \mbox{SCT}$ for the algebra of conformal symmetries acting on the spatial slices.

An alternate prescription for the soft limits was used in \cite{He:2015zea, Cheung:2016iub}, where the soft limit was taken first for gravitons of one helicity, and then for the other helicity.  As was the case the cosmological correlators, however, it should ultimately be checked whether a given prescription satisfies the adiabatic mode conditions; although we have not checked explicitly at second order in the metric perturbations, we expect that in the BMS case as well only the antisymmetrized consecutive double soft limit will satisfy appropriate adiabatic mode conditions.

\subsection{BMS commutator at leading order}

Examining the expression \eqref{subleading} in terms of the soft graviton amplitudes, the left hand side depends upon the antisymmetrized consecutive double-soft factor $S(q_1, q_2)$, and is therefore gauge invariant.  The subleading charge commutator then becomes
\begin{equation}\label{subleading2}
\begin{split}
&\hskip -1.5cm\langle out | \left[(\left[Q_{1H}, Q_{2S}\right] + \left[Q_{1S}, Q_{2H}\right]), \mathcal{S}\right] | in \rangle \supset\\
&-\frac{i}{4\pi^2 \kappa^2}\lim_{\omega_1 \to 0}\lim_{\omega_2 \to 0}\int d^2 z_1 d^2 z_2 \, \\
&\hskip 2.2cm\Big\{ D_{\bar{z}_1}^2 T_1 \partial_{\bar{z}_2}^3 Y_2^{\bar{z}_2} \omega_1 (1 + \omega_2 \partial_{\omega_2})S(q_1, q_2)\\
 &\hskip 2cm+  D_{\bar{z}_2}^2 T_2 \partial_{\bar{z}_1}^3 Y_1^{\bar{z}_1} \omega_2 (1 + \omega_1 \partial_{\omega_1})S(q_1, q_2)\Big\}\langle out|\mathcal{S}| in \rangle\,,
\end{split}
\end{equation}
where the ellipses indicate a sum over helicities (although we have shown only the $(1_{+}2_{+})$ term in the equation above).  The first set of terms in \eqref{subleading2} picks out the part of the amplitude proportional to $1/q_1$, and the second set picks out the terms proportional to $1/q_2$.  For our present discussion it will be convenient to break the antisymmetrized consecutive double-soft factor in \eqref{doublesoft} up into different contributions
\begin{equation}
S(q_1, q_2)=S_1(q_1, q_2)+S_2(q_1, q_2)+S_3(q_1, q_2)-(1\leftrightarrow 2)\,,
\end{equation}
with different pole structures. We have terms that are singular as $q_1$ or $q_2$ are taken to zero or become collinear with one of the hard momenta
\begin{equation}\label{doublesoft1}
\begin{split}
S_1(q_1, q_2) &=\frac{\kappa^2}{4}\sum_k \Bigg[\frac{(p_k \cdot \bar{\epsilon}_1)^2}{(p_k \cdot q_1)}\left(\frac{2(p_k \cdot \bar{\epsilon}_2)(q_1 \cdot \bar{\epsilon}_2)}{(p_k \cdot q_2)} - \frac{(p_k \cdot \bar{\epsilon}_2)^2}{(p_k \cdot q_2)^2}(q_1 \cdot q_2)\right)\\
&\hskip 2cm -(p_k \cdot \bar{\epsilon}_1)\left(\frac{2(p_k \cdot \bar{\epsilon}_2)(\bar{\epsilon}_1 \cdot \bar{\epsilon}_2)}{(p_k \cdot q_2)} - \frac{(p_k \cdot \bar{\epsilon}_2)^2}{(p_k \cdot q_2)^2}(\bar{\epsilon}_1 \cdot q_2)\right)\Bigg]\,,
\end{split}
\end{equation}
and terms that are singular as $q_1$ and $q_2$ become collinear
\begin{equation}\label{doublesoft2}
\begin{split}
S_2(q_1, q_2) &= -\frac{\kappa^2}{4}\sum_k \Bigg[\frac{(q_1 \cdot \bar{\epsilon}_2)^2(p_k \cdot \bar{\epsilon}_1)^2}{(q_1 \cdot q_2)(p_k \cdot q_1)}\Bigg]\,,
\end{split}
\end{equation}
and
\begin{equation}\label{doublesoft3}
S_3(q_1, q_2)= \frac{\kappa^2}{4}\sum_k \Bigg[ \frac{(q_2 \cdot \bar{\epsilon}_1)}{(q_1 \cdot q_2)}(\bar{\epsilon}_1 q_1 J_2)\left\{\frac{(p_k \cdot \bar{\epsilon}_2)^2}{(p_k \cdot q_2)}\right\}\Bigg] \,.
\end{equation}
The individual contributions to the amplitude are not gauge-invariant, but as explained in the previous section, they do not have to be. 

We transform equations \eqref{subleading2} and \eqref{doublesoft1}-\eqref{doublesoft3} to holomorphic coordinates as before using \eqref{Stromingergauge} and \eqref{basicExpressions}.  We first focus on the terms in \eqref{doublesoft2} proportional to $1/q_{2}$; the terms proportional to $1/q_{1}$ follow by interchanging the labels.

Expressing equation \eqref{doublesoft1} in holomorphic coordinates, we have terms which are singular as $z_1 \to z_k$ and as $z_2 \to z_k$
\begin{equation}\label{doublesoftpartone}
\begin{split}
S_1(q_1, q_2) = \begin{dcases} -\frac{\kappa^2}{4} \sum_k \frac{E_k}{\omega_2}\frac{(\bar{z}_1 - \bar{z}_2)(\bar{z}_1-\bar{z}_k)}{(z_1-z_k)(z_2-z_k)}\frac{(1+z_2\bar{z}_2)}{(1+z_k \bar{z}_k)} \hskip 2.95cm (1_{+}2_{+})\\
 \frac{\kappa^2}{4} \sum_k \frac{E_k}{\omega_2}\frac{(z_2-z_k)(\bar{z}_1 + \bar{z}_2 - 2\bar{z}_k)(\bar{z}_1-\bar{z}_k)}{(z_1-z_k)(\bar{z}_2-\bar{z}_k)^2}\frac{(1+z_2\bar{z}_2)}{(1+z_k \bar{z}_k)} \hskip 0.65cm (1_{+}2_{-})\,.\end{dcases}
\end{split}
\end{equation}
To flip the helicities, we can simply take the complex conjugates. The second contribution~\eqref{doublesoft2} is singular as $z_1 \to z_2$ and as $z_2 \to z_k$:
\begin{equation}\label{doublesoftparttwo}
\begin{split}
S_2(q_1, q_2) =\begin{dcases} 
 -\frac{\kappa^2}{4}\sum_k \frac{E_k}{\omega_2}\frac{(\bar{z}_1-\bar{z}_k)(\bar{z}_1-\bar{z}_2)(1+z_2 \bar{z}_2)}{(z_1 - z_k)(z_1 - z_2)(1+z_k \bar{z}_k)}\qquad (1_{+}2_{+})\\
 -\frac{\kappa^2}{4}\sum_k \frac{E_k}{\omega_2}\frac{(\bar{z}_1-\bar{z}_k)(z_1-z_2)(1+z_2 \bar{z}_2)}{(z_1-z_k)(\bar{z}_1-\bar{z}_2)(1+z_k \bar{z}_k)}\qquad (1_{+}2_{-})\,.\end{dcases}
\end{split}
\end{equation}
The remaining terms can be written more explicitly as
\begin{equation}
\begin{split}
S_3(q_1, q_2) =& -\frac{\kappa^2}{4}\Bigg[\frac{(q_2 \cdot \bar{\epsilon}_1)^2(p_k \cdot \bar{\epsilon}_2)^2}{(q_1 \cdot q_2)(p_k \cdot q_2)}\frac{(p_k \cdot q_1)}{(p_k \cdot q_2)}-(q_2 \cdot \bar{\epsilon}_1)\left(\frac{(p_k \cdot \bar{\epsilon}_2)^2(p_k \cdot \bar{\epsilon}_1)}{(p_k \cdot q_2)^2}\right)\\
&\hskip -1.5cm-\frac{(q_2 \cdot \bar{\epsilon}_1)(\bar{\epsilon}_1 \cdot \bar{\epsilon}_2)}{(q_1 \cdot q_2)}\left(\frac{2(p_k \cdot \bar{\epsilon}_2)(p_k \cdot q_1)}{(p_k \cdot q_2)}\right) +\frac{(q_2 \cdot \bar{\epsilon}_1)(q_1 \cdot \bar{\epsilon}_2)}{(q_1 \cdot q_2)}\left(\frac{2(p_k \cdot \bar{\epsilon}_2)(p_k \cdot \bar{\epsilon}_1)}{(p_k \cdot q_2)}\right)\Bigg]\,,
\end{split}
\end{equation}
and in holomorphic coordinates they become
\begin{equation}
\begin{split}\label{doublesoftpartthree}
S_3(q_1, q_2) &= \begin{dcases}\frac{\kappa^2}{4}\sum_k \frac{E_k}{\omega_2}\frac{(\bar{z}_1 - \bar{z}_2)}{(z_1 - z_2)}\frac{(\bar{z}_1 - \bar{z}_k)}{(z_2 - z_k)}\frac{(1+z_2 \bar{z}_2)}{(1+z_k \bar{z}_k)}\hskip 3.7cm (1_{+}2_{+})\\
 -\frac{\kappa^2}{4}\sum_k \frac{E_k}{\omega_2}\frac{(z_2 - z_k)(\bar{z}_1 + \bar{z}_2 - 2\bar{z}_k)(\bar{z}_1 - \bar{z}_k)}{(\bar{z}_2 - \bar{z}_k)^2(z_1 - z_2)}\frac{(1+z_2 \bar{z}_2)}{(1+z_k \bar{z}_k)}\qquad (1_{+}2_{-})\,.\end{dcases}
\end{split}
\end{equation}

To integrate over the moduli space of soft momentum directions, as for the single soft limits, we will integrate by parts in $z_1$ and $z_2$ and assume that there are no boundary terms at infinity, although whether this is ultimately justified will depend on our choice of fall-off conditions for $T$ and $Y^{A}$. In addition to the global structure of moduli space, we potentially need to consider the local structure arising at the loci where $z_1$, $z_2$, and $z_k$ all come together. Such multiple-collisions can be subtle and a different set of coordinates (conformal cross-ratios) may be required to obtain a correct local description\footnote{Na\"ively, the (compactified) moduli space of the $n$-punctured Riemann sphere looks like $(n-3)$ copies of $\mathbb{CP}^1$. But this picture breaks down near the boundary, when multiple punctures collide. The correct description of the boundary \cite{Keel:1989} requires a sequence of blowups of the na\"ive space, and the conformal cross-ratios provide good coordinates near the exceptional divisors.  In the case at hand, $\overline{\mathcal{M}}_{0,5}$ is not the na\"ive $\mathbb{CP}^1\times \mathbb{CP}^1$, parametrized by $z_1$ and $z_2$, but rather its blowup at 3 points (the del Pezzo${}_4$ surface).} --- e.g.~to show that certain terms will vanish upon integration. In our case, however, since the answer is finite, we can afford to ignore such subtleties and stick with $z_1$ and $z_2$ as coordinates in what follows.

To compute the BMS commutator at leading order, we insert the holomorphic expressions for the amplitude into the expressions for the charge in \eqref{subleading2}, summing over the helicities of both gravitons, and then integrate by parts in $z_1$ and $z_2$.  

\subsubsection{$\left[Q_{1H}, Q_{2S}\right]_{H} + \left[Q_{1S}, Q_{2H}\right]_{H}$ contact terms}
We start with the contact terms between single-soft limits, and begin with the contributions $(1_{+}2_{+})$ where both gravitons have the same (positive) helicity.  Plugging \eqref{doublesoftpartone} into \eqref{subleading2}, we have
\begin{eqnarray}
&&\hskip -3.3cm\langle out | \left[ (\left[Q_{1H}, Q_{2S}\right]_{H} + \left[Q_{1S}, Q_{2H}\right]_{H}), \mathcal{S}\right] | in \rangle \supset\nonumber\\
&&\hskip -1.3cm\frac{i}{16\pi^2}\int d^2 z_1 d^2 z_2\, \partial_{\bar{z}_1}^{3}Y_1^{\bar{z}_1}D_{\bar{z}_2}^{2}T_2 \\
&&\hskip .2cm\times \sum_k E_k\frac{(\bar{z}_1 - \bar{z}_2)(\bar{z}_1-\bar{z}_k)}{(z_1-z_k)(z_2-z_k)}\frac{(1+z_2\bar{z}_2)}{(1+z_k \bar{z}_k)}\langle out|\mathcal{S}|in\rangle\nonumber\\
&&\hskip -0.2cm-(1 \leftrightarrow 2)\,.\nonumber
\end{eqnarray}
Integrating by parts in $\bar{z}_1$ and making use of the Cauchy-Pompeiu formula $\partial_{\bar{z}_1}\left(\frac{1}{z_1-z_k}\right) = (2\pi)\delta^{(2)}(z_1-z_k)$, we have
\begin{equation}
\begin{split}
\langle out | \left[ \right.&\left.(\left[Q_{1H}, Q_{2S}\right]_{H} + \left[Q_{1S}, Q_{2H}\right]_{H}), \mathcal{S}\right] | in \rangle \supset\\
&\frac{i}{16\pi^2} \int d^2 z_2 \sum_k E_k \Bigg[-\frac{2\pi(\bar{z}_2-\bar{z}_k)(1+z_2 \bar{z}_2)}{(z_2-z_k)(1+z_k \bar{z}_k)}\partial_{\bar{z}_k}Y_1^{\bar{z}_k}D_{\bar{z}_{2}}^{2}T_2 \\
&\hskip 3.2cm- \frac{4\pi(1+z_2 \bar{z}_2)}{(z_2-z_k)(1+z_k \bar{z}_k)}Y_1^{\bar{z}_k}D_{\bar{z}_{2}}^{2}T_2\Bigg]\langle out|\mathcal{S}|in\rangle\\
&\hskip 8.5cm -(1\leftrightarrow 2)\,.
\end{split}
\end{equation}
Integrating by parts in $z_2$, we have
\begin{equation}
\begin{split}
\langle out | \left[ (\left[Q_{1H}, Q_{2S}\right]_{H} + \left[Q_{1S}, Q_{2H}\right]_{H}), \mathcal{S}\right] | in \rangle \supset\\
&\hskip -4cm\frac{i}{2}\sum_{k}E_k \left(Y_1^{\bar{z}_k}\partial_{\bar{z}_k}T_2 - \frac{1}{2}D_{\bar{z}_k}Y_1^{\bar{z}_k}T_2\right)\langle out|\mathcal{S}|in\rangle-(1\leftrightarrow 2)\,.
\end{split}
\end{equation}
We now consider the opposite helicity terms $(1_{+}2_{-})$.  Substituting \eqref{doublesoftpartone} into \eqref{subleading2}, we have
\begin{equation}
\begin{split}
&\hskip -2.5cm\langle out | \left[ (\left[Q_{1H}, Q_{2S}\right]_{H} + \left[Q_{1S}, Q_{2H}\right]_{H}), \mathcal{S}\right] | in \rangle \supset\\
-\frac{i}{16\pi^2}\int &d^2 z_1 d^2 z_2 \partial_{\bar{z}_1}^{3}Y_1^{\bar{z}_1} D_{z_2}^{2}T_2 \\ &\sum_k E_k\frac{(z_2 - z_k)(\bar{z}_1 + \bar{z}_2 - 2\bar{z}_k)(\bar{z}_1-\bar{z}_k)}{(z_1-z_k)(\bar{z}_2-\bar{z}_k)^2}\frac{(1+z_2\bar{z}_2)}{(1+z_k \bar{z}_k)}\langle out|\mathcal{S}|in\rangle\\
&\hskip 8cm-(1\leftrightarrow 2)\,.
\end{split}
\end{equation}
We can then integrate by parts in $z_1$:
\begin{equation}
\begin{split}
&\hskip -2 cm \langle out | \left[ (\left[Q_{1H}, Q_{2S}\right]_{H} + \left[Q_{1S}, Q_{2H}\right]_{H}), \mathcal{S}\right] | in \rangle \supset\\
&\frac{i}{16\pi^2} \int d^2 z_2 \sum_k E_k \Bigg[-\frac{2\pi(z_2-z_k)(1+z_2 \bar{z}_2)}{(\bar{z}_2-\bar{z}_k)(1+z_k \bar{z}_k)}\partial_{\bar{z}_k}Y_1^{\bar{z}_k}D_{z_{2}}^{2}T_2 \\
&\hskip 2.6cm+ \frac{4\pi(1+z_2 \bar{z}_2)(z_2-z_k)}{(\bar{z}_2-\bar{z}_k)^2(1+z_k \bar{z}_k)}Y_1^{\bar{z}_k}D_{z_{2}}^{2}T_2\Bigg]\langle out|\mathcal{S}|in\rangle\,\\
&\hskip 8.5cm-(1\leftrightarrow 2)\,,
\end{split}
\end{equation}
and then in $z_2$:
\begin{equation}
\begin{split}
\langle out | \left[ (\left[Q_{1H}, Q_{2S}\right]_{H} + \left[Q_{1S}, Q_{2H}\right]_{H}), \mathcal{S}\right] | &in \rangle \supset\\
&\hskip -4cm\frac{i}{2}\sum_{k}E_k \left(Y_1^{\bar{z}_k}\partial_{\bar{z}_k}T_2 - \frac{1}{2}D_{\bar{z}_k}Y_1^{\bar{z}_k}T_2\right)\langle out|\mathcal{S}|in\rangle-(1\leftrightarrow 2)\,,
\end{split}
\end{equation}
where we have differentiated the Cauchy-Pompeiu formula to find 
\begin{equation}
\partial_{\bar{z}}\left(\frac{1}{(z-z_k)^2}\right) = -(2\pi)\partial_{z}\delta^{(2)}(z-z_k)\,.
\end{equation}
Summing over all combinations of helicities, we have
\begin{equation}
\begin{split}
\langle out | \left[ (\left[Q_{1H}, Q_{2S}\right]_{H} + \left[Q_{1S}, Q_{2H}\right]_{H}), \mathcal{S}\right] | in \rangle \supset \\
&\hskip -3cm i\sum_{k}E_k \Bigg[Y_1^{A}\partial_{A}T_2 - \frac{1}{2}D_{A}Y_1^{A}T_2 -(1 \leftrightarrow 2)\Bigg]\langle out|\mathcal{S}|in\rangle\,,
\end{split}
\end{equation}
consistent with 
\begin{equation}
\langle out | [ [Q_{1H}, Q_{2S}]_{H} + [Q_{1S}, Q_{2H}]_{H}, \mathcal{S}] | in \rangle =\langle out | [i Q_{[1,2]H}, \mathcal{S}] | in \rangle\,.
\end{equation}

\subsubsection{$\left[Q_{1H}, Q_{2S}\right]_{S} + \left[Q_{1S}, Q_{2H}\right]_{S}$ commutator}

Let us now consider the soft parts of $\left[Q_{1H}, Q_{2S}\right] + \left[Q_{1S}, Q_{2H}\right]$. Substituting the $(1_{+}2_{+})$ contribution in \eqref{doublesoftparttwo} into equation~\ref{subleading2} leads to
\begin{equation}
\begin{split}
&\hskip -8.2cm\langle out | \left[ (\left[Q_{1H}, Q_{2S}\right]_{S} + \left[Q_{1S}, Q_{2H}\right]_{S}), \mathcal{S}\right] | in \rangle \supset\\
\frac{i}{16\pi^2}\int d^2 z_1 d^2 z_2\, \partial_{\bar{z}_{1}}^{3} Y_1^{\bar{z}_1}D_{\bar{z}_2}^{2}T_2 \sum_k &E_k \frac{(\bar{z}_1-\bar{z}_k)(\bar{z}_1-\bar{z}_2)(1+z_2 \bar{z}_2)}{(z_1 - z_k)(z_1 - z_2)(1+z_k \bar{z}_k)}\\
&\hskip 2cm\times \langle out|\mathcal{S}|in\rangle\\
&\hskip 3cm-(1\leftrightarrow 2)\,,
\end{split}
\end{equation}
which can be integrated by parts in $z_2$ to give
\begin{equation}
\begin{split}
&\hskip -2cm\langle out | \left[ (\left[Q_{1H}, Q_{2S}\right]_{S} + \left[Q_{1S}, Q_{2H}\right]_{S}), \mathcal{S}\right] | in \rangle \supset\\
&\frac{i}{8\pi}\int d^2 z_1\, \partial_{\bar{z}_{1}}^{3} Y_1^{\bar{z}_1} T_2(z_1)\sum_k E_k\frac{(\bar{z}_1-\bar{z}_k)(1+z_1 \bar{z}_1)}{(z_1-z_k)(1+z_k \bar{z}_k)}\langle out|\mathcal{S}|in\rangle\\
&\hskip 9cm-(1\leftrightarrow 2)\,.
\end{split}
\end{equation}
Similarly, the $(1_{+}2_{-})$ terms give
\begin{equation}
\begin{split}
\langle out | \left[ (\left[Q_{1H}, Q_{2S}\right]_{S} + \left[Q_{1S}, Q_{2H}\right]_{S}), \mathcal{S}\right] | in \rangle &\supset\\
\frac{i}{16\pi^2}\int d^2 z_1 d^2 z_2\, \partial_{\bar{z}_{1}}^{3}  Y_1^{\bar{z}_1} D_{z_2}^{2}T_2 \sum_k &E_k \frac{(\bar{z}_1-\bar{z}_k)(z_1-z_2)(1+z_2 \bar{z}_2)}{(z_1-z_k)(\bar{z}_1-\bar{z}_2)(1+z_k \bar{z}_k)} \langle out|\mathcal{S}|in\rangle\\
&\hskip 4cm-(1\leftrightarrow 2)\,,
\end{split}
\end{equation}
and integration by parts in $z_2$ leads to
\begin{equation}
\begin{split}
&\hskip -2cm\langle out | \left[ (\left[Q_{1H}, Q_{2S}\right]_{S} + \left[Q_{1S}, Q_{2H}\right]_{S}), \mathcal{S}\right] | in \rangle \supset\\
&\frac{i}{8\pi}\int d^2 z_1 \partial_{\bar{z}_{1}}^{3} Y_1^{\bar{z}_1} T_2(z_1)\sum_k E_k\frac{(\bar{z}_1-\bar{z}_k)(1+z_1 \bar{z}_1)}{(z_1-z_k)(1+z_k \bar{z}_k)}\langle out|\mathcal{S}|in\rangle\\
&\hskip 9cm-(1\leftrightarrow 2)\,.
\end{split}
\end{equation}
Finally, we need the terms in \eqref{doublesoftpartthree}.  The $(1_{+}2_{+})$ terms give
\begin{equation}
\begin{split}
&\langle out | \left[ (\left[Q_{1H}, Q_{2S}\right]_{S} + \left[Q_{1S}, Q_{2H}\right]_{S}), \mathcal{S}\right] | in \rangle \supset\\
&-\frac{i}{16\pi^2}\int d^{2}z_1 d^{2}z_2 \, \partial_{\bar{z}_1}^{3}Y_{1}^{\bar{z}_1} D_{\bar{z}_2}^{2}T_{2}\sum_k E_k \frac{(\bar{z}_1 - \bar{z}_2)}{(z_1 - z_2)}\frac{(\bar{z}_1 - \bar{z}_k)}{(z_2 - z_k)}\frac{(1+z_2 \bar{z}_2)}{(1+z_k \bar{z}_k)}\langle out | \mathcal{S} | in \rangle\\
&\hskip 9cm-(1\leftrightarrow 2)\,.
\end{split}
\end{equation}
Integrating by parts in $z_2$ turns this into 
\begin{equation}
\begin{split}
&\hskip -0.5cm\langle out | \left[ (\left[Q_{1H}, Q_{2S}\right]_{S} + \left[Q_{1S}, Q_{2H}\right]_{S}), \mathcal{S}\right] | in \rangle \supset\\
&\frac{i}{8\pi}\int d^{2}z_1 \, \partial_{\bar{z}_1}^{3}Y_{1}^{\bar{z}_1} \sum_k E_k\left(\partial_{\bar{z}_k}T_2(z_k)\frac{(\bar{z}_1 - \bar{z}_k)^2}{(z_1 - z_k)} + T_2 (z_k)\frac{(1+\bar{z}_1 z_k)}{(1+z_k \bar{z}_k)}\frac{(\bar{z}_1 - \bar{z}_k)}{(z_1 - z_k)}\right)\\
&\hskip 6cm\left.- T_{2}(z_1)\frac{(\bar{z}_1 - \bar{z}_k)}{(z_1 - z_k)}\frac{(1+z_1 \bar{z}_1)}{(1+z_k \bar{z}_k)}\right)\langle out | \mathcal{S} | in \rangle\\
&\hskip 9cm-(1\leftrightarrow 2)\,,
\end{split}
\end{equation}
and integration by parts in $z_1$ gives
\begin{equation}
\begin{split}
&\hskip -1cm\langle out | \left[ (\left[Q_{1H}, Q_{2S}\right]_{S} + \left[Q_{1S}, Q_{2H}\right]_{S}), \mathcal{S}\right] | in \rangle \supset\\
&\hskip 1cm-\frac{i}{2}\sum_k E_k\left(Y_{1}^{\bar{z}_k}\partial_{\bar{z}_k}T_{2} - \frac{1}{2}D_{\bar{z}_k}Y_{1}^{\bar{z}_k}T_2\right)\langle out | \mathcal{S} | in \rangle\\ 
&\hskip 1cm- \frac{i}{8\pi}\int d^{2}z_1 \, \partial_{\bar{z}_1}^{3}Y_{1}^{\bar{z}_1}T_{2}\sum_k E_k\frac{(\bar{z}_1-\bar{z}_k)}{(z_1-z_k)}\frac{(1+z_1\bar{z}_1)}{(1+z_k \bar{z}_k)}\langle out | \mathcal{S} | in \rangle\\
&\hskip 9cm-(1\leftrightarrow 2)\,.
\end{split}
\end{equation}
Similarly, the $(1_{+}2_{-})$ terms in  \eqref{doublesoftpartthree} lead to
\begin{equation}
\begin{split}
&\langle out | \left[ (\left[Q_{1H}, Q_{2S}\right]_{S} + \left[Q_{1S}, Q_{2H}\right]_{S}), \mathcal{S}\right] | in \rangle \supset\\
&\frac{i}{16\pi^2}\int d^{2}z_1 d^{2}z_2 \, \partial_{\bar{z}_1}^{3}Y_{1}^{\bar{z}_1} D_{z_2}^{2}T_{2} \sum_k E_k\frac{(z_2 - z_k)(\bar{z}_1 + \bar{z}_2 - 2\bar{z}_k)(\bar{z}_1 - \bar{z}_k)}{(\bar{z}_2 - \bar{z}_k)^2(z_1-z_2)}\frac{(1+z_2 \bar{z}_2)}{(1+z_k \bar{z}_k)}\\ &\times \langle out | \mathcal{S} | in \rangle\\
&\hskip 9cm-(1\leftrightarrow 2)\,.
\end{split}
\end{equation}
We can again integrate by parts in $z_1$ 
\begin{equation}
\begin{split}
&\hskip -1cm\langle out | \left[ (\left[Q_{1H}, Q_{2S}\right]_{S} + \left[Q_{1S}, Q_{2H}\right]_{S}), \mathcal{S}\right] | in \rangle \supset\\
&-\frac{i}{4\pi}\int d^{2}z_2  \, \partial_{\bar{z}_2}^{2}Y_{1}^{\bar{z}_2} D_{z_2}^{2}T_{2} \sum_k E_k(z_2 - z_k)\frac{(1+z_2 \bar{z}_2)}{(1+z_k \bar{z}_k)} \langle out | \mathcal{S} | in \rangle\\
&+\frac{i}{8\pi}\int d^{2}z_2  \, \partial_{\bar{z}_2}Y_{1}^{\bar{z}_2} D_{z_2}^{2}T_{2} \sum_k E_k\frac{3(z_2 - z_k)}{(\bar{z}_2 - \bar{z}_k)}\frac{(1+z_2 \bar{z}_2)}{(1+z_k \bar{z}_k)} \langle out | \mathcal{S} | in \rangle\\
&-\frac{i}{8\pi}\int d^{2}z_2  \, Y_{1}^{\bar{z}_2} D_{z_2}^{2}T_{2} \sum_k E_k\frac{2(z_2 - z_k)}{(\bar{z}_2 - \bar{z}_k)^2}\frac{(1+z_2 \bar{z}_2)}{(1+z_k \bar{z}_k)} \langle out | \mathcal{S} | in \rangle\\
&\hskip 9cm-(1\leftrightarrow 2)\,.
\end{split}
\end{equation}
as well as in $z_2$ to find
\begin{equation}
\begin{split}
&\langle out | \left[ (\left[Q_{1H}, Q_{2S}\right]_{S} + \left[Q_{1S}, Q_{2H}\right]_{S}), \mathcal{S}\right] | in \rangle\\ 
&\hskip 2cm\supset -\frac{i}{2}\sum_k E_k \left(Y_{1}^{\bar{z}_k}\partial_{\bar{z}_k}T_{2} - \frac{1}{2}D_{\bar{z}_k}Y_{1}^{\bar{z}_k}T_2 \right)\langle out | \mathcal{S} | in \rangle-(1\leftrightarrow 2)\,.
\end{split}
\end{equation}
Summing over all combinations of helicities, we have
\begin{equation}
\begin{split}
\langle &out | \left[ (\left[Q_{1H}, Q_{2S}\right]_{S} + \left[Q_{1S}, Q_{2H}\right]_{S}), \mathcal{S}\right] | in \rangle \supset \\
& -i\sum_{k}E_k \Bigg[Y_1^{A}\partial_{A}T_2 - \frac{1}{2}D_{A}Y_1^{A}T_2 - \frac{1}{8\pi}\int d^2 z \left(\frac{(1+z \bar{z})(\bar{z}-\bar{z}_k)}{(1+z_k \bar{z}_k)(z-z_k)}\partial_{\bar{z}}^{3}Y_1^{\bar{z}} T_2 + c.c.\right)\\
&\qquad \qquad -(1 \leftrightarrow 2)\Bigg]\langle out|\mathcal{S}|in\rangle\,.
\end{split}
\end{equation}
The first term is the commutator, and the second is the extension. So we see that at this order 
\begin{equation}
\langle out | [ [Q_{1H},Q_{2S}]_S +[Q_{1S},Q_{2H}]_S, \mathcal{S}] | in \rangle=i\langle out | [ (Q_{[1,2]S} + K_{(1,2)S}), \mathcal{S}] | in \rangle\,.
\end{equation}

\subsection{BMS commutator at subleading order}

We can similarly evaluate the commutator at subsubleading order using the expression for the soft terms in \eqref{subsubleading}:
\begin{equation}\label{subsubleading2}
\begin{split}
&\langle out| \left[ (\left[Q_{1H}, Q_{2S}\right] + \left[Q_{1S}, Q_{2H}\right]), \mathcal{S}\right] | in \rangle \supset\\
&\hskip 0.7cm-\frac{1}{4\pi^2 \kappa^2}\lim_{\left[\omega_2 \to 0\right.}\lim_{\left.\omega_1 \to 0\right]}\int d^2 z_1 d^2 z_2 \,  \\
&\hskip 1.8cm\Big\{\partial_{\bar{z}_1}^3 Y_1^{\bar{z}_1} \partial_{z_2}^3 Y_2^{z_2} (1+\omega_1 \partial_{\omega_1})(1+\omega_2 \partial_{\omega_2})\langle out | a_{+}a_{-}\mathcal{S} | in \rangle + \cdots \Big\}
\end{split}
\end{equation}
The antisymmetric consecutive double soft graviton factor $S(q_1, q_2)$ can be evaluated at subsubleading order either by explicit calculation using Feynman rules, by using the BCFW recursion relations at tree level, or by evaluating the contact terms in the antisymmetric consecutive double-soft limit (see \cite{Klose:2015xoa, Volovich:2015yoa}).  The last method is the quickest, and the relevant contact terms are:
\begin{equation}
\begin{split}
S^{{\rm NNLO}}(q_1, q_2) = 
\Big[S^{(2)}(q_1)\left\{S^{(0)}(q_2)\right\} + S^{(1)}(q_1)\left\{S^{(1)}(q_2)\right\} - (1 \leftrightarrow 2)\Big]\mathcal{M}
\end{split}
\end{equation}
where $S^{{\rm NNLO}}$ is the subleading part of the factor defined in \eqref{Sdef}, and the curly brackets denote that one or both derivatives act on the momenta in the other soft factor. Only the second set of terms are non-zero in the double-soft limit, so that these determine the commutator. We will further break this contribution to the soft factor up into contributions based on the pole structure
\begin{equation}
S^{{\rm NNLO}}(q_1, q_2))\mathcal{M}=\left(S_{1}^{\rm{NNLO}}(q_1, q_2)+S_{2}^{\rm{NNLO}}(q_1, q_2)-(1\leftrightarrow 2)\right)\mathcal{M}\,.
\end{equation}
where
\begin{eqnarray}
\left(S_{1}^{\rm{NNLO}}(q_1, q_2)-(1\leftrightarrow 2)\right)\mathcal{M}&=&\nonumber\\*
&&\hskip -3.5cm\frac{\kappa^2}{4}\sum_k\left[\frac{(p_k \cdot \bar{\epsilon}_1)(\bar{\epsilon}_1 q_1 J_k)}{(p_k \cdot q_1)}\left\{\frac{(p_k \cdot \bar{\epsilon}_2)(\bar{\epsilon}_2 q_2 J_k)}{(p_k \cdot q_2)}\right\}-(1\leftrightarrow 2)\right]\mathcal{M}\,,
\end{eqnarray}
and
\begin{eqnarray}
\left(S_{2}^{\rm{NNLO}}(q_1, q_2)-(1\leftrightarrow 2)\right)\mathcal{M}&=&\nonumber\\
&&\hskip -3.5cm\frac{\kappa^2}{4}\sum_k\left[\frac{(q_2 \cdot \bar{\epsilon}_1)(\bar{\epsilon}_1 q_1 J_2)}{(q_1 \cdot q_2)}\left\{\frac{(p_k \cdot \bar{\epsilon}_2)(\bar{\epsilon}_2 q_2 J_k)}{(p_k \cdot q_2)}\right\}-(1\leftrightarrow 2)\right]\mathcal{M}\,.
\end{eqnarray}
The first contribution will encode the hard part of the commutator $(\left[Q_{1H}, Q_{2S}\right]_{H} + \left[Q_{1S}, Q_{2H}\right]_{H})$.
It can be written more explicitly as
\begin{equation}\label{eq:S1NNLO}
\begin{split}
&\hskip -0.5cm S_{1}^{\rm{NNLO}}(q_1, q_2)=\\
&\frac{\kappa^2}{4}\sum_k\Bigg[\frac{(p_k \cdot \bar{\epsilon}_1)^2}{(p_k\cdot q_1)}\left(\frac{(q_1 \cdot \bar{\epsilon}_2)(\bar{\epsilon}_2 q_2 J_k)}{(p_k \cdot q_2)} - \frac{(p_k \cdot \bar{\epsilon}_2)(q_1 \cdot q_2)(\bar{\epsilon}_2 q_2 J_k)}{(p_k \cdot q_2)^2}\right)\\
&\hskip 1.1cm-(p_k \cdot \bar{\epsilon}_1)\;\left(\frac{(\bar{\epsilon}_1 \cdot \bar{\epsilon}_2)(\bar{\epsilon}_2 q_2 J_k)}{(p_k \cdot q_2)} - \frac{(p_k \cdot \bar{\epsilon}_2)(\bar{\epsilon}_1 \cdot q_2)(\bar{\epsilon}_2 q_2 J_k)}{(p_k \cdot q_2)^2}\right)\\
& \hskip 1.1cm -\frac12 \frac{(p_k \cdot \bar{\epsilon}_1)}{(p_k \cdot q_1)}\frac{(p_k \cdot \bar{\epsilon}_2)}{(p_k \cdot q_2)}\big((q_2 \cdot \bar{\epsilon}_1)(\bar{\epsilon}_2 q_1 J_k)-(q_1 \cdot \bar{\epsilon}_2)(\bar{\epsilon}_1 q_2 J_k)\\
&\hskip 4.5cm+(q_1 \cdot q_2)(\bar{\epsilon}_1 \bar{\epsilon}_2 J_k)+(\bar{\epsilon}_1\cdot \bar{\epsilon}_2)(q_1q_2 J_k)\big)\Bigg]\,.
\end{split}
\end{equation}
where we have used that the action of the angular momentum operator on the momentum is given by
\begin{equation}
J_k^{\mu\nu}p_k^\rho=\eta^{\nu\rho}p_k^\mu-\eta^{\mu\rho}p_k^\nu
\end{equation}
and that the the angular momentum operators obey 
\begin{equation}
[J_k^{\mu\nu},J_k^{\rho\sigma}]\mathcal{M}=\left(\eta^{\nu\rho}J^{\mu\sigma}_k+\eta^{\mu\sigma}J^{\nu\rho}_k-\eta^{\mu\rho}J^{\nu\sigma}_k-\eta^{\nu\sigma}J^{\mu\rho}_k\right)\mathcal{M}\,.
\end{equation}
The second line in equation~(\ref{eq:S1NNLO}) contains terms that are not doubly singular and therefore vanish when we integrate by parts.  As we did for the leading order calculation, we will nevertheless keep them around, because they tend to make the expression in terms of holomorphic coordinates simpler. We find
\begin{equation}\label{doublesoftsubleading1}
\begin{split}
&S^{{\rm NNLO}}_1(q_1, q_2) =\begin{dcases}-\frac{\kappa^2}{4}\sum_k\Bigg[\frac{(\bar{z}_1 - \bar{z}_2)((\bar{z}_1 -\bar{z}_k)(1+\bar{z}_2 z_k) + (\bar{z}_2 - \bar{z}_k)(1+\bar{z}_1 z_k))}{2(z_1 - z_k)(z_2 - z_k)(1+z_k \bar{z}_k)}\\
\hskip 0.5cm \times \left(E_k \partial_{E_k} + h_k\right) +\frac{(\bar{z}_1-\bar{z}_2)(\bar{z}_1 - \bar{z}_k)(\bar{z}_2-\bar{z}_k)}{(z_1 - z_k) (z_2 - z_k)}(\partial_{\bar{z}_k}+h_k\Omega_{\bar{z}_k})\Bigg]\quad (1_{+}2_{+}) \\
\frac{\kappa^2}{4}\sum_k\Bigg[\frac{(\bar{z}_1-\bar{z}_k)^2(z_2-z_k)(1+z_2 \bar{z}_k)}{(z_1-z_k)(\bar{z}_2-\bar{z}_k)^2(1+z_k\bar{z}_k)}\left(E_k \partial_{E_k}-h_k\right)\\
\hskip 4cm+
\frac{(\bar{z}_1 - \bar{z}_k)^2 (z_2 - z_k)^2}{(z_1 - z_k) (\bar{z}_2 - \bar{z}_k)^2}(\partial_{z_k}-h_k\Omega_{z_k})\Bigg]\hskip 1.05cm (1_{+}2_{-})\,.
\end{dcases}
\end{split}
\end{equation}
As before, the other helicity combinations are related to this by complex conjugation and sending $h_k$ to $-h_k$.  

The second contribution will encode the soft part of the commutator $(\left[Q_{1H}, Q_{2S}\right]_{S} + \left[Q_{1S}, Q_{2H}\right]_{S})$ and can be written as
\begin{equation}
\begin{split}
S_{2}^{\rm NNLO}(q_1, q_2) = &\frac{\kappa^2}{4}\sum_k\Bigg[\frac{(q_2 \cdot \bar{\epsilon}_1)^2}{(q_1 \cdot q_2)}\left(-\frac{(p_k \cdot \bar{\epsilon}_2)(p_k \cdot q_1)(\bar{\epsilon}_2 q_2 J_k)}{(p_k \cdot q_2)^2} + \frac{(p_k \cdot \bar{\epsilon}_2)}{(p_k \cdot q_2)}(\bar{\epsilon}_2 q_1 J_k)\right)\\
&\hskip 1.1cm+(q_2 \cdot \bar{\epsilon}_1)\left(\frac{(p_k \cdot \bar{\epsilon}_1)(p_k\cdot \bar{\epsilon}_2 )(\bar{\epsilon}_2 q_2 J_k)}{(p_k \cdot q_2)^2} + \frac{(p_k \cdot \bar{\epsilon}_2)}{(p_k \cdot q_2)}(\bar{\epsilon}_1\bar{\epsilon}_2 J_k)\right)  \\
&\hskip 1.1cm+\frac{(q_2 \cdot \bar{\epsilon}_1)(\bar{\epsilon}_1 \cdot \bar{\epsilon}_2)}{(q_1 \cdot q_2)}\left(\frac{(p_k \cdot q_1)(\bar{\epsilon}_2 q_2 J_k)}{(p_k \cdot q_2)}+\frac{(p_k \cdot \bar{\epsilon}_2)(q_1 q_2 J_k)}{(p_k \cdot q_2)}\right)  \\
&\hskip 1.1cm-\frac{(q_2 \cdot \bar{\epsilon}_1)(q_1 \cdot \bar{\epsilon}_2)}{(q_1 \cdot q_2)}\left(\frac{(p_k \cdot \bar{\epsilon}_1)(\bar{\epsilon}_2 q_2 J_k)}{(p_k \cdot q_2)}+\frac{(p_k \cdot \bar{\epsilon}_2)(\bar{\epsilon}_1 q_2 J_k)}{(p_k \cdot q_2)}\right)\Bigg]\,,
\end{split}
\end{equation}
which in holomorphic coordinates becomes
\begin{equation}\label{doublesoftsubleading2}
\begin{split}
S_{2}^{\rm NNLO}(q_1, q_2) =\begin{dcases}
\frac{\kappa^2}{4}\sum_k\Bigg[\frac{(\bar{z}_1 - \bar{z}_2)((\bar{z}_1 -\bar{z}_k)(1+\bar{z}_2 z_k) + (\bar{z}_2 - \bar{z}_k)(1+\bar{z}_1 z_k))}{(z_1 - z_2)(z_2 - z_k)(1+z_k \bar{z}_k)}\\
\hskip 0.5cm\times \left(E_k \partial_{E_k} + h_k\right) + \frac{2(\bar{z}_1 - \bar{z}_2)(\bar{z}_1 - \bar{z}_k)(\bar{z}_2 - \bar{z}_k)}{(z_1 - z_2)(z_2 - z_k)}(\partial_{\bar{z}_k}+h_k\Omega_{\bar{z}_k})\Bigg] \quad (1_{+}2_{+}) \\
-\frac{\kappa^2}{4}\sum_k\Bigg[ \frac{(\bar{z}_1 - \bar{z}_k)^2(z_2 -z_k) (1+ z_2 \bar{z}_k)}{(z_1 -z_2)(\bar{z}_2 - \bar{z}_k)^{2}(1+z_k \bar{z}_k)}\left(E_k \partial_{E_k} - h_k\right) \\ 
\hskip 4cm+ \frac{(\bar{z}_1 - \bar{z}_k)^2 (z_2 - z_k)^2 }{(z_1 - z_2)(\bar{z}_2 - \bar{z}_k)^2}(\partial_{z_k} -h_k \Omega_{z_k})\Bigg]\hskip 1.25cm (1_{+}2_{-})\,.\end{dcases}
\end{split}
\end{equation}

\subsubsection{$\left[Q_{1H}, Q_{2S}\right]_{H} + \left[Q_{1S}, Q_{2H}\right]_{H}$ commutator}
We will begin with the terms in \eqref{doublesoftsubleading1}, which represent contact terms between the single-soft factors acting on the hard momenta.  Integrating by parts is laborious but straightforward.  Starting with the contribution from two positive helicities $(1_{+}2_{+})$, we have
\begin{equation}
\begin{split}
&\langle out | \left[ (\left[Q_{1H}, Q_{2S}\right]_{H} + \left[Q_{1S}, Q_{2H}\right]_{H})), \mathcal{S}\right] | in \rangle \supset \frac{1}{16\pi^2}\int d^2 z_1 d^2 z_2 \, \partial_{\bar{z}_1}^{3}Y_1^{\bar{z}_1}\partial_{\bar{z}_2}^{3}Y_2^{\bar{z}_2} \\ &\hskip 1cm\times\sum_k \Bigg[\frac{(\bar{z}_1 - \bar{z}_2)((\bar{z}_1 -\bar{z}_k)(1+\bar{z}_2 z_k) + (\bar{z}_2 - \bar{z}_k)(1+\bar{z}_1 z_k))}{2(z_1 - z_k)(z_2 - z_k)(1+z_k \bar{z}_k)}\left(E_k \partial_{E_k} + h_k\right) \\ &\hskip 1cm \qquad + \frac{(\bar{z}_1 - \bar{z}_2)(\bar{z}_1 - \bar{z}_k)(\bar{z}_2 - \bar{z}_k)}{(z_1 - z_k)(z_2 - z_k)}(\partial_{\bar{z}_k} + h_k \Omega_{\bar{z}_k}) \Bigg]\langle out | \mathcal{S} | in \rangle\\
&\hskip 10cm-(1\leftrightarrow 2)\,.
\end{split}
\end{equation}
We can integrate contributions involving $(E_k \partial_{E_k} + h_k)$ by parts in $z_1$ to write it as 
\begin{equation}
\begin{split}
&\hskip -0.5cm\langle out | \left[ (\left[Q_{1H}, Q_{2S}\right]_{H} + \left[Q_{1S}, Q_{2H}\right]_{H}), \mathcal{S}\right] | in \rangle \supset\\
&\hskip -0cm\frac{1}{16\pi^2}\int d^2 z_2 \, \partial_{\bar{z}_2}^{3}Y_2^{\bar{z}_2}\sum_k \Bigg[\frac{2\pi (\bar{z}_2-\bar{z}_k)^2}{(z_2-z_k)}\partial_{\bar{z}_k}^{2}Y_{1}^{\bar{z}_k}-\frac{4\pi z_k (\bar{z}_2 - \bar{z}_k)^2}{(z_2-z_k)(1+z_k \bar{z}_k)}\partial_{\bar{z}_k}Y_1^{\bar{z}_k}\\
&\hskip 4.3cm-\frac{4\pi  (1+2\bar{z}_2z_k-z_k\bar{z}_k)}{(z_2 - z_k)(1+z_k \bar{z}_k)}Y_1^{\bar{z}_k}\Bigg]\left(E_k \partial_{E_k} + h_k\right)\langle out | \mathcal{S} | in \rangle\,,
\end{split}
\end{equation}
and integrating by parts in $z_2$ gives 
\begin{equation}
\begin{split}
&\hskip -2.2cm\langle out | \left[ (\left[Q_{1H}, Q_{2S}\right]_{H} + \left[Q_{1S}, Q_{2H}\right]_{H}), \mathcal{S}\right] | in \rangle \supset\\
\sum_k &\frac{1}{2}\left(Y_1^{\bar{z}_k}\partial_{\bar{z}_k}^2 Y_2^{\bar{z}_k}- \frac{2z_k}{1+z_k \bar{z}_k}Y_1^{\bar{z}_k}\partial_{\bar{z}_k}Y_2^{\bar{z}_k} 
 \right. \\
&\hskip 0.2cm  \left.-Y_2^{\bar{z}_k}\partial_{\bar{z}_k}^2 Y_1^{\bar{z}_k} + \frac{2 z_k}{1+z_k \bar{z}_k}Y_2^{\bar{z}_k}\partial_{\bar{z}_k}Y_1^{\bar{z}_k} \right) \left(E_k \partial_{E_k} + h_k\right)\langle out | \mathcal{S} | in \rangle\,,
\end{split}
\end{equation}
or more compactly
\begin{equation}
\begin{split}
&\hskip -2.2cm\langle out | \left[ (\left[Q_{1H}, Q_{2S}\right]_{H} + \left[Q_{1S}, Q_{2H}\right]_{H}), \mathcal{S}\right] | in \rangle \supset\\
\sum_k &\frac{1}{2}D_{\bar{z}_k}(Y_1^{A}\partial_{A} Y_2^{\bar{z}_k}-Y_2^{A}\partial_{A} Y_1^{\bar{z}_k} ) \left(E_k \partial_{E_k} + h_k\right)\langle out | \mathcal{S} | in \rangle\,.
\end{split}
\end{equation}
Next, let us consider the $(\partial_{\bar{z}_k} +h_k\Omega_{\bar{z}_k})$ terms. Including the contribution in which $1$ and $2$ are interchanged, they are given by
\begin{equation}
\begin{split}
&\hskip -0.2cm\langle out | \left[ (\left[Q_{1H}, Q_{2S}\right]_{H} + \left[Q_{1S}, Q_{2H}\right]_{H}), \mathcal{S}\right] | in \rangle \supset\\
&\frac{1}{16\pi^2}\int d^2 z_1 d^2 z_2 \,\partial_{\bar{z}_1}^{3}Y_1^{\bar{z}_1}\partial_{\bar{z}_2}^{3}Y_2^{\bar{z}_2}\sum_k\frac{2(\bar{z}_1-\bar{z}_2)(\bar{z}_1 - \bar{z}_k)(\bar{z}_2-\bar{z}_k)}{(z_1 - z_k) (z_2 - z_k)}(\partial_{\bar{z}_k} + h_k\Omega_{\bar{z}_k})\langle out | \mathcal{S} | in \rangle\,.
\end{split}
\end{equation}
Integrating by parts in $z_1$, we have
\begin{equation}
\begin{split}
&\hskip -0.2cm\langle out | \left[ (\left[Q_{1H}, Q_{2S}\right]_{H} + \left[Q_{1S}, Q_{2H}\right]_{H}), \mathcal{S}\right] | in \rangle \supset \frac{1}{16\pi^2}\int d^2 z_2 \, \partial_{\bar{z}_2}^{3}Y_2^{\bar{z}_2}\\
&\hskip 1cm\times\sum_k \Bigg[-\frac{4\pi (\bar{z}_2 - \bar{z}_k)^2}{(z_2 - z_k)}\partial_{\bar{z}_k} Y_1^{\bar{z}_k}- \frac{8\pi (\bar{z}_2 - \bar{z}_k)}{(z_2 - z_k)}Y_1^{\bar{z}_k}\Bigg](\partial_{\bar{z}_k} + h_k\Omega_{\bar{z}_k})\langle out | \mathcal{S} | in \rangle\,,
\end{split}
\end{equation}
and finally integrating by parts in $z_2$, we have
\begin{equation}
\begin{split}
&\hskip -0.2cm\langle out | \left[ (\left[Q_{1H}, Q_{2S}\right]_{H} + \left[Q_{1S}, Q_{2H}\right]_{H}), \mathcal{S}\right] | in \rangle \supset\\
&\hskip 2.2cm-\sum_k \left(Y_1^{\bar{z}_k}\partial_{\bar{z}_k}Y_2^{\bar{z}_k} - Y_2^{\bar{z}_k}\partial_{\bar{z}_k}Y_1^{\bar{z}_k} \right)(\partial_{\bar{z}_k} + h_k\Omega_{\bar{z}_k})\langle out | \mathcal{S} | in \rangle\,.
\end{split}
\end{equation}

We now turn to the  contribution denoted by $(1_{+}2_{-})$ in which the first graviton has positive helicity, and the second graviton has negative helicity
\begin{equation}
\begin{split}
\langle out &| \left[ (\left[Q_{1H}, Q_{2S}\right]_{H} + \left[Q_{1S}, Q_{2H}\right]_{H}), \mathcal{S}\right] | in \rangle \supset\\
&-\frac{1}{16\pi^2}\int d^2 z_1 d^2 z_2 \, \partial_{\bar{z}_1}^{3}Y_1^{\bar{z}_1}\partial_{z_2}^{3}Y_2^{z_2}\\
&\hskip 1.5cm\times \sum_k\Bigg[\frac{(\bar{z}_1-\bar{z}_k)^2(z_2-z_k)(1+\bar{z}_2 z_k)}{(z_1-z_k)(\bar{z}_2-\bar{z}_k)^2(1+z_k\bar{z}_k)}  \left(E_k \partial_{E_k}-h_k\right)\\
&\hskip 4.5cm+\frac{(\bar{z}_1 - \bar{z}_k)^2 (z_2 - z_k)^2}{(z_1 - z_k) (\bar{z}_2 - \bar{z}_k)^2}(\partial_{z_k}-h_k\Omega_{z_k})\Bigg]\langle out | \mathcal{S} | in \rangle\\
&\hskip 10cm-(1\leftrightarrow 2)\,.
\end{split}
\end{equation}
We can integrate by parts in $z_1$ to write it as
\begin{equation}
\begin{split}
\langle out &| \left[ (\left[Q_{1H}, Q_{2S}\right]_{H} + \left[Q_{1S}, Q_{2H}\right]_{H}), \mathcal{S}\right] | in \rangle \supset\frac{1}{4\pi}\int d^2 z_1\, Y_1^{\bar{z}_k}\partial_{z_2}^3Y_2^{z_2}\\
&\times \sum_k\Bigg[\frac{(z_2-z_k)(1+\bar{z}_2 z_k)}{(\bar{z}_2-\bar{z}_k)^2(1+z_k\bar{z}_k)}  \left(E_k \partial_{E_k}-h_k\right) +\frac{(z_2 - z_k)^2 }{(\bar{z}_2 - \bar{z}_k)^2 }(\partial_{z_k}-h_k\Omega_{z_k})\Bigg]\langle out | \mathcal{S} | in \rangle\\
&\hskip 10cm-(1\leftrightarrow 2)\,,
\end{split}
\end{equation}
and after integration by parts in $z_2$, we see that both terms are total $\bar{z}_2$-derivatives so that there is no contribution from the terms in the amplitude in which the two soft gravitons have opposite helicities.

Adding the remaining contributions from the terms in which both gravitons have negative helicity, we have
\begin{equation}
\begin{split}
\langle out &| \left[ (\left[Q_{1H}, Q_{2S}\right]_{H} + \left[Q_{1S}, Q_{2H}\right]_{H}), \mathcal{S}\right] | in \rangle_{subleading} \supset \\
&\sum_k \Bigg[\frac12 D_{A}(Y_1^{B}\partial_{B}Y_2^{A} - Y_2^{B}\partial_{B}Y_1^{A}) E_k \partial_{E_k} \\ 
&\hskip 0.5cm+ \frac{h_k}{2}\left[D_{\bar{z}_k}(Y_1^{A}\partial_{A}Y_2^{\bar{z}_k} - Y_2^{A}\partial_{A}Y_1^{\bar{z}_k}) - D_{z_k}(Y_1^{A}\partial_{A}Y_2^{z_k} - Y_2^{A}\partial_{A}Y_1^{z_k})\right]\\
&\hskip 0.5cm - \left(Y_1^{B}\partial_{B}Y_2^{z_k} - Y_2^{B}\partial_{B}Y_1^{z_k}\right)(\partial_{z_k}-h_k\Omega_{z_k})\\
&\hskip 0.5cm - \left(Y_1^{B}\partial_{B}Y_2^{\bar{z}_k} - Y_2^{B}\partial_{B}Y_1^{\bar{z}_k}\right)(\partial_{\bar{z}_k}+h_k\Omega_{\bar{z}_k})\Bigg]\langle out |\mathcal{S} |in \rangle\,,
\end{split}
\end{equation}
consistent with
\begin{equation}
\langle out | [ [Q_{1H}, Q_{2S}]_{H} + [Q_{1S}, Q_{2H}]_{H}, \mathcal{S}] | in \rangle =\langle out | [i Q_{[1,2]H}, \mathcal{S}] | in \rangle\,.
\end{equation}
As anticipated there is no extension term.  

\subsubsection{$\left[Q_{1H}, Q_{2S}\right]_{S} + \left[Q_{1S}, Q_{2H}\right]_{S}$ commutator}

Next we treat the terms in \eqref{doublesoftsubleading2}, in which one soft graviton operator is treated as hard by the other.  Starting with the $(1_{+}2_{+})$ terms, we then have
\begin{equation}
\begin{split}
&\langle out | \left[ (\left[Q_{1H}, Q_{2S}\right]_{S} + \left[Q_{1S}, Q_{2H}\right]_{S})), \mathcal{S}\right] | in \rangle \supset -\frac{1}{16\pi^2}\int d^2 z_1 d^2 z_2 \, \partial_{\bar{z}_1}^{3}Y_1^{\bar{z}_1}\partial_{\bar{z}_2}^{3}Y_2^{\bar{z}_2} \sum_k \\ &\times \Bigg[\frac{(\bar{z}_1 - \bar{z}_2)((\bar{z}_1 -\bar{z}_k)(1+\bar{z}_2 z_k) + (\bar{z}_2 - \bar{z}_k)(1+\bar{z}_1 z_k))}{(z_1 - z_2)(z_2 - z_k)(1+z_k \bar{z}_k)}\left(E_k \partial_{E_k} + h_k\right) \\ & \qquad + \frac{2(\bar{z}_1 - \bar{z}_2)(\bar{z}_1 - \bar{z}_k)(\bar{z}_2 - \bar{z}_k)}{(z_1 - z_2)(z_2 - z_k)}(\partial_{\bar{z}_k} + h_k \Omega_{\bar{z}_k}) \Bigg]\langle out | \mathcal{S} | in \rangle\\
&\hskip 10cm-(1\leftrightarrow 2)\,.
\end{split}
\end{equation}
Starting with the $(E_k \partial_{E_k} + h_k)$ terms, we can integrate by parts in $\bar{z}_1$ to find
\begin{equation}
\begin{split}
&\langle out | \left[ (\left[Q_{1H}, Q_{2S}\right]_{S} + \left[Q_{1S}, Q_{2H}\right]_{S}), \mathcal{S}\right] | in \rangle \supset -\frac{1}{4\pi}\int d^2 z_2 \, \partial_{\bar{z}_2}^{3}Y_{2}^{\bar{z}_2}\sum_k\\ 
&\times\Bigg[ \partial_{\bar{z}_2}Y_{1}^{\bar{z}_2} \frac{(\bar{z}_2 - \bar{z}_k)(1+\bar{z}_2 z_k)}{(z_2 - z_k)(1+z_k \bar{z}_k)}- Y_{1}^{\bar{z}_2}\frac{1+2\bar{z}_2z_k-z_k\bar{z}_k}{(z_2 - z_k)(1+z_k \bar{z}_k)}\Bigg]\left(E_k \partial_{E_k} + h_k\right) \langle out | \mathcal{S} | in \rangle\\
&\hskip 1cm -(1\leftrightarrow 2)\,.
\end{split}
\end{equation}
Integrating by parts in $\bar{z}_2$ finally leads us to
\begin{equation}
\begin{split}
&\hskip -1.4 cm \langle out | \left[ (\left[Q_{1H}, Q_{2S}\right]_{S} + \left[Q_{1S}, Q_{2H}\right]_{S}), \mathcal{S}\right] | in \rangle \supset \\
& -\sum_{k}\frac{1}{2}\Bigg[Y_{1}^{\bar{z}_k}\partial_{\bar{z}_k}D_{\bar{z}_k}Y_{2}^{\bar{z}_k} - Y_{2}^{\bar{z}_k}\partial_{\bar{z}_k}D_{\bar{z}_k}Y_{1}^{\bar{z}_k}\Bigg]\left(E_k \partial_{E_k} + h_k\right)\langle out | \mathcal{S} | in \rangle\,.
\end{split}
\end{equation}
Similarly, the $(\partial_{\bar{z}_k} + h_k\Omega_{\bar{z}_k})$ terms become
\begin{equation}
\begin{split}
&\hskip -0.3cm\langle out | \left[ (\left[Q_{1H}, Q_{2S}\right]_{S} + \left[Q_{1S}, Q_{2H}\right]_{S}), \mathcal{S}\right] | in \rangle \supset \\ &-\frac{1}{4\pi}\int d^{2}z_{2} \, \partial_{\bar{z}_2}^{3}Y_{2}^{\bar{z}_2}\sum_k \Bigg[\partial_{\bar{z}_2}Y_{1}^{\bar{z}_2}\frac{(\bar{z}_2 - \bar{z}_k)^2}{(z_2 - z_k)} - 2Y_{1}^{\bar{z}_2}\frac{(\bar{z}_2 - \bar{z}_k)}{(z_2 - z_k)}\Bigg](\partial_{\bar{z}_k} +h_k\Omega_{\bar{z}_k})\langle out | \mathcal{S} | in \rangle \\
&\hskip 12cm-(1\leftrightarrow 2)\\
&=  \sum_{k}\left(Y_{1}^{\bar{z}_k}\partial_{\bar{z}_k}Y_{2}^{\bar{z}_k} - Y_{2}^{\bar{z}_k}\partial_{\bar{z}_k}Y_{1}^{\bar{z}_k}\right)(\partial_{\bar{z}_k} + h_k\Omega_{\bar{z}_k})\langle out | \mathcal{S} | in \rangle\,.
\end{split}
\end{equation}
Consider now the $(1_{+}2_{-})$ terms. These contribute the following terms to the amplitude:
\begin{equation}
\begin{split}
&\langle out | \left[ (\left[Q_{1H}, Q_{2S}\right]_{S} + \left[Q_{1S}, Q_{2H}\right]_{S}), \mathcal{S}\right] | in \rangle \supset \frac{1}{16\pi^2}\int d^2 z_1 d^2 z_2 \, \partial_{\bar{z}_1}^{3}Y_1^{\bar{z}_1}\partial_{z_2}^{3}Y_2^{z_2}\sum_k  \\ &\hskip 1.5cm\times \Bigg[ \frac{(\bar{z}_1 - \bar{z}_k)^2(z_2 -z_k) (1+ z_2 \bar{z}_k)}{(z_1 -z_2)(\bar{z}_2 - \bar{z}_k)^{2}(1+z_k \bar{z}_k)}\left(E_k \partial_{E_k} - h_k\right) \\ 
&\hskip 5cm+ \frac{(\bar{z}_1 - \bar{z}_k)^2 (z_2 - z_k)^2 }{(z_1 - z_2)(\bar{z}_2 - \bar{z}_k)^2}(\partial_{z_k} -h_k \Omega_{z_k})\Bigg]\langle out | \mathcal{S} | in \rangle\\
&\hskip 10cm -(1\leftrightarrow 2)\,.
\end{split}
\end{equation} 
After integration by parts in $\bar{z}_1$, the $(E_k \partial_{E_k}-h_k)$ terms become
\begin{equation}
\begin{split}
&\hskip -0.5cm\langle out | \left[ (\left[Q_{1H}, Q_{2S}\right]_{S} + \left[Q_{1S}, Q_{2H}\right]_{S}), \mathcal{S}\right] | in \rangle \supset \frac{1}{8\pi}\int d^2 z_2  \, \partial_{z_2}^{3}Y_2^{z_2}\sum_k  \\ 
&\hskip 1.4cm\times\Bigg\{-\frac{(z_2-z_k)(1+z_2 \bar{z}_k)}{(1+z_k\bar{z}_k)}\partial_{\bar{z}_2}^2Y_1^{\bar{z}_2}+\partial_{\bar{z}_2} \Bigg[\frac{ 2(z_2 - z_k)(1+z_2 \bar{z}_k)}{(\bar{z}_2 - \bar{z}_k) (1+ z_k \bar{z}_k)}Y_1^{\bar{z}_2}\Bigg]\Bigg\}\\[.2cm]
&\hskip 9cm \times\left(E_k \partial_{E_k} - h_k\right)\langle out | \mathcal{S} | in \rangle\,,
\end{split}
\end{equation} 
and the terms involving $(\partial_{z_k} - h_k\Omega_{z_k})$ become
\begin{equation}
\begin{split}
&\hskip -0.5cm\langle out | \left[ (\left[Q_{1H}, Q_{2S}\right]_{S} + \left[Q_{1S}, Q_{2H}\right]_{S}), \mathcal{S}\right] | in \rangle \supset \frac{1}{8\pi}\int d^2 z_2  \, \partial_{z_2}^{3}Y_2^{z_2}\sum_k  \\ 
&\hskip 0.5cm\times\Bigg\{-(z_2-z_k)^2\partial_{\bar{z}_2}^2Y_1^{\bar{z}_2}+\partial_{\bar{z}_2} \Bigg[\frac{ 2(z_2 - z_k)^2}{(\bar{z}_2 - \bar{z}_k) }Y_1^{\bar{z}_2}\Bigg]\Bigg\}\left(\partial_{z_k} - h_k\Omega_{z_k}\right)\langle out | \mathcal{S} | in \rangle\,,
\end{split}
\end{equation} 
Provided we assume that the vector fields at most have poles at infinity, both contributions vanish so that as before only the amplitudes in which the two gravitons have the same helicities contribute. 

Putting everything together and summing over helicities, we then have
\begin{equation}
\begin{split}
\langle out &| \left[ (\left[Q_{1H}, Q_{2S}\right]_{S} + \left[Q_{1S}, Q_{2H}\right]_{S}), \mathcal{S}\right] | in \rangle_{subleading} = \\
&-\sum_k \Bigg[\frac12 D_{A}(Y_1^{B}\partial_{B}Y_2^{A} - Y_2^{B}\partial_{B}Y_1^{A}) E_k \partial_{E_k} \\ 
&\hskip 0.9cm+ \frac{h_k}{2}\left[D_{\bar{z}_k}(Y_1^{\bar{z}_k}\partial_{\bar{z}_k}Y_2^{\bar{z}_k} - Y_2^{\bar{z}_k}\partial_{\bar{z}_k}Y_1^{\bar{z}_k}) - D_{z_k}(Y_1^{z_k}\partial_{z_k}Y_2^{z_k} - Y_2^{z_k}\partial_{z_k}Y_1^{z_k})\right]\\
&\hskip 0.9cm - \left(Y_1^{z_k}\partial_{z_k}Y_2^{z_k} - Y_2^{z_k}\partial_{z_k}Y_1^{z_k}\right)(\partial_{z_k}-h_k\Omega_{z_k})\\
&\hskip 0.9cm - \left(Y_1^{\bar{z}_k}\partial_{\bar{z}_k}Y_2^{\bar{z}_k} - Y_2^{\bar{z}_k}\partial_{\bar{z}_k}Y_1^{\bar{z}_k}\right)(\partial_{\bar{z}_k}+h_k\Omega_{\bar{z}_k})\Bigg]\langle out |\mathcal{S} |in \rangle\,,\\
\end{split}
\end{equation}
consistent with
\begin{equation}
\langle out | \left[ (\left[Q_{1H}, Q_{2S}\right]_{S} + \left[Q_{1S}, Q_{2H}\right]_{S}), \mathcal{S}\right] | in \rangle=\langle out | \left[ iQ_{[1,2]S}, \mathcal{S} \right] | in \rangle\,.
\end{equation}

\subsection{Generalized cocycle condition for $K$}\label{cocyle}

In order for the algebra of charges to satisfy the Jacobi identity the extension terms must satisfy the cocycle condition
\begin{equation}
i[K_{[1,2]},Q_3]-K_{[[1,2],3]}+cyclic=0\,,
\end{equation}
for which we will need the commutator of $K$ with the charges.  Starting with the expression
\begin{equation}
K_{(1,2)S}=-\frac{1}{32\pi G}\int_{\mathcal{I}^+_\pm} d^2z\gamma_{z\bar{z}}\left[ C^{BC}(T_1D_BD_CD_AY_2^A)-(1 \leftrightarrow 2)\right]\,,
\end{equation}
and using the mode expansion for the Bondi news, we can write $K$ as
\begin{eqnarray}
K_{(1,2)S}&=&\frac{1}{8\pi \kappa} \int d^2z\left\{\bar{W}_{[1,2]}\lim_{\omega\to 0}\omega\left[a_+(\omega\hat{x})+a_-(\omega\hat{x})^\dagger\right] +h.c.\right\}\,,
\end{eqnarray}
where
\begin{equation}
\bar{W}_{[1,2]}=-4D_{\bar{z}}^2\bar{V}_{[1,2]}\,,
\end{equation}
with
\begin{equation}
\bar{V}_{[1,2]}=\frac{1}{8\pi}\int d^2w\frac{(1+w\bar{w})(\bar{w}-\bar{z})}{(1+z\bar{z})(w-z)}(T_2\partial^3_{\bar{w}}Y_1^{\bar{w}}-T_1\partial^3_{\bar{w}}Y_2^{\bar{w}})\,.
\end{equation}
This expression for $K$ also appeared in (\ref{algebraFromOperators}) when we found the commutator of the charges directly from the operators.  If $V$ were real, this could simply be a leading soft charge with $T=-4V$, but since it is complex we cannot write it in this way.

First, notice that $K$ only contains a soft piece so that this breaks up into two conditions
\begin{equation}
i[K_{[1,2]},Q_3]_S-K_{[[1,2],3]}+cyclic=0\,,
\end{equation}
and
\begin{equation}
i[K_{[1,2]},Q_3]_H+cyclic=0\,.
\end{equation}
Working with the operators, we only have enough information to compute the soft contribution, but we can find both by working directly with the soft limits of the scattering amplitudes. 

Let us begin with the commutator of $K$ with the soft charges and by recalling that the expressions for $K$ and $Q^{(0)}_S$ are
\begin{eqnarray}
K_{(1,2)S}&=&-\frac{1}{\kappa^2}\int du d^2z\gamma^{z\bar{z}}\left[ \bar{W}_{[1,2]} N_{zz}+W_{[1,2]} N_{\bar{z}\bar{z}}\right]\,,\\*
Q_S^{(0)}&=&-\frac{2}{\kappa^2}\int du d^2 z \, \gamma^{z\bar{z}}\Big[D_{\bar{z}}^2T N_{zz}+D_z^2T N_{\bar{z}\bar{z}}\Big]\,.
\end{eqnarray}
The commutators of $N_{zz}$ and $N_{\bar{z}\bar{z}}$ are given by
\begin{equation}
[N_{z_1z_1},N_{\bar{z}_2\bar{z}_2}]=\frac{\kappa^2}{4\pi}\gamma_{z_1\bar{z}_1}\delta(z_1-z_2)\int_0^\infty dq \,q \left(e^{-iq(u_1-u_2)}-e^{iq(u_1-u_2)}\right)\,,
\end{equation}
so that
\begin{equation}
\begin{split}
[K_{(1,2)S},Q_{3S}^{(0)}]&=\frac{1}{2\pi\kappa^2}\int du_1\int du_2\int dq\,q\int d^2z \gamma^{z\bar{z}}\\
&\hskip 1cm\times\left[\bar{W}_{[1,2]}D_z^2T_3-W_{[1,2]}D_{\bar{z}}^2T_3\right]\left(e^{-iq(u_1-u_2)}-e^{iq(u_1-u_2)}\right)\\
&=0.
\end{split}
\end{equation}
The commutator with the subleading contribution to the charge can be obtained by replacing $D_z^2T_3$ by $u D_z^3Y^z_3$ and similarly vanishes. So we only have to consider the commutator of $K$ with the hard charges. With the leading hard charge 
\begin{equation}
Q_H^{(0)}=\frac{1}{16\pi^3}\int d^2 z \, \gamma_{z\bar{z}}T \int_0^\infty dq\, q^2\left[a_+(q\hat{x})^\dagger a_+(q\hat{x})+a_-(q\hat{x})^\dagger a_-(q\hat{x})\right]\,,
\end{equation}
the commutator can be written as
\begin{equation}
[K_{[1,2]},Q_{3H}^{(0)}]_S=\frac{1}{\kappa^2}\int du d^2z \gamma^{z\bar{z}}\left[\bar{W}_{[1,2]}[Q_{3H}^{(0)},N_{zz}]+W_{[1,2]}[Q_{3H}^{(0)},N_{\bar{z}\bar{z}}]\right]\,.
\end{equation}
With the commutators
\begin{equation}
\begin{split}
[a_+^\dagger(\omega\hat{x}_1)a_+(\omega\hat{x}_1),N_{z_2z_2}]&=2\pi\kappa\,  \delta(z_1-z_2)a_+(\omega\hat{x}_2)e^{-i\omega u}\,,\\
{[}a_-^\dagger(\omega\hat{x}_1)a_-(\omega\hat{x}_1),N_{z_2z_2}]&=-2\pi \kappa\,  \delta(z_1-z_2)a_-(\omega\hat{x}_2)^\dagger e^{i\omega u}\,,\\
{[}a_+^\dagger(\omega\hat{x}_1)a_+(\omega\hat{x}_1),N_{\bar{z}_2\bar{z}_2}]&=-2\pi \kappa \,\delta(z_1-z_2)a_+(\omega\hat{x}_2)^\dagger e^{i\omega u}\,,\\
{[}a_-^\dagger(\omega\hat{x}_1)a_-(\omega\hat{x}_1),N_{\bar{z}_2\bar{z}_2}]&=2\pi \kappa  \,\delta(z_1-z_2)a_-(\omega\hat{x}_2) e^{-i\omega u}\,,
\end{split}
\end{equation}
we find
\begin{equation}
\begin{split}
[Q_{3H}^{(0)},N_{zz}]&=\frac{\kappa}{8\pi^2}\gamma_{z\bar{z}}T_3\int_0^\infty dq\,q^2\left[a_+(q\hat{x})e^{-iq u}-a_-(q\hat{x})^\dagger e^{iq u}\right]\,,\\
{[}Q_{3H}^{(0)},N_{{\bar{z}}{\bar{z}}}]&=\frac{\kappa}{8\pi^2}\gamma_{z\bar{z}}T_3\int_0^\infty dq\,q^2\left[a_-(q\hat{x})e^{-iq u}-a_+(q\hat{x})^\dagger e^{iq u}\right]\,.
\end{split}
\end{equation}
The soft part of the commutator between $K$ and the charge is then given by
\begin{equation}
[K_{[1,2]},Q_{3H}^{(0)}]_S=\frac{1}{8\pi\kappa}\int d^2z T_3 W_{[1,2]}\lim_{q\to0}q^2\left[a_-(q\hat{x})-a_+(q\hat{x})^\dagger \right]+h.c.\,.
\end{equation}
This does not lead to a contribution in the soft limit, so $K$ commutes with supertranslations at the level of soft scattering amplitudes.

For the subleading piece we can use the results derived below for the commutators of the hard charge with the leading soft piece. We find
\begin{eqnarray}
[Q_{3H}^{(1)}, K_{[1,2]}]_S &=& \nonumber\\*
&&\hskip -2.0cm\frac{i}{8\pi \kappa}\lim_{\omega \to 0}\omega  \int d^{2}z\, \Big[\bar{W}_{[1,2]}\Big(\frac12D_{A}Y_{3}^{A}\omega \partial_{\omega} + ( D_{\bar{z}}Y_{3}^{\bar{z}}-D_{z}Y^{z}_{3}) - Y_{3}^{A}D_{A}\Big)a_{+}\nonumber\\* 
&&\hskip 0.9cm - \bar{W}_{[1,2]}\Big(\frac12 D_{A}Y_{3}^{A}\omega \partial_{\omega} + (D_{\bar{z}}Y_{3}^{\bar{z}}-D_{z}Y^{z}_{3}) + Y_{3}^{A}D_{A}\Big)a_{-}^\dagger\nonumber\\*
&&\hskip 0.9cm + W_{[1,2]}\Big(\frac12 D_{A}Y_{3}^{A}\omega \partial_{\omega} - (D_{\bar{z}}Y_{3}^{\bar{z}}-D_{z}Y^{z}_{3}) - Y_{3}^{A}D_{A}\Big)a_{-}\nonumber\\*
&&\hskip 0.9cm - W_{[1,2]}\Big(\frac12 D_{A}Y_{3}^{A}\omega \partial_{\omega} - (D_{\bar{z}}Y_{3}^{\bar{z}}-D_{z}Y^{z}_{3}) + Y_{3}^{A}D_{A}\Big)a_{+}^{\dagger}\Big]\,.
\end{eqnarray}
The commutator of this with $\mathcal{S}$ is then given by
\begin{eqnarray}
\langle out|[[Q_{3H}^{(1)},  K_{[1,2]}]_S,\mathcal{S}]|in\rangle&=&\frac{i}{4\pi \kappa}\lim_{\omega \to 0}\omega  \int d^{2}z\,\nonumber \\
&&\hskip -3cm\times\Big[\bar{W}_{[1,2]}\Big(\frac12D_{A}Y_{3}^{A}\omega \partial_{\omega} + ( D_{\bar{z}}Y_{3}^{\bar{z}}-D_{z}Y^{z}_{3}) - Y_{3}^{A}D_{A}\Big)\langle out|a_{+}\mathcal{S}|in\rangle\\
&&\hskip -2.75cm +W_{[1,2]}\Big(\frac12D_{A}Y_{3}^{A}\omega \partial_{\omega} - ( D_{\bar{z}}Y_{3}^{\bar{z}}-D_{z}Y^{z}_{3}) - Y_{3}^{A}D_{A}\Big)\langle out|a_{-}\mathcal{S}|in\rangle\Big]\nonumber\,.
\end{eqnarray}
Using the soft graviton theorem, this becomes
\begin{eqnarray}
\langle out|[[Q_{3H}^{(1)},  K_{[1,2]}]_S,\mathcal{S}]|in\rangle&=&\frac{i}{8\pi}\sum_k E_k \int d^{2}z\, \\
&&\hskip -4cm\times\Big\{\Big[W_{[1,2]}\Big(\frac12 D_AY_3^A-\partial_{z}Y^{z}_{3} + \partial_{\bar{z}}Y_{3}^{\bar{z}} + Y_{3}^{A}\partial_{A}\Big)\frac{(1+z \bar{z})(z-z_k)}{(1+z_k\bar{z}_k)(\bar{z}-\bar{z}_k)}\Big]+c.c.\Big\}\langle out|\mathcal{S}|in\rangle\nonumber\,.
\end{eqnarray}
We can write this more explicitly as
\begin{eqnarray}
\langle out|[[Q_{3H}^{(1)},  K_{[1,2]}]_S,\mathcal{S}]|in\rangle&=&\frac{i}{8\pi}\sum_k E_k \int d^{2}z\,\nonumber \\*
&&\hskip -4cm\times\Big\{\Big[\frac{(1+z \bar{z})}{(1+z_k\bar{z}_k)(\bar{z}-\bar{z}_k)}(T_1\partial_z^3Y^z_2-T_2\partial_z^3Y^z_1)Y_3^z\nonumber\\*
&&\hskip -3.75cm-\frac{(1+z \bar{z})(z-z_k)}{(1+z_k\bar{z}_k)(\bar{z}-\bar{z}_k)^2} (T_1\partial_z^3Y^z_2-T_2\partial_z^3Y^z_1)Y_3^{\bar{z}}\nonumber\\
&&\hskip -3.75cm-\frac12 \frac{(1+z \bar{z})(z-z_k)}{(1+z_k\bar{z}_k)(\bar{z}-\bar{z}_k)}(T_1\partial_z^3Y^z_2-T_2\partial_z^3Y^z_1)\partial_z Y_3^z\nonumber\\
&&\hskip -3.75cm+\frac32 \frac{(1+z \bar{z})(z-z_k)}{(1+z_k\bar{z}_k)(\bar{z}-\bar{z}_k)}(T_1\partial_z^3Y^z_2-T_2\partial_z^3Y^z_1)\partial_{\bar{z}} Y_3^{\bar{z}}\Big]+c.c.\Big\}\langle out|\mathcal{S}|in\rangle\,.
\end{eqnarray}
The generalized cocycle condition then becomes
\begin{equation}
\begin{split}
&\langle out|[K_{[1,[2,3]]}+K_{[2,[3,1]]}+K_{[3,[1,2]]},\mathcal{S}]|in \rangle\\
&\qquad-i\langle out|[Q_3,K_{[1,2]}]+[Q_1,K_{[2,3]}]+[Q_2,K_{[3,1]}],\mathcal{S}]|in \rangle=0\,.
\end{split}
\end{equation}
This can only be nontrivial if two of the transformations are superrotations and one is a supertranslation. Without loss of generality, let us take the first two to be the superrotations corresponding to $Y_1$, $Y_2$, and the third to be the supertranslation associated with $T_3$. In this case we find
\begin{eqnarray}
\langle out|[K_{[1,[2,3]]},\mathcal{S}]|in \rangle&=&\frac{1}{8\pi}\int d^2z\sum_k E_k\nonumber\\
&&\hskip -2.75cm\times\Big\{\Big[\frac{(1+z\bar{z})(z-z_k)}{(\bar{z}-\bar{z}_k)(1+z_k\bar{z}_k)} Y^z_2\partial_z T_3\partial_{z}^3 Y_1^{z}\nonumber\\
&&\hskip -2.25cm+\frac{(1+z\bar{z})(z-z_k)}{(\bar{z}-\bar{z}_k)(1+z_k\bar{z}_k)} Y^{\bar{z}}_2\partial_{\bar{z}} T_3\partial_{z}^3 Y_1^{z}\nonumber\\
&&\hskip -2.25cm-\frac12\frac{(1+z\bar{z})(z-z_k)}{(\bar{z}-\bar{z}_k)(1+z_k\bar{z}_k)}  T_3D_zY^z_2\partial_{z}^3 Y_1^{z}\nonumber\\
&&\hskip -2.25cm-\frac12\frac{(1+z\bar{z})(z-z_k)}{(\bar{z}-\bar{z}_k)(1+z_k\bar{z}_k)}  T_3D_{\bar{z}}Y^{\bar{z}}_2\partial_{z}^3 Y_1^{z}\Big]+c.c.\Big\}\langle out|\mathcal{S}|in\rangle\,.
\end{eqnarray}
After integration by parts we can write this as
\begin{eqnarray}
\langle out|[K_{[1,[2,3]]},\mathcal{S}]|in \rangle&=&-\frac{1}{8\pi}\int d^2z\sum_k E_k\\
&&\hskip -0.75cm\times\Big\{\Big[\frac{(1+z\bar{z})}{(1+z_k\bar{z}_k)(\bar{z}-\bar{z}_k)} T_3 Y^z_2 \partial_{z}^3 Y_1^{z}\nonumber\\
&&\hskip -0.3cm+\frac{3}{2}\frac{(1+z\bar{z})(z-z_k)}{(\bar{z}-\bar{z}_k)(1+z_k\bar{z}_k)}T_3 \partial_z Y^z_2 \partial_{z}^3 Y_1^{z}\nonumber\\
&&\hskip -0.3cm+\frac{(1+z\bar{z})(z-z_k)}{(\bar{z}-\bar{z}_k)(1+z_k\bar{z}_k)} T_3 Y^z_2 \partial_{z}^4 Y_1^{z}\nonumber\\
&&\hskip -0.3cm-\frac{(1+z\bar{z})(z-z_k)}{(\bar{z}-\bar{z}_k)^2(1+z_k\bar{z}_k)} T_3 Y^{\bar{z}}_2 \partial_{z}^3 Y_1^{z}\nonumber\\
&&\hskip -0.3cm +\frac32\frac{(1+z\bar{z})(z-z_k)}{(\bar{z}-\bar{z}_k)(1+z_k\bar{z}_k)} T_3 \partial_{\bar{z}}Y^{\bar{z}}_2 \partial_{z}^3 Y_1^{z}\Big]+c.c.\Big\}\langle out|\mathcal{S}|in\rangle\,.
\end{eqnarray}
We similarly have
\begin{eqnarray}
\langle out|[K_{[2,[3,1]]},\mathcal{S}]|in \rangle&=&\frac{1}{8\pi}\int d^2z\sum_k E_k\\
&&\hskip -0.75cm\times\Big\{\Big[\frac{(1+z\bar{z})}{(1+z_k\bar{z}_k)(\bar{z}-\bar{z}_k)} T_3 Y^z_1 \partial_{z}^3 Y_2^{z}\nonumber\\
&&\hskip -0.3cm+\frac{3}{2}\frac{(1+z\bar{z})(z-z_k)}{(\bar{z}-\bar{z}_k)(1+z_k\bar{z}_k)}T_3 \partial_z Y^z_1 \partial_{z}^3 Y_2^{z}\nonumber\\
&&\hskip -0.3cm+\frac{(1+z\bar{z})(z-z_k)}{(\bar{z}-\bar{z}_k)(1+z_k\bar{z}_k)} T_3 Y^z_1 \partial_{z}^4 Y_2^{z}\nonumber\\
&&\hskip -0.3cm-\frac{(1+z\bar{z})(z-z_k)}{(\bar{z}-\bar{z}_k)^2(1+z_k\bar{z}_k)} T_3 Y^{\bar{z}}_1 \partial_{z}^3 Y_2^{z}\nonumber\\
&&\hskip -0.3cm +\frac32\frac{(1+z\bar{z})(z-z_k)}{(\bar{z}-\bar{z}_k)(1+z_k\bar{z}_k)} T_3 \partial_{\bar{z}}Y^{\bar{z}}_1 \partial_{z}^3 Y_2^{z}\Big]+c.c.\Big\}\langle out|\mathcal{S}|in\rangle\,.
\end{eqnarray}
and finally
\begin{eqnarray}
\langle out|[K_{[3,[1,2]]},\mathcal{S}]|in \rangle&=&-\frac{1}{8\pi}\int d^2z\sum_k E_k\\
&&\hskip -2.5cm\times\Big\{\Big[\frac{(1+z\bar{z})(z-z_k)}{(\bar{z}-\bar{z}_k)(1+z_k\bar{z}_k)}T_3\partial_z^3(Y_1^z\partial_z Y^z_2-Y_2^z\partial_z Y^z_1)\Big]+c.c.\Big\}\langle out|\mathcal{S}|in\rangle\,.
\end{eqnarray}
which we can equivalently write as
\begin{eqnarray}
\langle out|[K_{[3,[1,2]]},\mathcal{S}]|in \rangle&=&-\frac{1}{8\pi}\int d^2z\sum_k E_k\\
&&\hskip -2.5cm\times\Big\{\Big[\frac{(1+z\bar{z})(z-z_k)}{(\bar{z}-\bar{z}_k)(1+z_k\bar{z}_k)}2T_3(\partial_zY_1^z\partial_z^3 Y^z_2-\partial_zY_2^z\partial_z^3 Y^z_1)\nonumber\\
&&\hskip -2.0cm+\frac{(1+z\bar{z})(z-z_k)}{(\bar{z}-\bar{z}_k)(1+z_k\bar{z}_k)}T_3(Y_1^z\partial_z^4 Y^z_2-Y_2^z\partial_z^4 Y^z_1)\Big]+c.c.\Big\}\langle out|\mathcal{S}|in\rangle\,.
\end{eqnarray}
We will also need
\begin{eqnarray}
-i\langle out|[[Q_{1H}^{(1)},  K_{[2,3]}]_S,\mathcal{S}]|in\rangle&=&-\frac{1}{8\pi}\sum_k E_k \int d^{2}z\,\nonumber \\*
&&\hskip -2cm\times\Big\{\Big[\frac{(1+z \bar{z})}{(1+z_k\bar{z}_k)(\bar{z}-\bar{z}_k)}T_3\partial_z^3Y^z_2Y_1^z\nonumber\\*
&&\hskip -1.75cm-\frac{(1+z \bar{z})(z-z_k)}{(1+z_k\bar{z}_k)(\bar{z}-\bar{z}_k)^2} T_3\partial_z^3Y^z_2 Y_1^{\bar{z}}\nonumber\\
&&\hskip -1.75cm-\frac12 \frac{(1+z \bar{z})(z-z_k)}{(1+z_k\bar{z}_k)(\bar{z}-\bar{z}_k)}T_3\partial_z^3Y^z_2\partial_z Y_1^z\nonumber\\
&&\hskip -1.75cm +\frac32 \frac{(1+z \bar{z})(z-z_k)}{(1+z_k\bar{z}_k)(\bar{z}-\bar{z}_k)}T_3\partial_z^3Y^z_2\partial_{\bar{z}} Y_1^{\bar{z}}\Big]+c.c.\Big\}\langle out|\mathcal{S}|in\rangle\,,
\end{eqnarray}
as well as 
\begin{eqnarray}
-i\langle out|[[Q_{2H}^{(1)},  K_{[3,1]}]_S,\mathcal{S}]|in\rangle&=&\frac{1}{8\pi}\sum_k E_k \int d^{2}z\,\nonumber \\*
&&\hskip -2cm\times\Big\{\Big[\frac{(1+z \bar{z})}{(1+z_k\bar{z}_k)(\bar{z}-\bar{z}_k)}T_3\partial_z^3Y^z_1Y_2^z\nonumber\\*
&&\hskip -1.75cm-\frac{(1+z \bar{z})(z-z_k)}{(1+z_k\bar{z}_k)(\bar{z}-\bar{z}_k)^2} T_3\partial_z^3Y^z_1 Y_2^{\bar{z}}\nonumber\\
&&\hskip -1.75cm-\frac12 \frac{(1+z \bar{z})(z-z_k)}{(1+z_k\bar{z}_k)(\bar{z}-\bar{z}_k)}T_3\partial_z^3Y^z_1\partial_z Y_2^z\nonumber\\
&&\hskip -1.75cm +\frac32 \frac{(1+z \bar{z})(z-z_k)}{(1+z_k\bar{z}_k)(\bar{z}-\bar{z}_k)}T_3\partial_z^3Y^z_1\partial_{\bar{z}} Y_2^{\bar{z}}\Big]+c.c.\Big\}\langle out|\mathcal{S}|in\rangle\,,
\end{eqnarray}
and of course
\begin{equation}
-i\langle out|[[Q_{3H}^{(1)},  K_{[1,2]}]_S,\mathcal{S}]|in\rangle=0\,.
\end{equation}
Combining the different contributions, we see that
\begin{equation}
\begin{split}
&\langle out|[K_{[1,[2,3]]}+K_{[2,[3,1]]}+K_{[3,[1,2]]},\mathcal{S}]|in \rangle\\
&\qquad-i\langle out|[Q_3,K_{[1,2]}]_S+[Q_1,K_{[2,3]}]_S+[Q_2,K_{[3,1]}]_S,\mathcal{S}]|in \rangle=0\,,
\end{split}
\end{equation}
so that the generalized cocycle condition indeed holds as expected.

\subsection{Summary of results}\label{summary}

To summarize the results of this section, we have shown that the antisymmetrized double consecutive soft graviton amplitude contains information about the commutator of the BMS algebra.  The soft parts of the commutator can be found at the level of the operators, though the hard parts of the operator still remain to be computed explicitly.  Both the hard and soft parts can be found at the level of scattering amplitudes.  Splitting the amplitude by pole structures, the individual pieces of the commutator become
\begin{equation}
\langle out | \left[(\left[Q_{1H}, Q_{2S}\right]_{S} + \left[Q_{1S}, Q_{2H}\right]_{S}), \mathcal{S}\right] | in \rangle = \langle out | \left[ (iQ_{[1,2]S} + iK_{(1,2)S}), \mathcal{S}\right] | in \rangle
\end{equation}
for the soft charges, and
\begin{equation}
\langle out | \left[ (\left[Q_{1H}, Q_{2S}\right]_{H} + \left[Q_{1S}, Q_{2H}\right]_{H}), \mathcal{S}\right] | in \rangle = \langle out | \left[ iQ_{[1,2]H}, \mathcal{S}\right] | in \rangle
\end{equation}
for the commutator of the hard parts.  While we have worked with the tree-level amplitudes, the commutator should hold at the quantum level as well once the subleading part of the charge is dressed with the appropriate one-loop correction terms needed to preserve the Virasoro symmetry of the single-soft theorem.

The extension appears in the soft charges but not in the hard ones, consistent with the intuition from the Chern-Simons example in \cite{Banados:2016zim}, where the extension piece has a boundary contribution and no corresponding contribution in the bulk.  
The extension term therefore means that the BMS symmetry is broken when the local transformations are included.  

The algebra closes because the extension terms satisfy the generalized cocycle condition -- although this can in principle be derived from triple-soft amplitudes, we have derived it from a transformed single-soft amplitude.

\section{2d algebra and operators}\label{Jacobi}

We have just shown that scattering amplitudes and soft theorems realize an extension of the BMS charge algebra. 
In this section we consider the 2d structure of the operator algebra and its implications for defining a dual description for the 4d scattering amplitudes.  In the notation of \cite{Barnich:2011mi}, the (unextended) BMS charge algebra is realized in terms of the fields on the 2-sphere as
\begin{equation}
\begin{split}
T_{\left[s_1, s_2\right]} &= Y_1^{A}\partial_{A}T_2 - \frac{1}{2}D_{A}Y_1^{A}T_2 - Y_2^{A}\partial_{A} T_1 + \frac{1}{2}D_{A}Y_2^{A}T_1\,, \\
Y_{\left[s_1, s_2\right]}^{A} &=Y_1^{B}\partial_{B}Y_2^{A} - Y_2^{B}\partial_{B}Y_1^{A}\,,
\end{split}
\end{equation}
where $s_{1,2} = (T_{1,2}, Y^{A}_{1,2})$.  
Expanding $T(z,\bar{z})$ and $Y^z(z)$ in the basis $t_{m,n} = \frac{z^{m}\bar{z}^{n}}{(1+ z \bar{z})}$, $l_{m} = -z^{m+1}$, we see that this leads to an algebra for the associated operators $T_{m,n}$ and $L_m$ of the form
\begin{equation}\label{algebra}
\begin{split}
\left[T_{m,n}, T_{p,q}\right] &= \left[L_m, \bar{L}_n\right] = 0\,,\\
[L_{l}, T_{m, n}] &=i \left(\frac{l+1}{2} -m \right)T_{l+m,n}\,,\\
[L_{m}, L_{n}] &= i(m-n)L_{m+n}\,,
\end{split}
\end{equation}
and similarly for the antiholomorphic generators.  The last term is the Virasoro algebra.  Including the extension term calculated above, the BMS algebra is realized on the fields as
\begin{eqnarray}
T_{[1,2]}&=&Y^A_1\partial_{A} T_2-\frac12T_2D_AY^A_1-Y^A_2\partial_{A} T_1+\frac12T_1D_AY^A_2\,,\nonumber\\[0.2cm]
Y_{[1,2]}^B&=&Y_{1}^{A}\partial_AY_{2}^{B}-Y_{2}^{A}\partial_{A}Y_{1}^{B}\,,\nonumber\\[.1cm]
V_{[1,2]}&=&\frac{1}{8\pi}\int d^2w \frac{(1+w \bar{w})(w-z)}{(1+z\bar{z})(\bar{w}-\bar{z})}(T_2\partial_{w}^3Y_1^{w}-T_1\partial_{w}^3Y_2^{w}) \,, \bar{V}_{[1,2]} = h.c.
\end{eqnarray}
where we have introduced the fields $V, \bar{V}$ representing the (generalized) 2-cocycle. 
In \cite{Barnich:2011mi, Barnich:2017ubf}, the extension term is interpreted as a field-dependent central extension, with a new local field representing the complex shear of the boundary data.  It is suggestive that we can keep the description local by working in terms of the fields $\phi(z, \bar{z}) = D_{z}^2 V, \bar{\phi} = D_{\bar{z}}^{2}\bar{V}$, although it is not clear whether this construction is unique.
These fields contain the structure of the 2-cocycle, and while the calculation in \S \ref{cocyle} shows that the generalized cocycle condition is satisfied, it is not clear whether the fields $\phi, \bar{\phi}$ can be understood as independent (unconstrained) degrees of freedom.  In other words, although general $\phi, \bar{\phi}$ independent of the cocycle condition can be defined at the level of the operators, it is not as clear whether they can be accessed at the level of on-shell scattering amplitudes.  It is nonetheless interesting that they indicate the existence of a nontrivial Lie algebra extension to the BMS4 algebra: applying the calculation in \S \ref{cocyle} to general $\phi, \bar{\phi}$ and using the the basis elements $\phi_{m,n} = \frac{z^m \bar{z}^{n}}{(1+ z \bar{z})^2}$, we find that the algebra can be extended to
\begin{equation}\label{extendedalgebra}
\begin{split}
\left[T_{m,n}, T_{p,q}\right] &= \left[L_m, \bar{L}_n\right] = 0\,,\\
[L_{l}, T_{m, n}] &=i \left(\frac{l+1}{2} -m \right)T_{l+m,n} - i\frac{l(l^2 - 1)}{4}\Phi_{l+m-2,n}\,,\\
[L_{m}, L_{n}] &= i(m-n)L_{m+n}\,,\\
[L_{l}, \Phi_{m,n}] &= i\left(-\frac{3}{2}(l+1) - m\right)\Phi_{l+m,n}\,,\\
[L_{l}, \bar{\Phi}_{l+m,n}] &= i\left(\frac{1}{2}(l+1) - m\right)\bar{\Phi}_{l+m,n}
\end{split}
\end{equation}
and it is straightforward to check that the extended algebra still satisfies the Jacobi identity.  The operator $\Phi$ scales like a primary operator of dimension $(-1/2, 3/2)$ under the action of the Virasoro generators.  Note that the fields $\phi, \bar{\phi}$ are very similar to the field $\sigma$ defined in \cite{Barnich:2017ubf}; however, we have integrated out the $u$-direction, so the behavior of the fields under the BMS algebra is not the same.
While we stress that it is still unclear whether this construction is unique, or whether the interpretation of the BMS operator algebra in terms of a 2d CFT structure with the operators and scaling dimensions above is sensible or not, it is nevertheless interesting that the BMS4 algebra admits this modification.  It would be interesting to know whether this structure can be used to make further predictions, e.g. about the behavior of graviton amplitudes off-shell, or about the behavior of higher-point correlators.  

Could there be central extensions that we have overlooked in calculating the BMS algebra? Indeed, an arbitrary central charge can be added to the Virasoro commutator without altering the closure of the Jacobi identity. However, our calculation and the assumption that the $Y^A$ are regular everywhere except perhaps at infinity do not allow us to settle this question. We can try to search for a central charge term in the four-dimensional calculation, arising directly from the commutator algebra for the Virasoro parts of the charge operators $Q = Q_{S} + Q_{H}$ in terms of creation and annihilation operators, provided that we have taken the constraints into account correctly.  Our preliminary attempts to do so suggest that the answer is zero, which is sensible if the dual description is coupled to dynamical gravity; however, this calculation is not always straightforward in known field theoretic examples unless the regulator is well understood.  Since the central charge comes from Schwinger terms proportional to the derivatives of delta functions, furthermore, it is certainly possible that we have missed important information by integrating by parts.  To resolve this question one should consider the pole structure of the terms in the integrated charges more carefully, or begin from a purely local prescription for the Noether currents -- we leave this for future work.

A great deal of recent work has focused on searches for 2d CFT structure in the behavior of 4d scattering amplitudes \cite{Lipstein:2015rxa, Kapec:2016jld, Pasterski:2016qvg, Cheung:2016iub}.  In addition to the existence of the Virasoro symmetry, it is also of interest to define local operators in the 2d picture based on the 4d soft fields.  As discussed in \cite{Kapec:2016jld}, the combination
\begin{equation}
\begin{split}\label{Tdef}
t(z) &= \frac{i}{8\pi G}\int d^2 w \frac{1}{z-w}D_{w}^{2}D^{\bar{w}}N^{(1)}_{\bar{w}\bar{w}}\\
&= -\frac{\kappa}{64\pi^2 G}\lim_{\omega \to 0}(1+\omega \partial_{\omega})\int d^2 w \, \gamma_{w \bar{w}}\frac{1}{z-w}D_{w}^2 D^{\bar{w}}\Big[a_{-}(\omega \hat{x}) - a_{+}(\omega \hat{x})^{\dagger}\Big]\\
&\qquad + \textrm{1-loop corrections}\,,
\end{split}
\end{equation}
where $N^{(1)}_{\bar{w}\bar{w}} = \int du \, u N_{\bar{w}\bar{w}}$, and the 1-loop corrections preserve the tree-level Virasoro symmetry, acts upon local operators like a 2d stress tensor.  Up to integration by parts this is the subleading soft charge with $Y^{w} = \frac{1}{(z-w)}, Y^{\bar{w}} = 0$, and so using the single-soft theorem reviewed in \S \ref{singleSoft}, the single-soft limit acts like the OPE of a holomorphic stress-tensor operator $t(z)$ with the local operators.  In the notation of \cite{Kapec:2016jld},:
\begin{equation}
\begin{split}
\langle t(z)&\mathcal{O}_1 \cdots \mathcal{O}_n\rangle =\\ &\sum_{k=1}^{n}\Bigg[\frac{c_k}{(z - z_k)^2} + \frac{\Gamma_{z_k z_k}^{z_k}}{(z-z_k)}c_k + \frac{1}{(z-z_k)}(\partial_{z_k}-h_k\Omega_{z_k})\Bigg]\langle\mathcal{O}_1 \cdots \mathcal{O}_n\rangle
\end{split}
\end{equation}
where $c_k = \left(-\frac{h_k}{2} - \frac{1}{2}E_{k}\partial_{E_k}\right)$ is the holomorphic conformal weight of the $k^{th}$ operator, and $\Omega_{z_k}$ is the spin connection.  The charge $Q_{\mathcal{C}}\left[Y\right] = -i \int \frac{dz}{2\pi i}Y^{z}t(z)$, where the curve $\mathcal{C}$ encloses the points $z_k$, and $Y$ is chosen to be nonsingular in the interior of $\mathcal{C}$, corresponds to the part of the charge $Q_{S}$ that creates a soft outgoing graviton with negative helicity.

A difficulty with this definition for the local operator, however, is that the OPE $t(z_1)t(z_2)$ should contain terms that are singular as $z_1 \to z_2$.  We see that this does not occur 
because $t(z_1)$ is the same as the soft charge $Q_{S}$ for the superrotations, with the particular choice of $Y^{A} = Y^{z} = \frac{1}{(z_1 -z)}$.  Applying two such charges inside a correlator and taking the double-soft limit, 
\begin{equation}
\langle t(z_1)t(z_2) \mathcal{O}_1 \cdots \mathcal{O}_n\rangle\,,
\end{equation}
we will find terms with powers of $(z_1 - z_k)$ and $(z_2 - z_k)$ in the denominator, but not $(z_1 - z_2)$.\footnote{Note that the order of soft limits is irrelevant since both gravitons have the same helicity.}  To address this problem, we can amend the definition of the operator to include terms that generate a linear rotation for hard gravitons:
\begin{equation}
\begin{split}\label{Tdef2}
T(z)
&= -\frac{\kappa}{64\pi^2 G}\lim_{\omega \to 0}(1+\omega \partial_{\omega})\int d^2 w \, \gamma_{w \bar{w}}\frac{1}{z-w}D_{w}^2 D^{\bar{w}}\Big[a_{-}(\omega \hat{x}) - a_{+}(\omega \hat{x})^{\dagger}\Big]\\
&-\frac{i}{16\pi^3}\int d^{2}w \, \gamma_{w\bar{w}}\int_{0}^{\infty}d\omega \, \Bigg[ \\
& \left(\frac{1}{(z-w)^2} - \frac{2\bar{w}}{1+w\bar{w}}\frac{1}{z-w}\right)\left[\left(-\frac{1}{2}\omega\partial_{\omega} + 1\right) a_{+}(\omega \hat{x})\right] \omega a_{+}(\omega \hat{x})^{\dagger}\\
&+ \left(\frac{1}{(z-w)^2} - \frac{2\bar{w}}{1+w\bar{w}}\frac{1}{z-w}\right)\left[\left(-\frac{1}{2}\omega\partial_{\omega} - 1\right) a_{-}(\omega \hat{x})\right] \omega a_{-}(\omega \hat{x})^{\dagger}\\
&+ \frac{1}{z-w}D_{w}\left[a_{-}(\omega \hat{x})\right]\omega a_{-}(\omega \hat{x})^{\dagger} +\frac{1}{z-w}D_{w}\left[a_{+}(\omega \hat{x})\right]\omega a_{+}(\omega \hat{x})^{\dagger}
\Bigg]\\
&\qquad + \textrm{1-loop corrections}\,.
\end{split}
\end{equation}
This is the expression for $Q$ with $Y^{w} = (z-w)^{-1}$, and there will also be a matter contribution depending on the fields present.  Using \eqref{Tdef2} and taking the consecutive double-soft limit for the graviton insertions now implies the OPE
\begin{equation}
T(z) T(w) = \frac{2}{(z-w)^2}T(w) + \frac{\partial T(w)}{(z-w)} + \cdots\,.
\end{equation}
This is equivalent to the third line in \eqref{algebra}, which is the Virasoro algebra familiar from the study of 2d CFTs.  Since it is irrelevant which graviton is taken to be soft first when the gravitons have equal helicity, the OPE will be symmetric.  
Note also that the Christoffel term cancels against a corresponding term from the spin connection.  There will also be a second copy $\bar{T}$ corresponding to the opposite helicity, and
\begin{equation}
T(z)\bar{T}(\bar{w}) = 0 + \cdots
\end{equation}
since they are holomorphic and antiholomorphic respectively.  From the definition \eqref{Tdef2} it is clear that $T$ generates the expected transformations $\left[T, \mathcal{O}\right]$ for local operators.  It will commute, however, with the $\mathcal{S}$-matrix itself.

We can also define an operator $J$ corresponding to supertranslations, plus local fields $\phi, \bar{\phi}$ corresponding to the extension term.  These carry both holomorphic and antiholomorphic indices, and can in principle be defined using the charge operators in a similar manner, although it is less clear which values of the fields $T(z, \bar{z}), V(z, \bar{z}), \bar{V}(z, \bar{z})$ we should choose. 
The extended commutator algebra \eqref{extendedalgebra} is then equivalent to the following set of OPEs:
\begin{equation}
\begin{split}
T(z)T(w) &= \frac{2}{(z-w)^2}T(w) + \frac{\partial T(w)}{(z-w)} + \cdots\\
T(z)J(w, \bar{w}) &= \frac{\phi(w, \bar{w})}{6(z-w)^4} + \frac{3}{2}\frac{J(w, \bar{w})}{(z-w)^2} + \frac{\hat{\partial}J(w, \bar{w})}{(z-w)} + \cdots\\
T(z)\phi(w, \bar{w}) &= -\frac{1}{2}\frac{\phi(w, \bar{w})}{(z-w)^2} + \frac{\hat{\partial}\phi(w, \bar{w})}{(z-w)} + \cdots\\
T(z)\bar{\phi}(w, \bar{w}) &= \frac{3}{2}\frac{\bar{\phi}(w, \bar{w})}{(z-w)^2} + \frac{\hat{\partial}\bar{\phi}(w, \bar{w})}{(z-w)} + \cdots
\end{split}
\end{equation}
Here $T$ is the local operator corresponding to superrotations, $J$ generates supertranslations, $\phi$ is the field appearing in the extension term, $\hat{\partial}\mathcal{O} = \partial \mathcal{O} - \frac{\bar{w}}{1+w \bar{w}}\mathcal{O}$, and all other OPEs are nonsingular.  We emphasize once again that allowing $\phi, \bar{\phi}$ to be unconstrained degrees of freedom (as opposed to a generalized 2-cocycle constructed from supertranslations and superrotations) appears to involve operators beyond those accessible to the on-shell scattering amplitudes.  Because of the extension term, $J$ cannot be a primary operator.  Furthermore, even if there is a nonzero central charge present, the negative operator dimension of $\phi$ seems to indicate that unitarity is violated.
As before, we emphasize that it is not clear whether there is a well-defined 2d CFT structure in the BMS charges, or whether we have identified the correct prescriptions for defining this structure; nevertheless, this seems suggestive.
A more thorough interpretation of this theory and whether it can be made well defined may have to await a better understanding of the dual of flat space, if such a dual exists, and we leave this to future work.

\section{Discussion and further directions}\label{discussion}

In this paper we have shown how the BMS charge algebra in four dimensions is realized at the level of the operator algebra as well as in terms of the double soft graviton amplitudes.  Our results are a check of the algebra derived in \cite{Barnich:2011mi}; we agree with the form of the algebra and with the form of the leading part of the Lie algebroid extension term as well, which vanishes in the global subalgebra of BMS.  In 4d the extension term means that the BMS symmetry is broken, similar to the breaking of a conformal symmetry by a central charge.  The extension term itself contains a soft graviton insertion, and while its interpretation is still unclear, it seems to indicate the existence of a nontrivial extension to the group algebra structure in either a 4d or a 2d description.
Whether the suitably extended BMS algebra has implications either for quantum gravity in flat space or for flat space holography deserves further study.  

Our derivation of the commutator algebra from the contact terms in the consecutive double soft amplitudes also makes it manifest that the BMS algebra is already guaranteed by the single-soft limits, even though we had to consider more than one soft graviton.  This means that the results here are robust and the only potential quantum corrections either arise as Schwinger terms, which do not contribute to the integrated charges, or via one-loop corrections to the subleading soft theorem that arise in the collinear limit.  As in \cite{He:2017fsb} the divergent one-loop contributions can be redefined away in the definition of the charges, and if finite corrections are also present at one-loop, they may be fixable as well.  On the one hand this is encouraging, since it implies that the symmetry is robust even in the presence of quantum corrections, but on the other hand, being fixed by the single-soft limits also means that the commutator is determined by Poincar\'e and gauge invariance, and it is therefore not clear whether we have really learned anything new about quantum gravity that was not already guaranteed by the known symmetries.  

There are a number of possible avenues of study for using BMS to learn more about the structure of gravitational scattering amplitudes.  The question whether there is a central charge in the Virasoro subalgebra deserves further study, and it would be interesting to study this from a local expression for the Noether current.  It would also be interesting to further explore aspects of the charge algebra which are not fixed by the symmetry, such as different combinations of the charges or correlators involving an arbitrary number of soft modes.  From the study of scattering amplitudes it is known that gravitational amplitudes behave in many situations like a product of gauge theory amplitudes -- can this observation help guide us, and does this product structure appear somehow in the asymptotic charges?

While the symmetry algebra derived here has interesting hints of a 2d CFT structure, it is still not clear whether the amplitudes can be understood in terms of a dual CFT interpretation.
We have tried to be clear about the choices leading to our prescription, but it is certainly possible that there exists a different prescription for defining the charges and local operators which leads to more sensible physics.  It is also possible that the BMS symmetries make more sense physically in the context of black hole horizons than they do for asymptotic ones, where they correspond to the symmetries of a compressible fluid living on the horizon \cite{Penna:2017bdn}.  

The role of asymptotic symmetries in gravity and gauge theory is surprisingly subtle, and it remains to be fully understood exactly how much information about quantum gravity is contained in the Ward identities of BMS symmetry.  We hope that the current work helps clarify some of the subtleties in this problem, and will help develop our understanding of the role that asymptotic symmetries play in the scattering of physical gravitons.  


\section*{Acknowledgments}

We thank Steven Avery, Cliff Cheung, Kurt Hinterbichler, Lam Hui, Austin Joyce, John McGreevy, Mehrdad Mirbabayi, Robert Penna and Rachel Rosen for very helpful discussions.  This work is supported in part by the National Science Foundation under grant PHY-1620610 and by the Department of Energy under grant DE-SC0009919.  JD would like to acknowledge the hospitality of the Aspen Center for Physics, supported by NSF grant PHY–1066293, while this work was being completed. RF also acknowledges support by the Alfred P. Sloan Foundation and a grant from the Simons Foundation/SFARI 560536.

\begin{appendices}

\section{Soft factors at subleading order}\label{app:soft}
Since various expressions for soft factors at subleading order, not all consistent with each other, have appeared in the literature, we collect our conventions in this appendix. The soft factor in our conventions is given by
\begin{equation}
S^{(\lambda)}_1(q,p_k)=\frac{\kappa}{2}\frac{\bar{\epsilon}_{\mu\nu}(\mathbf{q},\lambda)p_k^{\nu}q_\rho J^{\mu\rho}_k}{p_k\cdot q-i\epsilon}
\end{equation}
If the amplitude is expressed in terms of spinor helicity variables, the angular momentum operator can be written as
\begin{equation}
 J^{\mu\rho}_k=-{{\Sigma^{\mu\rho}}_{\alpha}}^\beta u^\alpha_k\frac{\partial}{\partial u^\beta_k}-{{\overline\Sigma^{\mu\rho\,\dot\beta}}}_{\dot\alpha} \bar{u}_{k\,\dot\beta}\frac{\partial}{\partial \bar{u}_{k\,\dot\alpha}}\,,
\end{equation}
where
\begin{equation}
\Sigma^{\mu\nu}=\frac14\left(\sigma^\mu\bar\sigma^\nu-\sigma^\nu\bar\sigma^\mu\right)\qquad\text{and}\qquad \overline\Sigma^{\mu\nu}=\frac14\left(\bar\sigma^\mu\sigma^\nu-\bar\sigma^\nu\sigma^\mu\right)\,,
\end{equation}
where $\sigma^0$ and $\bar\sigma^0$ are the identity matrix, and $\sigma^i$ and $-\bar\sigma^i$ are the Pauli matrices.

The two component spinors are related to the stereographic coordinates we use in the main text  according to
\begin{equation}
u^\alpha(\mathbf{p}_k)=\left(\begin{array}{c}z_k\sqrt{\frac{2E_k}{1+z_k \bar{z}_k}}\\-\sqrt{\frac{2E_k}{1+z_k \bar{z}_k}}\end{array}\right)e^{-\frac{i}{2}\phi_k}\,,\qquad\text{and}\qquad \bar{u}_{\dot\alpha}(\mathbf{p}_k)=\left(\begin{array}{c}\sqrt{\frac{2E_k}{1+z_k \bar{z}_k}}\\\bar{z}_k\sqrt{\frac{2E_k}{1+z_k \bar{z}_k}}\end{array}\right)e^{\frac{i}{2}\phi_k}\,,
\end{equation}
where $\phi_k$ is some arbitrary phase that depends on the choice of standard Lorentz transformation to take the standard 4-vector to the 4-momentum of the particle. The phase is typically taken to be zero for convenience.

We can solve these equations for $E_k$, $z_k$, $\bar{z}_k$, and the phase $\phi_k$. In this way we find that the subleading soft factor for positive helicity gravitons in the stereographic coordinates is given by
\begin{equation}
S_1^{(+)}(q,p_k)=\frac{\kappa}{2}\left[-\frac{(1+\bar{z}_q z_k)(\bar{z}_q-\bar{z}_k)}{(z_q-z_k)(1+z_k \bar{z}_k)}E_k\frac{\partial}{\partial E_k}-\frac{(\bar{z}_q-\bar{z}_k)^2}{z_q-z_k}\frac{\partial}{\partial \bar{z}_k}+i\frac{\bar{z}_q-\bar{z}_k}{z_q-z_k}\frac{\partial}{\partial\phi_k}\right]\,.
\end{equation}

To simplify this further, note that the amplitude for an outgoing particle with helicity $h_k$ is proportional to $\exp(ih_k\phi_k)$ (consistent with the explicit expressions for the two-component spinors.) This implies that the soft factor can equivalently be written as
\begin{equation}
S_1^{(+)}(q,p_k)=\frac{\kappa}{2}\left[-\frac{(1+\bar{z}_q z_k)(\bar{z}_q-\bar{z}_k)}{(z_q-z_k)(1+z_k \bar{z}_k)}E_k\frac{\partial}{\partial E_k}-\frac{(\bar{z}_q-\bar{z}_k)^2}{z_q-z_k}\frac{\partial}{\partial \bar{z}_k}-h_k\frac{\bar{z}_q-\bar{z}_k}{z_q-z_k}\right]\,.
\end{equation}

The subleading soft factor for emission of a negative helicity graviton is similarly given by
\begin{equation}
S_1^{(-)}(q,p_k)=\frac{\kappa}{2}\left[-\frac{(1+z_q \bar{z}_k)(z_q-z_k)}{(\bar{z}_q-\bar{z}_k)(1+z_k \bar{z}_k)}E_k\frac{\partial}{\partial E_k}-\frac{(z_q-z_k)^2}{\bar{z}_q-\bar{z}_k}\frac{\partial}{\partial z_k}+h_k\frac{z_q-z_k}{\bar{z}_q-\bar{z}_k}\right]\,.
\end{equation}
We have verified these expressions with explicit perturbative calculations for hard particles of spin-$\frac12$ and spin-$1$.

When discussing photons and gravitons, it is also helpful to have expressions of the angular momentum operator and soft factors at hand that are expressed in terms of polarization vectors rather than the spinor variables. To find the relevant expression, we will imagine that we have expressed the amplitude in terms of momenta and polarization vectors and in turn now express them in terms of spinor-helicity variables. Given the spinor variables
\begin{equation}
u^\alpha(\mathbf{p}_k)=\left(\begin{array}{c}z_k\sqrt{\frac{2E_k}{1+z_k \bar{z}_k}}\\-\sqrt{\frac{2E_k}{1+z_k \bar{z}_k}}\end{array}\right)\,,\qquad\text{and}\qquad \bar{u}_{\dot\alpha}(\mathbf{p}_k)=\left(\begin{array}{c}\sqrt{\frac{2E_k}{1+z_k \bar{z}_k}}\\\bar{z}_k\sqrt{\frac{2E_k}{1+z_k \bar{z}_k}}\end{array}\right)\,,
\end{equation}
we can write the momenta as
\begin{equation}
p^\mu=-\frac12 \bar{u}_{\dot{\alpha}}(\mathbf{p}_k)\bar{\sigma}^{\mu \dot{\alpha}\alpha}u_\alpha(\mathbf{p}_k)\qquad\text{or}\qquad p^\mu=-\frac12 u^\alpha(\mathbf{p}_k)\sigma^\mu_{\alpha\dot{\alpha}}\bar{u}^{\dot{\alpha}}(\mathbf{p}_k)\,, 
\end{equation}
and the polarization vectors as
\begin{equation}
\bar{\epsilon}^\mu(\mathbf{p}_k,+)=\frac{1}{\sqrt{2}}\frac{\bar{u}_{\dot{\alpha}}(\mathbf{p}_k)\bar{\sigma}^{\mu \dot{\alpha}\alpha}n_\alpha}{u^\beta(\mathbf{p}_k) n_\beta}\qquad\text{and}\qquad\bar{\epsilon}^\mu(\mathbf{p}_k,-)=-\frac{1}{\sqrt{2}}\frac{u^\alpha(\mathbf{p}_k)\sigma^\mu_{\alpha\dot{\alpha}}\bar{n}^{\dot{\alpha}}}{\bar{u}_{\dot{\beta}}(\mathbf{p}_k)\bar{n}^{\dot{\beta}}}\,,
\end{equation}
where $n_\alpha$ is an arbitrary reference vector. The choice that corresponds to the polarization vectors used in the main text is
\begin{equation}
n_\alpha=\left(\begin{array}{c}0\\1\end{array}\right)\,.
\end{equation}
The angular momentum operator in terms of derivatives with respect to momenta and polarization vectors is then
\begin{eqnarray}
J^{\rho\sigma}&=&-{{\Sigma^{\rho\sigma}}_{\alpha}}^\beta u^\alpha\frac{\partial}{\partial u^\beta}-{{\overline\Sigma^{\rho\sigma\,\dot\alpha}}}_{\dot\beta}\bar{u}_{\dot\alpha}\frac{\partial}{\partial \bar{u}_{\dot\beta}}\nonumber\\
&=&-{{\Sigma^{\rho\sigma}}_{\alpha}}^\beta u^\alpha\frac{\partial p^\mu }{\partial u^\beta}\frac{\partial}{\partial p^\mu}-{{\overline\Sigma^{\rho\sigma\,\dot\alpha}}}_{\dot\beta}\bar{u}_{\dot\alpha}\frac{\partial p^\mu }{\partial \bar{u}_{\dot\beta}}\frac{\partial}{\partial p^\mu }\nonumber\\
&&-{{\Sigma^{\rho\sigma}}_{\alpha}}^\beta u^\alpha\frac{\partial \bar{\epsilon}_+^\mu }{\partial u^\beta}\frac{\partial}{\partial \bar{\epsilon}_+^\mu}-{{\Sigma^{\rho\sigma}}_{\alpha}}^\beta u^\alpha\frac{\partial \bar{\epsilon}_-^\mu }{\partial u^\beta}\frac{\partial}{\partial \bar{\epsilon}_-^\mu}\nonumber\\
&&-{{\overline\Sigma^{\rho\sigma\,\dot\alpha}}}_{\dot\beta}\bar{u}_{\dot\alpha}\frac{\partial \bar{\epsilon}_+^\mu }{\partial \bar{u}_{\dot\beta}}\frac{\partial}{\partial \bar{\epsilon}_+^\mu }-{{\overline\Sigma^{\rho\sigma\,\dot\alpha}}}_{\dot\beta}\bar{u}_{\dot\alpha}\frac{\partial \bar{\epsilon}_-^\mu }{\partial \bar{u}_{\dot\beta}}\frac{\partial}{\partial \bar{\epsilon}_-^\mu }\,,
\end{eqnarray}
where the derivatives with respect to the momenta only act on the explicit momentum dependence of the amplitude, not the momentum dependence of the polarization vectors.
To evaluate this we will need the derivatives of the momenta with respect to the spinor-helicity variables
\begin{equation}
\begin{split}
\frac{\partial p^\mu}{\partial \bar{u}_{\dot{\alpha}}}&=-\frac{1}{2}\bar{\sigma}^{\mu \dot{\alpha}\alpha}u_\alpha\,,\\
\frac{\partial p^\mu}{\partial u^\alpha}&=-\frac{1}{2} \sigma^{\mu}_{\alpha \dot{\alpha}}\bar{u}^{\dot{\alpha}}\,,
\end{split}
\end{equation}
as well as the derivatives of the polarization vectors
\begin{equation}
\begin{split}
\frac{\partial \bar{\epsilon}_+^\mu}{\partial \bar{u}_{\dot{\alpha}}}&=\frac{1}{\sqrt{2}}\frac{\bar{\sigma}^{\mu \dot{\alpha}\alpha}n_\alpha}{u^\beta n_\beta}\,,\\
\frac{\partial \bar{\epsilon}_-^\mu}{\partial u^\alpha}&=-\frac{1}{\sqrt{2}}\frac{\sigma^{\mu}_{\alpha \dot{\alpha}}\bar{n}^{\dot{\alpha}}}{\bar{u}_{\dot{\gamma}}\bar{n}^{\dot{\gamma}}}\,,\\
\frac{\partial \bar{\epsilon}_+^\mu}{\partial u^\alpha}&=-\frac{1}{\sqrt{2}}\frac{\bar{u}\bar{\sigma}^{\mu}n}{u^\beta n_\beta}\frac{n_\alpha}{u^\gamma n_\gamma}=-\bar{\epsilon}_+^\mu\frac{n_\alpha}{u^\gamma n_\gamma}\,,\\
\frac{\partial \bar{\epsilon}_-^\mu}{\partial \bar{u}_{\dot{\alpha}}}&=\frac{1}{\sqrt{2}}\frac{u \sigma^{\mu}\bar{n}}{\bar{u}_{\dot{\beta}}\bar{n}^{\dot{\beta}}}\frac{\bar{n}_{\dot{\alpha}}}{\bar{u}_{\dot{\gamma}}\bar{n}^{\dot{\gamma}}}\hskip .3cm =-\bar{\epsilon}_-^\mu\frac{\bar{n}^{\dot{\alpha}}}{\bar{u}_{\dot{\gamma}}\bar{n}^{\dot{\gamma}}}\,.
\end{split}
\end{equation} 
The angular momentum operator then takes the form
\begin{eqnarray}
J^{\rho\sigma}&=&\frac{1}{2} u^\alpha{{\Sigma^{\rho\sigma}}_{\alpha}}^\beta\sigma^{\mu}_{\beta \dot{\alpha}}\bar{u}^{\dot{\alpha}}\frac{\partial}{\partial p^\mu}+\frac{1}{2}\bar{u}_{\dot\alpha}{{\overline\Sigma^{\rho\sigma\,\dot\alpha}}}_{\dot\beta}\bar{\sigma}^{\mu \dot{\beta}\alpha}u_\alpha\frac{\partial}{\partial p^\mu }\nonumber\\
&&\hskip -.3cm+\frac{u^\alpha{{\Sigma^{\rho\sigma}}_{\alpha}}^\beta n_\beta}{u^\gamma n_\gamma}\bar{\epsilon}_+^\mu \frac{\partial}{\partial \bar{\epsilon}_+^\mu}+\frac{1}{\sqrt{2}} \frac{u^\alpha{{\Sigma^{\rho\sigma}}_{\alpha}}^\beta\sigma^{\mu}_{\beta \dot{\alpha}}\bar{n}^{\dot{\alpha}}}{\bar{u}_{\dot{\gamma}}\bar{n}^{\dot{\gamma}}}\frac{\partial}{\partial \bar{\epsilon}_-^\mu}\\
&&\hskip -.3cm-\frac{1}{\sqrt{2}}\frac{\bar{u}_{\dot\alpha}{{\overline\Sigma^{\rho\sigma\,\dot\alpha}}}_{\dot\beta}\bar{\sigma}^{\mu \dot{\beta}\alpha}n_\alpha}{u^\beta n_\beta}\frac{\partial}{\partial \bar{\epsilon}_+^\mu }+\frac{\bar{u}_{\dot\alpha}{{\overline\Sigma^{\rho\sigma\,\dot\alpha}}}_{\dot\beta}\bar{n}^{\dot{\beta}}}{\bar{u}_{\dot{\gamma}}\bar{n}^{\dot{\gamma}}}\bar{\epsilon}_-^\mu\frac{\partial}{\partial \bar{\epsilon}_-^\mu }\,.\nonumber
\end{eqnarray}
To simplify this we can use
\begin{equation}
\begin{split}
\Sigma^{\mu\nu}\sigma^\rho=\frac12\left(\eta^{\mu\rho}\sigma^\nu-\eta^{\nu\rho}\sigma^\mu\right)+\frac{i}{2}\epsilon^{\mu\nu\rho\kappa} \sigma_\kappa\,,\\
\overline{\Sigma}^{\mu\nu}\bar{\sigma}^\rho=\frac12\left(\eta^{\mu\rho}\bar{\sigma}^\nu-\eta^{\nu\rho}\bar{\sigma}^\mu\right)-\frac{i}{2}\epsilon^{\mu\nu\rho\kappa} \bar{\sigma}_\kappa\,,
\end{split}
\end{equation}
where $\epsilon^{0123}=1$, and write it as
\begin{eqnarray}
J^{\rho\sigma}&=&p^\rho\frac{\partial}{\partial p_\sigma}-p^\sigma\frac{\partial}{\partial p_\rho}\nonumber\\
&&\hskip -.3cm+\frac{u^\alpha{{\Sigma^{\rho\sigma}}_{\alpha}}^\beta n_\beta}{u^\gamma n_\gamma}\bar{\epsilon}_+^\mu \frac{\partial}{\partial \bar{\epsilon}_+^\mu}-\frac12\left(\eta^{\rho\mu}\bar{\epsilon}_-^\sigma-\eta^{\sigma\mu}\bar{\epsilon}_-^\rho\right)\frac{\partial}{\partial \bar{\epsilon}_-^\mu}-\frac{i}{2}\epsilon^{\rho\sigma\mu\kappa}\bar{\epsilon}_{-\,\kappa}\frac{\partial}{\partial \bar{\epsilon}_-^\mu}\nonumber\\
&&\hskip -.3cm+\frac{\bar{u}_{\dot\alpha}{{\overline\Sigma^{\rho\sigma\,\dot\alpha}}}_{\dot\beta}\bar{n}^{\dot{\beta}}}{\bar{u}_{\dot{\gamma}}\bar{n}^{\dot{\gamma}}}\bar{\epsilon}_-^\mu\frac{\partial}{\partial \bar{\epsilon}_-^\mu }-\frac12\left(\eta^{\rho\mu}\bar{\epsilon}_+^\sigma-\eta^{\sigma\mu}\bar{\epsilon}_+^\rho\right)\frac{\partial}{\partial \bar{\epsilon}_+^\mu }+\frac{i}{2}\epsilon^{\rho\sigma\mu\kappa}\bar{\epsilon}_{+\,\kappa}\frac{\partial}{\partial \bar{\epsilon}_+^\mu}\,.
\end{eqnarray}

We will ultimately be interested in the soft factors, in which the angular momentum operator always appears in the combination $\bar{\epsilon}_\rho J^{\rho\sigma}q_\sigma$. Let us consider the coefficients of the derivatives with respect to the positive and negative helicity particles separately. After some algebra, one finds that the positive helicity coefficients are related by
\begin{eqnarray}
\bar{\epsilon}_{-\,\rho}\left[ \frac{u_k^\alpha{{\Sigma^{\rho\sigma}}_{\alpha}}^\beta n_\beta}{u_k^\gamma n_\gamma}\bar{\epsilon}_{k\,+}^\mu +\frac{i}{2}\epsilon^{\rho\sigma\mu\kappa}\bar{\epsilon}_{k+\,\kappa} \right] q_\sigma&=&\bar{\epsilon}_{-\,\rho}\left[-\frac12\left(\eta_k^{\rho\mu}\bar{\epsilon}_{k\,+}^\sigma-\eta_k^{\sigma\mu}\bar{\epsilon}_{k\,+}^\rho\right)\right]q_\rho-\frac{E_q+q^3}{E_k+p_k^3}p_k^\mu\nonumber\,,\\
\bar{\epsilon}_{+\,\rho}\left[ \frac{u_k^\alpha{{\Sigma^{\rho\sigma}}_{\alpha}}^\beta n_\beta}{u_k^\gamma n_\gamma}\bar{\epsilon}_{k\,+}^\mu +\frac{i}{2}\epsilon^{\rho\sigma\mu\kappa}\bar{\epsilon}_{k\,+\,\kappa} \right] q_\sigma&=&\bar{\epsilon}_{+\,\rho}\left[-\frac12\left(\eta^{\rho\mu}\bar{\epsilon}_{k\,+}^\sigma-\eta^{\sigma\mu}\bar{\epsilon}_{k\,+}^\rho\right)\right]q_\rho\,.
\end{eqnarray}
and similarly for the coefficients of the derivatives with respect to the negative helicity
\begin{eqnarray}
\bar{\epsilon}_{-\,\rho}\left[\frac{\bar{u}_{k\,\dot\alpha}{{\overline\Sigma^{\rho\sigma\,\dot\alpha}}}_{\dot\beta}\bar{n}^{\dot{\beta}}}{\bar{u}_{k\,\dot{\gamma}}\bar{n}^{\dot{\gamma}}}\bar{\epsilon}_{k\,-}^\mu -\frac{i}{2}\epsilon^{\rho\sigma\mu\kappa}\bar{\epsilon}_{k-\,\kappa} \right] q_\sigma&=&\bar{\epsilon}_{-\,\rho}\left[ -\frac12\left(\eta^{\rho\mu}\bar{\epsilon}_{k\,-}^\sigma-\eta^{\sigma\mu}\bar{\epsilon}_{k\,-}^\rho\right) \right]q_\rho\,,\\
\bar{\epsilon}_{+\,\rho}\left[\frac{\bar{u}_{k\,\dot\alpha}{{\overline\Sigma^{\rho\sigma\,\dot\alpha}}}_{\dot\beta}\bar{n}^{\dot{\beta}}}{\bar{u}_{k\,\dot{\gamma}}\bar{n}^{\dot{\gamma}}}\bar{\epsilon}_{k\,-}^\mu -\frac{i}{2}\epsilon^{\rho\sigma\mu\kappa}\bar{\epsilon}_{k-\,\kappa} \right] q_\sigma&=&\bar{\epsilon}_{+\,\rho}\left[ -\frac12\left(\eta^{\rho\mu}\bar{\epsilon}_{k\,-}^\sigma-\eta^{\sigma\mu}\bar{\epsilon}_{k\,-}^\rho\right) \right]q_\rho-\frac{E_q+q^3}{E_k+p_k^3}p_k^\mu\nonumber\,.
\end{eqnarray}
For a gauge invariant amplitude, we see that
\begin{equation}
p_k^\mu\frac{\partial}{\partial \bar{\epsilon}_{k+}^\mu}\mathcal{M}= 0\,,
\end{equation}
so that in a soft factor the action of the angular momentum operator is equivalent to
\begin{equation}
J^{\rho\sigma}\simeq p^\rho\frac{\partial}{\partial p_\sigma}-p^\sigma\frac{\partial}{\partial p_\rho}+\left(\bar{\epsilon}^\rho\frac{\partial}{\partial \bar{\epsilon}_\sigma}-\bar{\epsilon}^\sigma\frac{\partial}{\partial \bar{\epsilon}_\rho}\right)\,.
\end{equation}

\section{Amplitudes and charge algebra for pions}\label{softpions}

In this appendix we review single- and double-soft pion amplitudes, and provide a dictionary between the standard discussion and notation of \cite{Weinberg:1966gjf,Weinberg:1966kf} and the analysis and notation used in our paper. 

A current that corresponds to a spontaneously broken symmetry has non-trivial matrix elements between 1-particle states that carry the same charges, i.e., pions. As a consequence we can write it as
\begin{equation}
J^{\mu}_{a} = J^{\mu }_{S\,a}+J^{\mu }_{H\,a}\qquad\text{\rm with}\qquad J^{\mu }_{S\,a}=-f\partial^{\mu}\pi_{a}\,,
\end{equation}
where $\pi_{a}$ is the pion field.

In the standard discussion of soft pion theorems the central quantities are the Fourier transforms of matrix elements of time ordered products of these currents. By Lorentz invariance they must be of the form
\begin{multline}\label{eq:Jn}
 \int d^{4}x_1\cdots\int d^{4}x_n\,e^{iq_1x_1}\cdots e^{iq_nx_n}\langle \beta |T( J^{\mu_1}_{a_1}(x_1)\cdots J^{\mu_n}_{a_n}(x_n)) | \alpha \rangle=\\(2\pi)^{4}i\delta^{4}(p_{\beta} + q - p_{\alpha})\mathcal{M}^{\mu_1\dots\mu_n}_{a_1\dots a_n\,\beta\alpha}(q_1,\dots,q_n) \,.
\end{multline}
Soft pion theorems for amplitudes in which $n$ soft pions are emitted can be derived by evaluating the matrix elements in~(\ref{eq:Jn}), or rather its divergence, in two different ways. On the one hand we can evaluate them using current conservation, and on the other hand we can use decomposition of the currents into soft and hard pieces. Following~\cite{Weinberg:1966gjf}, we will denote the time ordered products of the hard parts of the current as
\begin{multline}\label{eq:Jn2}
 \int d^{4}x_1\cdots\int d^{4}x_n\,e^{iq_1x_1}\cdots e^{iq_nx_n}\langle \beta |T( J^{\mu_1}_{H\,a_1}(x_1)\cdots J^{\mu_n}_{H\,a_n}(x_n)) | \alpha \rangle=\\(2\pi)^{4}i\delta^{4}(p_{\beta} + q - p_{\alpha})\mathcal{N}^{\mu_1\dots\mu_n}_{a_1\dots a_1\,\beta\alpha}(q_1,\dots,q_n) \,.
\end{multline}
For a single current we simply have
\begin{equation}
(2\pi)^{4}i\delta^{4}(p_{\beta} + q - p_{\alpha})\mathcal{M}^{\mu}_{a}(q) = \int d^{4}x\,e^{iqx}\langle \beta | J^{\mu}_{a}(x) | \alpha \rangle\,.
\end{equation}
Decomposing the current into its soft and hard piece, we know that this is given by
\begin{equation}\label{currentexpanded}
\mathcal{M}^{\mu}_{a}(q) = -\frac{f q^{\mu}}{q^2}\mathcal{M}_{\beta \pi^{a}, \alpha} + \mathcal{N}^{\mu}_{a\beta \alpha}\,.
\end{equation}
where $\mathcal{M}_{\beta \pi^{a}, \alpha}$ is the Feynman amplitude for a process $\alpha\to\beta$ in which a single pion is emitted.
The current is conserved, and so $q_{\mu}\mathcal{M}^{\mu}_{a} = 0$ implies that
\begin{equation}\label{eq:singlesoft}
\mathcal{M}_{\beta \pi^{a}, \alpha} = \frac{1}{f}q_{\mu}\mathcal{N}^{\mu}_{\beta \alpha}\,.
\end{equation}
So far this is exact. If $\mathcal{N}$ is regular as $q \to 0$, as in the case where the theory consists only of pions and there are no cubic vertices, then the amplitude for the process in which a pion is emitted vanishes in the soft limit.  This is known as ``Adler's zero.''  If the theory contains nucleons, or other fields that have a 3-point interaction with pions, the Fourier transform of the hard part of the current contains poles associated with insertions of the hard part of the current in external nucleon lines. In this case 
\begin{equation}\label{eq:1pionNN}
\lim_{q \to 0}\mathcal{M}_{\beta \pi^{a}, \alpha} = -\frac{1}{f}\sum_{j}\frac{p_j \cdot q}{p_j \cdot q}T^{j}\mathcal{M}_{\beta, \alpha} = -\frac{1}{f}\sum_{j}T^{j}\mathcal{M}_{\beta, \alpha}
\end{equation}
where the generator $T^j$ acts on the $j^{th}$ nucleon, and we see that the emission of a single soft pion is dominated by emission from external lines in the diagram.

To make contact with the notation in the main text, let us also rewrite equation~\ref{eq:singlesoft} as
\begin{equation}\label{eq:1pionJH}
\langle\beta;\pi^a,q|\alpha\rangle=\frac{i}{f}\int d^4 x e^{iqx}\partial_\mu\langle\beta| J^\mu_{H_a}(x)|\alpha\rangle\,.
\end{equation}
As we take $q\to 0$, the integrand becomes a total divergence, and the equation becomes
\begin{equation}
\lim_{q\to 0}\langle\beta;\pi^a,q|\alpha\rangle=\frac{i}{f}\langle\beta|Q_{H\,a}^+-Q_{H\,a}^-|\alpha\rangle\,.
\end{equation}
As written here $Q_{H\,a}^\pm$ are the integral over the hard part of the current over space as $t\to \pm\infty$, but for massless states this is the same as the integrals of $*J_{H a}$ over $\mathscr{I}^\pm$. In the main text we also denote this as
\begin{equation}
\lim_{q\to 0}\langle\beta;\pi^a,q|\alpha\rangle=\frac{i}{f}\langle\beta|[Q_{H\,a},\mathcal{S}]|\alpha\rangle\,,
\end{equation}
which is, of course, equivalent to equation~\ref{eq:1pionNN}. We can go slightly further by formally defining the soft charge
\begin{equation}
Q_{S a}^+=\int_{\mathscr{I}^+} *J_{S}=\frac{i}{2}f\lim_{q\to 0}\int \frac{d^2\hat{q}}{4\pi}\left(a^{out}_a(\mathbf{q})-a_a^{\dagger\,out}(\mathbf{q})\right)\,.
\end{equation}
and similarly for $Q_{S a}^-$. Of course, as usual for spontaneously broken symmetries these charges create states whose norm diverges like the volume, and they are not well-defined operators of the theory. However, since their commutators with local operators are well defined operators, they are still of some use. Making use of crossing symmetry to relate the matrix element for the process in which a pion is absorbed to the matrix element in which it is emitted, we will here use these charges to write the S-matrix element as
\begin{equation}
\lim_{q\to 0}\int \frac{d^2 \hat{q}}{4\pi}\langle\beta;\pi^a,q|\alpha\rangle=-\frac{i}{f}\langle\beta|[Q_{S\,a},\mathcal{S}]|\alpha\rangle\,,
\end{equation}
so that the soft theorem formally simply becomes
\begin{equation}
\langle\beta|[Q_a,\mathcal{S}]|\alpha\rangle=0\,.
\end{equation}
Note that in the case of the BMS symmetry the integral of the amplitude over the angular directions is further weighted by functions $\Psi$ of $T, Y^{A}$ on the 2-sphere, the explicit form of which is given in the text, and involves a sum over graviton helicities as well. 

Our main interest here is  the double-soft pion theorem. In this case, the decomposition into soft and hard pieces implies

\begin{equation}
\mathcal{M}^{\mu \nu}_{ab}(q_1, q_2) = \frac{f^2 q_{1}^{\mu}q_{2}^{\nu}}{q_1^2 q_2^2}\mathcal{M}_{\beta \pi^{a}\pi^{b},\alpha} -\frac{f q_1^{\mu}}{q_1^2}\mathcal{N}^{\nu}_{b\beta \pi^{a}, \alpha} - \frac{f q_2^{\nu}}{q_2^2}\mathcal{N}^{\mu}_{a\beta \pi^{b}, \alpha} + \mathcal{N}^{\mu \nu}_{ab \beta, \alpha}\,.
\end{equation}
We can eliminate the factors $\mathcal{N}^{\mu}_{a\beta \pi^{b}, \alpha}$ with the help of
\begin{equation}
\mathcal{M}^{\mu}_{a\beta \pi^{b},\alpha}(q_1) = -\frac{f q_1^{\mu}}{q_1^2}\mathcal{M}_{\beta \pi^{a} \pi^{b},\alpha} + \mathcal{N}^{\mu}_{a\beta \pi^{b},\alpha}\,,
\end{equation}
and we then have
\begin{equation}
q_{1\mu}q_{2\nu}\mathcal{M}^{\mu \nu}_{ab}(q_1, q_2) = q_{1\mu}q_{2\nu}\mathcal{N}^{\mu \nu}_{ab}(q_1, q_2) - f^2 \mathcal{M}_{\beta \pi^{a}\pi^{b}, \alpha}\,.
\end{equation}
The traditional way to evaluate the left hand side is to note that it corresponds to taking derivatives of the matrix element with two current insertions and evaluating one of the derivatives. Because the currents are conserved, the only non-zero contribution arises when derivatives act on the theta functions associated with the time ordering. We see that the left hand side then becomes
\begin{equation}
\begin{split}
q_{1\mu}q_{2\nu}\int d^{4}x d^{4}y\, &e^{i q_1 x}e^{i q_2 y}\langle \beta | T(J^{\mu}_{a}(x) J^{\nu}_{b}(y)) | \alpha \rangle \\
&= i q_1^{\mu}\int d^{4}x d^{4}y\, e^{i q_1 x} e^{i q_2 y} \langle \beta | \delta(x_0 - y_0)\left[J_{0}^{b}(x), J_{\mu}^{a}(y)\right] | \alpha \rangle\\
&= - q_1^{\mu} f^{abc} \int d^{4}x \, e^{i(q_1 + q_2)x} \langle \beta | J^{\mu}_{c}(x) | \alpha \rangle\\
&= -i f^{abc} q_{1\mu} \mathcal{M}^{\mu}_{c\beta, \alpha}(q_1+q_2)(2\pi)^4 \delta^{4}(p_{\beta} + q_1 + q_2 - p_{\alpha})\,,
\end{split}
\end{equation}
so that
\begin{equation}
\begin{split}
f^2 \mathcal{M}_{\beta \pi^{a}\pi^{b}, \alpha}&=if^{abc}q_{1\mu} \mathcal{M}^{\mu}_{c\beta, \alpha}(q_1+q_2)+q_{1\mu}q_{2\nu}\mathcal{N}^{\mu \nu}_{ab\beta,\alpha}(q_1, q_2)\\
&= -if^{abc}q_{2\mu} \mathcal{M}^{\mu}_{c\beta, \alpha}(q_1+q_2)+q_{1\mu}q_{2\nu}\mathcal{N}^{\mu \nu}_{ab\beta,\alpha}(q_1, q_2)\,,
\end{split}
\end{equation}
where we can use the first or the second expression without loss of generality.  So far this is exact, and we see that the matrix element for two soft pions knows about the current commutator~\cite{Weinberg:1966gjf}. Using \eqref{currentexpanded} to expand the current $\mathcal{M}^{\mu}_{c\beta,\alpha}$, we have
\begin{equation}
\begin{split}\label{commutatoralternate}
f^2 \mathcal{M}_{\beta \pi^{a}\pi^{b}, \alpha}&= -\frac{if}{2} f^{abc}\mathcal{M}_{\beta \pi^c, \alpha}(q_1+q_2) + i f^{abc}q_{1\mu}\mathcal{N}^{\mu}_{c \beta \alpha}(q_1 + q_2)+q_{1\mu}q_{2\nu}\mathcal{N}^{\mu \nu}_{ab\beta,\alpha}(q_1, q_2)\\
&= \frac{if}{2} f^{abc}\mathcal{M}_{\beta \pi^c, \alpha}(q_1+q_2) - i f^{abc}q_{2\mu}\mathcal{N}^{\mu}_{c \beta \alpha}(q_1 + q_2)+q_{1\mu}q_{2\nu}\mathcal{N}^{\mu \nu}_{ab\beta,\alpha}(q_1, q_2)\,.
\end{split}
\end{equation}
Here we have taken $q_1$ and $q_2$ to be on-shell.  As both $q_1$ and $q_2$ are taken to zero, this will be dominated by diagrams in which the current is inserted in external lines, which shows that the amplitude in which two soft pions are emitted is given in terms of the amplitude for the underlying hard process with external lines rotated by the commutator of the generators associated with the soft pions.  Taking the antisymmetric double consecutive soft limit $\lim_{[q_1 \to 0}\lim_{q_2 \to 0]}$ of both sides of \eqref{commutatoralternate}, we find
\begin{equation}
\begin{split}
\lim_{[q_1 \to 0}\lim_{q_2 \to 0]}f^2 \mathcal{M}_{\beta \pi^{a}\pi^{b}, \alpha} &= -if f^{abc}\lim_{q \to 0}\mathcal{M}_{\beta \pi^c, \alpha}(q) + 2i f^{abc}\lim_{q \to 0}q_{\mu}\mathcal{N}^{\mu}_{c\beta\alpha}(q)\\ & \qquad + \lim_{[q_1 \to 0}\lim_{q_2 \to 0]}q_{1\mu}q_{2\nu}\mathcal{N}^{\mu \nu}_{ab\beta,\alpha}(q_1, q_2)\\
& = i f^{abc}\lim_{q \to 0}q_{\mu}\mathcal{N}^{\mu}_{c\beta\alpha}(q) + \lim_{[q_1 \to 0}\lim_{q_2 \to 0]}q_{1\mu}q_{2\nu}\mathcal{N}^{\mu \nu}_{ab\beta,\alpha}(q_1, q_2)\,,
\end{split}
\end{equation}
where we have used current conservation in going from the first to the second equality.  Note that there is an order of limits issue here, and had we kept the soft momenta off shell, so that $q_1^2, q_2^2 \neq 0$ and put them on-shell only after taking the soft limits, the result would be
\begin{equation}
\begin{split}
\lim_{[q_1 \to 0}\lim_{q_2 \to 0]}f^2 \mathcal{M}_{\beta \pi^{a}\pi^{b}, \alpha} &= -2if f^{abc}\lim_{q \to 0}\mathcal{M}_{\beta \pi^c, \alpha}(q) + 2i f^{abc}\lim_{q \to 0}q_{\mu}\mathcal{N}^{\mu}_{c\beta\alpha}(q)\\ & \qquad + \lim_{[q_1 \to 0}\lim_{q_2 \to 0]}q_{1\mu}q_{2\nu}\mathcal{N}^{\mu \nu}_{ab\beta,\alpha}(q_1, q_2)\\
& = \lim_{[q_1 \to 0}\lim_{q_2 \to 0]}q_{1\mu}q_{2\nu}\mathcal{N}^{\mu \nu}_{ab\beta,\alpha}(q_1, q_2)\,.
\end{split}
\end{equation}

In symmetric spaces, if the generators $T^{a}$ and $T^{b}$ correspond to broken symmetries, their commutators $f^{abc}$ are only nonzero with unbroken generators $T^{c}$.  In this case we can replace $\mathcal{M}^{\mu}_{c\beta,\alpha}$ with $\mathcal{N}^{\mu}_{c\beta, \alpha}$, and using the pion - nucleon vertex from before, we can find 
\begin{equation}\label{twosoftpions}
\mathcal{M}_{\beta \pi^{a}\pi^{b}, \alpha} = \frac{1}{2f^2}f^{abc}\sum_{j}\frac{p_{j} \cdot (q_1 - q_2)}{p_j \cdot (q_1 + q_2)}T_{c}\mathcal{M}_{\beta,\alpha} - \frac{1}{f^2}\sum_{i,j}\frac{1}{2}\left\{(T^{a})_{i}(T^{b})_{j}\right\}\mathcal{M}_{\beta, \alpha}
\end{equation}
using Feynman diagrams.  (See also \cite{ArkaniHamed:2008gz}.)  
The momentum prefactor means that the limit depends on the order in which the soft momenta are taken to zero, and the antisymmetrized consecutive double-soft limit picks out the commutator.

For a non-symmetric space, however, the commutator can contain a broken generator,
\begin{equation}
\begin{split}
\mathcal{M}_{\beta \pi^{a}\pi^{b}, \alpha}(q_1,q_2) &= -\frac{i}{2f}f^{abc}\mathcal{M}_{\beta \pi^{c},\alpha} + \frac{i}{f^2}f^{abc}q_{1\mu}\mathcal{N}^{\mu}_{c \beta,\alpha}(q_1 + q_2) \\
&\qquad - \frac{1}{f^2}q_{1\mu}q_{2\nu}\mathcal{N}^{\mu \nu}_{ab\beta, \alpha}(q_1,q_2)\,,
\end{split}
\end{equation}
which contributes an additional piece
\begin{equation}
\begin{split}
\mathcal{M}_{\beta \pi^{a}\pi^{b}, \alpha} &= - \frac{1}{2f^2}f^{abc}\sum_{j}T_{c}\mathcal{M}_{\beta, \alpha} + \frac{1}{2f^2}f^{abc}\sum_{j}\frac{p_{j} \cdot (q_1 - q_2)}{p_j \cdot (q_1 + q_2)}T_{c}\mathcal{M}_{\beta,\alpha}\\ & \qquad - \frac{1}{f^2}\sum_{i,j}\frac{1}{2}\left\{(T^{a})_{i}(T^{b})_{j}\right\}\mathcal{M}_{\beta, \alpha}
\end{split}
\end{equation}
up to terms arising from collinear divergences.  Taking the antisymmetric double soft consecutive limit, we find that the first and second terms separately know about the commutator, and will cancel.

To relate this to the notation in the main text, we will evaluate the double-soft limit of this expression differently, just like we evaluated equation~\ref{eq:1pionJH}. If we first take the limit $q_1\to 0$ and then $q_2\to 0$, we find 
\begin{eqnarray}
&&\hskip -1.5cm-\lim_{q_2\to 0}\lim_{q_1\to 0}\int d^{4}x d^{4}y e^{i q_1 x}e^{i q_2 y}\partial_\mu\partial_\nu \langle \beta | T(J^{\mu}_{a}(x) J^{\nu}_{b}(y)) | \alpha \rangle \nonumber\\
&&=-\lim_{q_2\to 0}\int d^{4}y e^{i q_2 y}\partial_\nu \langle \beta | Q^+_{a } J^{\nu}_{b}(y)-J^{\nu}_{b}(y)Q^-_{a }| \alpha \rangle \nonumber\\
&&=-\langle \beta | Q^+_{a } Q^+_{b}-Q^+_{a } Q^-_{b}-Q^+_{b}Q^-_{a }+Q^-_{b}Q^-_{a }| \alpha \rangle\,.
\end{eqnarray}
Taking the limits in the opposite order, we see that the anti-symmetric consecutive double-soft limit of this expression is simply
\begin{equation}
\lim_{[q_2\to 0}\lim_{q_1\to 0]}\int d^{4}x d^{4}y e^{i q_1 x}e^{i q_2 y}\partial_\mu\partial_\nu \langle \beta | T(J^{\mu}_{a}(x) J^{\nu}_{b}(y)) | \alpha \rangle=\langle \beta | [[Q_{a }, Q_{b}],\mathcal{S}]| \alpha \rangle\,.
\end{equation}
The Fourier transform of the divergence of the time ordered product of the hard parts of the current can be evaluated in the same way, so that the consecutive double-soft limit of the S-matrix element is given by
\begin{equation}
\begin{split}
\lim_{[q_2\to 0}\lim_{q_1\to 0]}f^2\int \frac{d^2 \hat{q_1}}{4\pi} \frac{d^2 \hat{q_2}}{4\pi}\langle\beta ;&\pi^{a},q_1,\pi^{b},q_2| \alpha\rangle \\ &=\langle \beta | [[Q_{S\,a }, Q_{S\,b}]+[Q_{S\,a }, Q_{H\,b}]+[Q_{H\,a }, Q_{S\,b}],\mathcal{S}]| \alpha \rangle\,.
\end{split} 
\end{equation}
The first term can at most contribute a Schwinger term, but for pions this contribution vanishes on-shell.  The consecutive anti-symmetrized double-soft limit then simplifies to
\begin{equation}\label{doublesoftmastereqn1}
\lim_{[q_2\to 0}\lim_{q_1\to 0]}f^2\int \frac{d^2 \hat{q_1}}{4\pi} \frac{d^2 \hat{q_2}}{4\pi}\langle\beta ;\pi^{a},q_1,\pi^{b},q_2| \alpha\rangle=\langle \beta | [[Q_{S\,a }, Q_{H\,b}]+[Q_{H\,a }, Q_{S\,b}],\mathcal{S}]| \alpha \rangle\,.
\end{equation}

Applying the same soft limits to \eqref{commutatoralternate}, we have
\begin{equation}\label{doublesoftmastereqn2}
\lim_{[q_2\to 0}\lim_{q_1\to 0]}f^2\int \frac{d^2 \hat{q_1}}{4\pi} \frac{d^2 \hat{q_2}}{4\pi}\langle\beta ;\pi^{a},q_1,\pi^{b},q_2| \alpha\rangle=\langle \beta | [iQ_{[a,b]S} + 2iQ_{[a,b]H} - [Q_{Ha},Q_{Hb}],\mathcal{S}]| \alpha \rangle\,,
\end{equation}
and equating the two expressions \eqref{doublesoftmastereqn1} and \eqref{doublesoftmastereqn2} and comparing the soft and hard parts of the charges, we have that
\begin{equation}
\begin{split}
\langle \beta | \left[ \left[Q_{Ha},Q_{Sb}\right]_{S} - \left[Q_{Hb},Q_{Sa}\right]_{S}, \mathcal{S} \right] | \alpha \rangle & = i\langle \beta | [Q_{[a,b]S},\mathcal{S}]| \alpha \rangle\,,\\
\langle \beta | \left[ \left[Q_{Ha},Q_{Sb}\right]_{H} - \left[Q_{Hb},Q_{Sa}\right]_{H}, \mathcal{S} \right] | \alpha \rangle & = \langle \beta | [2iQ_{[a,b]H} - [Q_{Ha},Q_{Hb}],\mathcal{S}]| \alpha \rangle\\
&= i\langle \beta | [Q_{[a,b]H},\mathcal{S}]| \alpha \rangle
\end{split}
\end{equation}
where the charge algebra for the hard operators is guaranteed by considering their action on other operators.

The reader may be puzzled why the charge algebra is realized by the commutator of $\left[Q_{Ha},Q_{Sb}\right] - \left[Q_{Hb},Q_{Sa}\right]$ with the S-matrix, instead of by the commutator $\left[\left[ Q_{a}, Q_{b}\right], \mathcal{S}\right]$.  We can repeat the derivation above while keeping the soft momenta $q_1, q_2$ off-shell and setting them on shell only at the end.  In this case, we indeed recover the result $\langle \beta | \left[\left[ Q_a, Q_b \right],\mathcal{S}\right] | \alpha \rangle = i f^{abc} \langle \beta | \left[Q_{c}, \mathcal{S}\right] | \alpha \rangle$.  Since the charges are formally undefined for spontaneously broken symmetries, it is perhaps not surprising that there are order of limits issues when computing their commutator.  We work with scattering amplitudes involving physical on-shell gravitons in the main text, and therefore it is $\left[Q_{Ha},Q_{Sb}\right] - \left[Q_{Hb},Q_{Sa}\right]$ that contains the commutator.  Had we tried working with the off-shell amplitude instead, calculating the antisymmetrized double soft limit would involve the subtraction of two divergent quantities.

If the coset is a symmetric space, the inversion symmetry guarantees that the broken generators consist of an odd number of creation- and annihilation operators. As a consequence their commutators contain an even number of creation and annihilation operators and the commutator does not contain a soft piece. This implies that the double-soft pion amplitude is related to the amplitude of the underlying hard process with external lines rotated by an infinitesimal amount. 

For cosets that are not symmetric spaces, as in the BMS case, the commutators of the soft and hard parts of the charges will contain contributions that are soft and create a single soft pion, as well as hard parts that rotate the external lines. The information about the charge algebra is entirely contained in the infinitesimal rotations of the external lines of the underlying hard process, but to extract it one must then take appropriate linear combinations of single and double-soft limits.

The case of two soft pions makes it clear that the backreaction terms in the Dirac brackets are not surprising -- they appear simply because the space is not symmetric and the Goldstone bosons are charged under the broken symmetry.  This is easy to see from the Noether current, since the currents both create the soft pion and perform the linear rotation on the hard modes.  The extension is absent in the case of soft pions.  

 Working with the currents is more straightforward from the perspective of quantum field theory for several reasons.  Firstly, the charges do not, strictly speaking, exist when the symmetries are broken, since their matrix elements with physical states are not always normalizable (though the matrix elements of their commutators with local operators are).  Second, the matrix element with multiple currents may have Schwinger terms as two insertion points approach one another.  These correspond to derivatives acting on delta functions in the current algebra and will disappear when we work with the integrated charges.  In typical field theory examples the Schwinger terms are canceled by the seagull diagrams.  We do not have a general proof that these terms always cancel in linearized gravity, but if there are uncanceled terms present they could be analyzed in a diagrammatic calculation of two local current insertions.  

 
\section{Single and double-soft gluons}\label{YangMills}

Although the focus of the present work is on the structure of the BMS charges, the same formalism applies to other asymptotic theories as well.  The reader may prefer to see the calculation for soft gluons and asymptotic Yang-Mills charges as a warm-up, before wading through the heavy algebra of Section 4.  The connection between the single-soft gluon theorems and asymptotic gauge charges is derived in \cite{Strominger:2013lka, He:2014cra}.  The asymptotic Yang-Mills charge at future null infinity is given by
\begin{equation}
\begin{split}
Q &= \frac{1}{e^2}\int_{\mathscr{I}^{+}_{\pm}} d^2 z \, \gamma_{z\bar{z}}\bar{\epsilon} F_{ru}\\
&= \frac{1}{e^2}\int_{\mathscr{I}^{+}} d^2 z du\, \bar{\epsilon}(z,\bar{z})\Big[\partial_{u}(\partial_{z}A_{\bar{z}}+\partial_{\bar{z}}A_{z}) + e^2 \gamma_{z\bar{z}}j_u]\,.
\end{split}
\end{equation}
Here $e$ is the Yang-Mills charge, and $\epsilon(z, \bar{z})$ is an arbitrary test function.  Following the discussion in \cite{He:2014cra}, we work in retarded radial gauge, and in the second line we have used the Maxwell equations $\nabla^{\mu}F_{\mu \nu} = e^2j_{\nu}$, rescaled by an overall factor of $r^2$ so the integral over the sphere will be finite.

Expressing the asymptotic field in terms of creation and annihilation operators and using the stationary phase approximation, we have
\begin{equation}
A_{z} = -\frac{i}{8\pi^2}\frac{\sqrt{2}e}{(1+z \bar{z})}\int_{0}^{\infty} d\omega_q \, \omega_q \Big[a_{+}(\omega_q \hat{x})e^{-i\omega_q u}+ a_{-}(\omega_q \hat{x})^{\dagger}e^{+i\omega_q u}\Big]
\end{equation}
and similarly for $A_{\bar{z}}$.  The soft charge operator insertion is then
\begin{equation}
\begin{split}
\langle out | &\left[Q_{S}, \mathcal{S}\right]| in \rangle = \\
&-\frac{\sqrt{2}}{4\pi e}\lim_{\omega \to 0}\int d^2 z\, \epsilon(z,\bar{z})\Bigg[\partial_{\bar{z}}\left(\frac{1}{(1+z\bar{z})}\omega \langle out|a_{+}(\omega \hat{x})\mathcal{S}|in\rangle\right)+\\
&\qquad + \partial_{z}\left(\frac{1}{(1+z\bar{z})}\omega \langle out|a_{-}(\omega \hat{x})\mathcal{S}|in\rangle\right)\Bigg]\,.
\end{split}
\end{equation}
Applying the soft gluon theorem,
\begin{equation}
\begin{split}
\lim_{\omega \to 0}\bar{\epsilon}^{\mu}\mathcal{M}_{\mu}(q,&a; p_1,i_1; \cdots ; p_n, i_n) \\ &= -\sum_k \frac{e (p_k \cdot \bar{\epsilon}) T_{a}^{i_k j_k}}{(p_k \cdot q)}\mathcal{M}(p_1, i_1 ; \cdots; p_k, j_k; \cdots ; p_n, i_n)\,,
\end{split}
\end{equation}
where all the momenta are taken to be outgoing, and we assume that the hard particles transform in the fundamental representation.  In holomorphic coordinates, the soft factor becomes
\begin{equation}
S^{(0)}(q) = -e \frac{(p_k \cdot \bar{\epsilon}^{+})}{(p_k \cdot q)}T_{a} = -\frac{e}{\sqrt{2}\omega}\frac{(1+ z\bar{z})}{(z-z_k)}T_{a}\,,
\end{equation}
and integrating by parts, we have
\begin{equation}
\langle out | \left[ Q_{S}, \mathcal{S} \right] | in \rangle = \sum_k \epsilon(z_k)T_{a}\langle out | \mathcal{S} | in \rangle = -\langle out | \left[ Q_{H}, \mathcal{S} \right] | in \rangle\,,
\end{equation}
where the symmetry generator $T_a$ acts on the $k^{th}$ particle and $a$ is the color index.  Note that similar to the case of supertranslations in gravity, keeping both helicities was important for the factors of two to come out correctly.  In the case of QED, the factor $T_{a}^{i_{k}j_{k}}$ is replaced by $Q_k$, where $e Q_k$ is the charge of the $k^{th}$ particle.

We can also study the charge algebra, by checking the expressions
\begin{equation}
\begin{split}
&\langle out | \left[(\left[Q_{1S}, Q_{2S}\right]+\left[Q_{1H},Q_{2S}\right]_{S}-\left[Q_{2H},Q_{1S}\right]_{S}), \mathcal{S}\right] | in \rangle\\
&\qquad = i\langle out | \left[ Q_{[1,2]S},\mathcal{S}\right] | in \rangle\,,\\
&\langle out | \left[(\left[Q_{1H}, Q_{2H}\right]_{H} - \left[Q_{2H}, Q_{1S}\right]_{H}), \mathcal{S}\right] | in \rangle\\
&\qquad = i\langle out | \left[ Q_{[1,2]H},\mathcal{S}\right] | in \rangle\,.
\end{split}
\end{equation}
The terms $\left[Q_{1S}, Q_{2S}\right]$ will vanish, and the rest are contained in the antisymmetrized consecutive limit of the double-soft amplitude.  The antisymmetric double-soft factor for Yang-Mills is
\begin{equation}
\begin{split}
&S^{(0)}(q_1)S^{(0)}(q_2) - S^{(0)}(q_2)S^{(0)}(q_1) \\&= ie^{2} \frac{(p_k \cdot \bar{\epsilon}_1)}{(p_k \cdot q_1)}\frac{(p_k \cdot \bar{\epsilon}_2)}{(p_k \cdot q_2)}f^{a_1 a_2 c}T_{c} \\& \qquad - i e^2 \frac{(q_2 \cdot \bar{\epsilon}_1)}{(q_1 \cdot q_2)}\frac{(p_k \cdot \bar{\epsilon}_2)}{(p_k \cdot q_2)}f^{a_1 a_2 c}T_{c} - i e^2 \frac{(q_1 \cdot \bar{\epsilon}_2)}{(q_1 \cdot q_2)}\frac{(p_k \cdot \bar{\epsilon}_1)}{(p_k \cdot q_1)}f^{a_1 a_2 c}T_{c}
\end{split}
\end{equation} 
The first term will be associated with $(\left[Q_{1H}, Q_{2S}\right]_{H} + \left[Q_{1S}, Q_{2H}\right]_{H}$ and the last two terms will be associated with $(\left[Q_{1H}, Q_{2S}\right]_{S} + \left[Q_{1S}, Q_{2H}\right]_{S}$.  Starting with the $(\left[Q_{1H}, Q_{2S}\right]_{H} + \left[Q_{1S}, Q_{2H}\right]_{H}$ terms, the amplitude becomes
\begin{equation}\label{doublesoftgluons}
\begin{split}
&\langle out | \left[(\left[Q_{1H}, Q_{2S}\right]_{H} + \left[Q_{1S}, Q_{2H}\right]_{H}), \mathcal{S}\right] | in \rangle \\&= \frac{i}{16\pi^2}\int d^2 z_1 d^2 z_2 \, \partial_{\bar{z}_1}\epsilon_1\partial_{\bar{z}_2}\epsilon_2 \frac{1}{(z_1 - z_k)(z_2 - z_k)}f^{abc}T_{c}\langle out | \mathcal{S} | in \rangle + \cdots\\
&= i\sum_k \epsilon_1(z_k)\epsilon_2 (z_k)f^{abc}T_{c}\langle out | \mathcal{S} | in \rangle\,\\
&= i\langle out | \left[Q_{[1,2]H}, \mathcal{S}\right] | in \rangle
\end{split}
\end{equation}
where the symmetry generator acts on the index of the $k^{th}$ particle, and the first line on the right hand side contains a sum over helicities, of which we have written out only the $(1_{+}2_{+})$ term.
This is the same as a single asymptotic gauge transformation with parameter $\epsilon_1 \epsilon_2$, and charge given by the commutator.  Note that this was much simpler than for gravity because the charge algebra is already reflected in the leading order soft factors.  

For the $(\left[Q_{1H}, Q_{2S}\right]_{S} + \left[Q_{1S}, Q_{2H}\right]_{S}$ terms, we have 
\begin{equation}\label{doublesoftgluons2}
\begin{split}
&\langle out | \left[(\left[Q_{1H}, Q_{2S}\right]_{S} + \left[Q_{1S}, Q_{2H}\right]_{S}), \mathcal{S}\right] | in \rangle \\&= \frac{1}{16\pi^2}\int d^2 z_1 d^2 z_2 \, \partial_{\bar{z}_1}\epsilon_1\partial_{\bar{z}_2}\epsilon_2 \frac{1}{(z_1 - z_2)(z_2 - z_k)}f^{abc}T_{c}\langle out | \mathcal{S} | in \rangle + \cdots\\
&= \frac{i}{8\pi}\sum_k \epsilon_{1}(z_1)\partial_{\bar{z}_1}\epsilon_2(z_1) \frac{1}{(z_1 - z_k)} + \cdots\\
&= -i\sum_k \epsilon_1(z_k)\epsilon_2 (z_k)f^{abc}T_{c}\langle out | \mathcal{S} | in \rangle\,\\
&= i\langle out | \left[Q_{[1,2]S}, \mathcal{S}\right] | in \rangle\,,
\end{split}
\end{equation}
where in going from the second to the third line we have taken the sum over helicities and collected the terms proportional to $\partial_{\bar{z}_1}(\epsilon_1 \epsilon_2)$ and $\partial_{z_1}(\epsilon_1 \epsilon_2)$.

We should emphasize that this prescription is different from that of \cite{He:2015zea}: we are taking the antisymmetrized consecutive double-soft limit of the soft gravitons instead of sending the soft momenta of gravitons with one helicity to zero first, and our definition of the charge contains an integral over local currents of both helicities.


\section{Double-soft photons}

The calculation in the previous Appendix also applies to photons.  Since QED is an abelian gauge theory, the commutator of two soft photon charges vanishes at leading order.  We can calculate the subleading piece from the contact terms,
\begin{equation}
\lim_{\omega_1 \to 0}\lim_{\omega_2 \to 0}\bar{\epsilon}_{1}^{\mu}\bar{\epsilon}_{2}^{\nu}\mathcal{M}_{\mu \nu}(q_1 ; q_2; p_1, \cdots p_n)=\Bigg[S^{(1)}(q_1)\left\{S^{(0)}(q_2)\right\} - S^{(1)}(q_2)\left\{S^{(0)}(q_1)\right\}\Bigg]\mathcal{M}\,,
\end{equation}
which corresponds to a commutator between the charge operator and a dipole charge operator of the kind described in \cite{Lysov:2014csa, Campiglia:2016hvg}. The dipole operator is built out of the subleading factor $S^{(1)}(q) = -\sum_k \frac{e Q_k \bar{\epsilon}_{\mu}q_{\nu}J_{k}^{\mu \nu}}{(p_k \cdot q)}$.  Since this can receive non-universal corrections due to the anomalous magnetic moment, the resulting charge commutator will be sensitive to quantum corrections; however, we are free to perform the calculation and see the result at tree level.  

The subleading soft charge at future null infinity is given by 
\begin{equation}
\begin{split}
Q &= \frac{1}{e^2}\int_{\mathscr{I}^{+}_{\pm}} d^2 z \, \gamma_{z\bar{z}}D_{A}Y^{A} A_{u}\\&= \frac{1}{e^2}\int_{\mathscr{I}^{+}} d^2 z du \,uD_{A}Y^{A}\Big[\partial_{u}(\partial_{z}A_{\bar{z}}+\partial_{\bar{z}}A_{z}) + e^2 \gamma_{z\bar{z}}j_u]\,,
\end{split}
\end{equation}
and the corresponding operator insertion (at tree level and for charged scalar hard operators) is 
\begin{equation}
\begin{split}
\langle &in | \left[Q_S, \mathcal{S}\right] | in \rangle =\\
&-\frac{i\sqrt{2}}{4\pi e}\lim_{\omega \to 0}(1+\omega \partial_{\omega})\int d^2 z\, \Bigg[D_{\bar{z}}Y^{\bar{z}}\partial_{\bar{z}}\left(\frac{1}{(1+z\bar{z})}\langle out|a_{+}(\omega \hat{x})\mathcal{S}|in\rangle\right)+\\
&\qquad + D_{z}Y^{z}\partial_{z}\left(\frac{1}{(1+z\bar{z})}\langle out|a_{-}(\omega \hat{x})\mathcal{S}|in\rangle\right)\Bigg]\\
&=\left(\frac{i\sqrt{2}}{4\pi}\right)\int d^2 z \, \Bigg[D_{\bar{z}}Y^{\bar{z}}\\
\times &\sum_k Q_k \left(\frac{(1+ \bar{z}z_k)}{\sqrt{2}(z - z_k)(1+z \bar{z})}\partial_{E_k} + \frac{(\bar{z} - \bar{z}_k)(1+z_k \bar{z}_k)}{\sqrt{2}(z - z_k)(1+ z \bar{z})}E_{k}^{-1} \partial_{\bar{z}_k}\right) + h.c. \Bigg]\\
& \times \langle out | \mathcal{S} | in \rangle\,,
\end{split}
\end{equation}
which integrates by parts to 
\begin{equation}
\langle in | \left[Q_S, \mathcal{S}\right] | in \rangle = -i \sum_k Q_k \left(D_{A}Y^{A} \partial_{E_k} - \frac{1}{E_k}Y^{A}\partial_{A} \right)\langle out | \mathcal{S} | in \rangle\,.
\end{equation}

To calculate the charge commutator, we need only to consider the terms
\begin{equation}
\langle out | \left[ (\left[Q_{1H}, Q_{2S}\right]_{H} + \left[Q_{1S}, Q_{2H}\right]_{H}), \mathcal{S}\right] | in \rangle
\end{equation}
and the terms $\left[Q_{H},Q_{S}\right]_{S}$ will be absent at tree level because the photon is not itself charged, which means that the combination $\left[Q_{H}, Q_{S}\right]$ does not commute with the S-matrix.  Focusing first on the terms proportional to $1/q_2,$ the soft factor is given by the contact terms
\begin{equation}
\begin{split}
S^{(1)}(q_1)\left\{S^{(0)}(q_2)\right\} &= \sum_{k}e^2 Q_{k}^{2}\Bigg[\frac{(p_k \cdot \bar{\epsilon}_1)}{(p_k \cdot q_1)}\left(\frac{(q_1 \cdot \bar{\epsilon}_2)}{(p_k \cdot q_2)} - \frac{(p_k \cdot \bar{\epsilon}_2)}{(p_k \cdot q_2)^2}(q_1 \cdot q_2)\right) \\
& \qquad - \left(\frac{(\bar{\epsilon}_1 \cdot \bar{\epsilon}_2)}{(p_k \cdot q_2)} - \frac{(p_k \cdot \bar{\epsilon}_2)}{(p_k \cdot q_2)^2}(\bar{\epsilon}_1 \cdot q_2)\right) \Bigg]\\
&= 0 \qquad (1_{+}2_{+})\\
&= \sum_k \frac{e^2 Q_k^2(\bar{z}_1 - \bar{z}_k)(1+z_k \bar{z}_k)}{2(1+z_1 \bar{z}_1)(z_1 - z_k)(\bar{z}_2 - \bar{z}_k)^2}\qquad (1_{+}2_{-})
\end{split}
\end{equation}
and calculating the $(1_{+}2_{-})$ terms first, we have
\begin{equation}
\begin{split}
&\hskip -3.4 cm \langle out | \left[ (\left[Q_{1H}, Q_{2S}\right]_{H} + \left[Q_{1S}, Q_{2H}\right]_{H}), \mathcal{S}\right] | in \rangle \supset\\
\left(\frac{i}{4\pi^2}\right)\int &d^2 z_1 d^2 z_2 \, \partial_{z_2}\bar{\epsilon}_2 D_{\bar{z}_1}^{2}Y_1^{\bar{z}_2}\sum_k \frac{Q_k^2 (\bar{z}_1 - \bar{z}_k)(1+z_k \bar{z}_k)}{2(1+z_1 \bar{z}_1)(z_1 - z_k)(\bar{z}_2 - \bar{z}_k)^2}\\
&= -\left(\frac{i}{4\pi^2}\right)\int d^2 z_1 \, D_{\bar{z}_1}^{2}Y_1^{\bar{z}_2} \sum_k Q_k^2 \partial_{\bar{z}_k}\bar{\epsilon}_2 \frac{\pi (\bar{z}_1 - \bar{z}_k)(1+z_k \bar{z}_k)}{(1+z_1 \bar{z}_1)(z_1 - z_k)}\\
&= -\frac{i}{2} \sum_k Q_k^2 Y_{1}^{\bar{z}_k}\partial_{\bar{z}_k}\bar{\epsilon}_2\,.
\end{split}
\end{equation}
Adding back the other helicities and antisymmetrizing, we find that the commutator of the dipole and monopole charges generates the linear shift
\begin{equation}
i\langle out | \left[ Q_{\left[1,2\right]H}, \mathcal{S}\right] | in \rangle =  -\frac{i}{2} \sum_k Q_k^2 \left(Y_1^{A}\partial_{A}\bar{\epsilon}_2 - Y_2^{A}\partial_{A}\bar{\epsilon}_1\right)\,.
\end{equation}
This is not fixed by symmetry, since the dipole operator can receive quantum corrections, and is therefore sensitive to the full dynamics of the theory.

\end{appendices}

\vspace{1cm}

\bibliographystyle{JHEP}
\renewcommand{\refname}{Bibliography}
\addcontentsline{toc}{section}{Bibliography}
\providecommand{\href}[2]{#2}\begingroup\raggedright

\end{document}